%% file: achirateCapa4.tex
\documentclass[12pt,journal,onecolumn]{IEEEtran}
\usepackage{amsmath,amssymb, dsfont,bbm,tabularx,marvosym}
\usepackage{epsfig,latexsym,amssymb,amsmath,graphics,color}
\usepackage{graphicx,subfigure}
\usepackage{extarrows}

\usepackage{float}
\usepackage[linesnumbered,ruled]{algorithm2e}
\usepackage{epsfig,latexsym,amssymb,amsmath,graphics}
\usepackage{graphicx,morefloats}
\usepackage[linesnumbered,ruled]{algorithm2e}
\usepackage{cite,color}
\usepackage{url,epstopdf,multirow}

\newtheorem{theorem}{Theorem}
\newtheorem{lemma}{Lemma}
\newtheorem{remark}{Remark}
\newtheorem{proposition}{Proposition}

\input{pream_bm}

\renewcommand{\baselinestretch}{1}

\def\IR{\mathbb R}

\renewcommand{\baselinestretch}{1.6}

\title{On the Achievable Rate and Capacity for a Sample-based Practical Photon-counting  Receiver}

\author{Zhimeng Jiang, Chen Gong, Guanchu Wang, and Zhengyuan Xu
		\thanks{This work was supported by Key Program of National Natural Science Foundation of China (Grant No. 61631018) and Key Research Program of Frontier Sciences of CAS (Grant No. QYZDY-SSW-JSC003). This work has been partially accepted by Optical Wireless Communication Workshop. IEEE ICC 2019 \cite{jiang2019achievable}.}
		\thanks{The authors are with Key Laboratory of Wireless-Optical Communications, Chinese Academy of Sciences, University of Science and Technology of China, Hefei, Anhui 230027, China. 
			Email: \{zhimengj, hegsns\}@mail.ustc.edu.cn, \{cgong821, xuzy\}@ustc.edu.cn.}}
\date{}

\begin{document}
\maketitle{}

\renewcommand{\baselinestretch}{1.3}
\begin{abstract}
	We investigate the achievable rate and capacity of a non-perfect photon-counting receiver. For the case of long symbol duration, the achievable rate under on-off keying modulation is investigated based on Kullback-Leibler (KL) divergence and Chernoff $\alpha$-divergence. We prove the tightness of the derived bounds for large peak power with zero background radiation with exponential convergence rate, and for low peak power of order two convergence rate. For large peak power with fixed background radiation and low background radiation with fixed peak power, the proposed bound gap is a small positive value for low background radiation and large peak power, respectively. Moreover, we propose an approximation on the achievable rate in the low background radiation and long symbol duration regime, which is more accurate compared with the derived upper and lower bounds in the medium signal to noise ratio (SNR) regime. For the symbol duration that can be sufficiently small, the capacity and the optimal duty cycle are is investigated. We show that the capacity approaches that of continuous Poisson capacity as $T_s=\tau\to0$. The asymptotic capacity is analyzed for low and large peak power. Compared with the continuous Poisson capacity, the capacity of a non-perfect receiver is almost lossless and loss with attenuation for low peak power given zero background radiation and nonzero background radiation, respectively. For large peak power, the capacity with a non-perfect receiver converges, while that of continuous Poisson capacity channel linearly increases. The above theoretical results are extensively validated by numerical results.
\end{abstract}

\renewcommand{\baselinestretch}{1.4}
\section{Introduction}
On some specific occasions where the conventional RF is prohibited and direct link transmission cannot be guaranteed, non-line-of-sight (NLOS) optical scattering communication can be adopted to provide certain information transmission rate \cite{xu2008ultraviolet}. Optical scattering communication is typically developed in the ultraviolet (UV) spectrum due to solar blind region (200nm-280nm) where the solar background radiation is negligible. On the UV scattering communication channel characterization, extensive studies based on Monte Carlo simulation \cite{ding2009modeling,zhang2012charac}, theoretical analysis \cite{gupta2012NLOS,zuo2013closed,sun2016closed} and experimental results \cite{chen2014expe,liao2015uv,raptis2016power} show that the atmospheric attenuation among scattering channel can be extremely large, especially for long-range transmission. Hence, it is difficult to detect the received signals using conventional continuous waveform receiver, such as photon-diode (PD) and avalanche photon-diode (APD). Instead, a photon-counting receiver is widely deployed.

For photon-counting receiver, the received signals are usually characterized by discrete photoelectrons, whose number in a certain interval satisfies a Poisson distribution. For such a Poisson channel, recent works mainly focus on point-to-point single-user channel, such as single transmitter \cite{wyner1988capacity,frey1991information}, multiple transmitters \cite{chakraborty2008outage} in continuous-time \cite{davis1980capacity,shamai1993bounds} and discrete-time
\cite{lapidoth2009capacity,lapidoth2011discrete,wang2014refined,cao2014capacity}, as well as the Poisson interference channel capacity \cite{lai2015capacity}. For multiple users scenario, \cite{lapidoth2003wide,kim2016superposition} focus
on the Poisson broadcast channel, \cite{lapidoth1998poisson} investigates the Poisson
multiple-access channel (MAC).  Besides, the system characterization and optimization, as well as the signal processing \cite{el2012binary,jiang2019clipping,gong2015non,ardakani2017relay,ardakani2017performance} have also been extensively studied from the receiver side.

Most information theoretical and signal processing works assume perfect photon-counting receiver, which is difficult to realize. A practical photon-counting receiver typically consists of a photomultiplier tube (PMT) as well as the subsequent sampling and processing blocks \cite{becker2005advanced}. Recently, extensive efforts have been made to design
and characterize practical photon-counting receivers, such as single
photon avalanche diode (SPAD), which has been applied in
many optical communication scenarios \cite{chitnis2014spad,gnecchi2016analysis}. In optical scattering communication, we consider a practical photon-counting receiver typically consisting of a photomultiplier tube
(PMT) and the subsequent pulse-holding circuits to generate
a series of rectangular pulses with certain width. The square pulses generated by pulse-holding circuits typically have positive width that incurs dead time effect \cite{cherry2012physics}, where a photon arriving during the pulse duration of the previous photon cannot be detected due to the merge of two pulses. The dead time effect and the model of sub-Poisson distribution for the photon-counting processing have been investigated in \cite{Omote1990dead,daniel2000mean}, where the variance is lower than the mean. The photon-counting system with dead time effect for infinite sampling rate has been investigated in optical communication for channel characterizations \cite{Drost2015dead,sarbazi2015detection}, optical wireless communications using SPAD detector \cite{sarbazi2016information,sarbazi2018statistical} and experimental implementation \cite{chitnis2014spad,shentu2013217}. The photon-counting system with dead time effect for finite sampling rate with shot noise of PMTs is investigated in \cite{zou2018characterization} based on a rising-edge detector. However, the performance analysis for a sampling-based detector focusing on the achievable transmission rate and channel capacity are still missing.

In this work, we analyze the achievable rate and capacity of a sampling-based detector under positive dead time and finite sampling rate, assuming negligible electrical thermal noise and shot noise. For the symbol duration that cannot be small, we first derive the upper and lower bounds on the achievable rate based on Kullback-Leibler (KL) divergence and Chernoff $\alpha$-divergence respectively. We also investigate the convergence rate of the proposed upper and lower bounds, and demonstrate that the bound gap converges to zero with exponential rate for large sampling number $L$, large peak power $A$ and zero background radiation $\Lambda_0$. For low peak power $A$, the bound gap converges to zero with order $A^2$. For large peak power $A$ with fixed background radiation $\Lambda_0$ and low background radiation $\Lambda_0$ with fixed peak power $A$, the bound
gap converges to certain small positive value for low background radiation $\Lambda_0$ and large peak power $A$, respectively.

For the symbol duration that cannot be arbitrarily small, we derive the capacity-achieving distribution and corresponding capacity. We show that continuous Poisson capacity equals to that of non-perfect receiver as $T_s=\tau\to0$. Furthermore, we characterize the capacity loss from the continuous Poisson channel for low and large peak power. We demonstrate negligible and significant capacity loss for low peak power given zero background radiation and nonzero background radiation, respectively. The capacity with non-perfect receiver approaches a certain value, while that of continuous Poisson channel increases linearly.

The remainder of this paper is organized as follows. In
Section \ref{sect.model}, we provide the system model of a practical
photon-counting receiver, along with the achievable rate with on-off keying (OOK) modulation for long symbol duration and the capacity for the symbol duration that cannot be sufficiently small. In
Section \ref{sect.bound}, we derive the upper and lower bounds on the maximum achievable rate and provide an approximation for the medium signal to noise ratio (SNR) regime. In Section \ref{sect.asym}, we investigate the asymptotic tightness of the upper and lower bounds for five scenarios. In Sections \ref{sect.sisocapa} and \ref{sect.asympcapa}, we investigate the capacity and the corresponding asymptotic properties, respectively. The theoretical analysis results are extensively validated by numerical results
in Section \ref{sect.numer}. Finally, we conclude this paper in Section \ref{sect.conc}.

\section{System Model}\label{sect.model}
\subsection{Signal Model}
We introduce the following notations that will be adopted throughout this paper. Random variables and vectors are denoted by upper-case letters and bold uppercase letters, respectively. We use notation $Z_{[j]}$ to denote a sequence of random variables $\{Z_1, Z_{2},\cdots,Z_j\}$. Realizations of random variables are denoted in lowercase letters, and follow the above notation conventions.

\begin{figure}
	\setlength{\abovecaptionskip}{-0.1cm}
	\centering
	{\includegraphics[angle=0, width=0.8\textwidth]{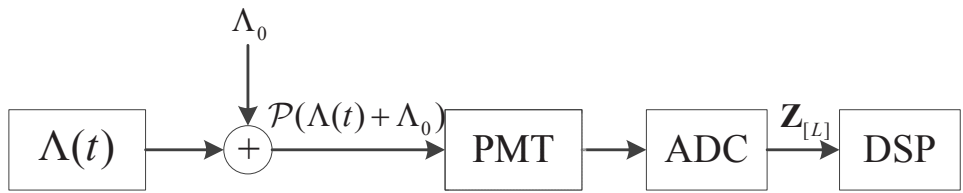}}
	\caption{The system diagram under consideration.}
	\label{fig.sysm}
\end{figure}

Consider single-user communicating to a single non-perfect receiver. The system model is shown in Fig.~\ref{fig.sysm}. Let $\Lambda(t)$ denote the $\mathbb{R}_0^+$-valued photon arrival rate at time $t$, and $Y(t)$ denote the Poisson photon arrival process observed at the receiver and
\be 
Y(t)=\mathcal{P}\Big(\Lambda(t)+\Lambda_0\Big),
\ee 
where $\Lambda_0$ is the dark current at receiver, and $\mathcal{P}(\cdot)$ is the Poisson arrival process that records the time instants and the number of photon arrivals. In particular, for any time interval $[t-\tau, t]$, the probability of $k$ photons arriving at the receiver is given by
\be 
\mathbb{P}\{Y(t)-Y(t-\tau)=k\}=\frac{1}{k!}e^{-X_t}(X_t)^k,\quad k=0,1,\cdots,
\ee 
where $X_t=\int_{t-\tau}^{t}\Lambda(t^{'})+\Lambda_0\mathrm{d}t^{'}$; the arrival rate $\Lambda$ is given by $\Lambda=\frac{P}{h\nu_0}$, and $P$, $h$ and $\nu_0$ denote the transmitted optical power,
the Planck’s constant and the optical spectrum frequency, respectively, such that the energy per photon is given by $h\nu_0$. Thus, the photon arrival rate $\Lambda(t)$ must satisfy the following constraint:
\be\label{eq.const1}
0\leq\Lambda(t)\leq A, 
\ee
where $A$ is related to the corresponding maximum power allowed. In practice, LEDs or lasers are adopted as the transmitter with limited the peak power, such that the peak power constraint is more of interest than the average power constraint. 

Assuming perfect photon-counting receiver, each photon and the corresponding arrival time can be detected without error. However, perfect photon-counting receiver is difficult to realize, and a non-perfect receiver with finite sampling rate consisting of a PMT detector, an ADC, and a digital signal processor (DSP) unit is of more interest. When a photon arrives, the PMT detector generates a pulse with certain width, which causes the merge of two pulses if the interval of two photon arrival is shorter than the pulse width. The maximum arrival time interval where the two pulses merge is called dead time, denoted as $\tau$. Denote $T_s$ as the ADC sampling interval and assume low to medium sampling rate such that $T_s\geq\tau$. Considering the PMT sampling sequence in a symbol interval $Z_{[L]}=\{Z_1,\cdots,Z_L\}$, where $L\dff\lfloor\frac{T}{T_s}\rfloor$, $\lfloor\cdot\rfloor$ is the lower rounding function. Note that for any $\tau>0$, the number of photon arrivals $N_{0,\tau}$ on
$[0,\tau]$ together with the corresponding (ordered) arrival time instants $\mathbb{T}^{N_{Y}}=(T_1,\cdots,T_{N_{\tau}})$ are complete descriptions of random process $Y_{0,\tau}$. 

For the practical photon-counting receiver under consideration, assume zero shot noise, thermal noise and finite dead time. 
For one or multiple photons arriving at the photon-counting receiver at $(iT_s-\tau,iT_s]$, the sampling value $Z_i$ is the same due to the self-sustaining avalanche in SPAD or the shaping circuit that converts bell-shaped response into rectangular response for photon-counting \cite{sarbazi2018statistical,zou2018characterization}. According to above statement, we have 
\be 
Z_i&=&\left\{\begin{array}{ll}
	0,&T_j\notin(iT_s-\tau,iT_s],\quad\forall\quad j=1,\cdots,N_T; \\
	1,&\text{otherwise};  
\end{array}\right.
\ee 
where $\mathbb{P}(Z_i=1)=1-e^{-(X_iA+\Lambda_0)\tau}$ and $Z_i$ and $Z_j$ are independent identically distributed for $i\neq j$ due to the property of independent increment for Poisson process. In other words, $Z_i$ is an indicator on whether one or more photons arrive within $\tau$ prior to the sampling instant.

\subsection{The Achievable Rate on Long Symbol Duration}
Assume OOK modulation with symbol interval $T_b$, where $\Lambda(t) = A$ for symbol one and $\Lambda(t) = 0$ for symbol zero. Let $X_i \in \{0,1\}$ denote the symbol in the $i^{th}$ slot, which is independent across different time slots. Then, the arrival rate $\Lambda(t)=\sum_{i=0}^{+\infty}X_iA\cdot\mathbbm{1}\{(i-1)T_b\leq t<iT_b\}$, where $\mathbbm{1}\{\cdot\}$ is an indicator function. Further assume that $X_i$ is independent and identically distributed for each $i$ with probability $\mathbb{P}(X_i = 1) = \mu$. In the remainder of this paper, since we are interested in the achievable rate and symbols $X_i$ are independent, we consider one symbol interval and omit subscript $i$.

Consider the achievable rate for the above communication system assuming long symbol duration $T_b$ that cannot be shortened to the sampling duration. This corresponds to practical application scenarios where the transmitter adopts an external modulator with certain maximum modulation rate. Let $p_0$ and $p_1$ denote probability $\mathbb{P}(Z_i = 1|X = 0)$ and $\mathbb{P}(Z_i = 1|X = 1)$, respectively.  As the sum of variables with i.i.d. binary distribution is a sufficient statistic for these variables, we define summation $\hat{N}\dff\sum_{i=1}^{L}Z_i$ and the achievable rate is given as follows,
\be 
I(X;\hat{N})=H(\hat{N})-H(\hat{N}|X),
\ee 
where $\hat{N}$ follows binomial distributions $\mathbb{B}(p_0,L)$ and $\mathbb{B}(p_1,L)$ for symbol $X=0$ and $X=1$, respectively, and $\mathbb{B}(p,L)$ denotes binomial distribution with probability $p$ for each trial and $L$ trials.
\subsection{The Capacity on Arbitrarily Small Symbol Duration}
Recall that the Poisson channel capacity is defined as
\be
C_{T_s,\tau}=\lim\limits_{T\to\infty}\max\limits_{\Lambda^T\in[0,A]}\frac{1}{T}I(\Lambda^T;\mathbf{Z}_{[L_c]}).
\ee
Here we assume that the symbol duration can be arbitrarily small. Since $\Lambda^T\rightarrow(N_T,\mathbf{T}^{N_T})\rightarrow\mathbf{Z}_{[L_c]}$ forms a Markov chain, we have $I(\Lambda^T;\mathbf{Z}_{[L_c]})\leq I(\Lambda^T;N_T,\mathbf{T}^{N_T})$, which shows that the Poisson channel capacity with non-perfect receiver is lower than or equal to that of continuous Poisson channel. 

According to the chain rule for mutual information, we have
\be 
\frac{1}{T}I(\Lambda^T;\mathbf{Z}_{[L_c]})&=&\frac{1}{T}\sum_{l=1}^{L_c}I(\Lambda_{(l-1)T_s}^{lT_s};Z_{l}|\Lambda^{(l-1)T_s};\mathbf{Z}_{[l-1]})\nonumber\\
&=&\frac{1}{T}\sum_{l=1}^{L_c}H(Z_{l}|\Lambda^{(l-1)T_s};\mathbf{Z}_{[l-1]})-H(Z_{l}|\Lambda^{lT_s};\mathbf{Z}_{[l-1]})\nonumber\\
&\overset{(a)}{=}&\frac{1}{T}\sum_{l=1}^{L_c}H(Z_{l}|\Lambda^{(l-1)T_s};\mathbf{Z}_{[l-1]})-H(Z_{l}|\Lambda_{(l-1)T_s}^{lT_s})\nonumber\\
&\leq&\frac{1}{T}\sum_{l=1}^{L_c}H(Z_{l})-H(Z_{l}|\Lambda_{(l-1)T_s}^{lT_s})=\frac{1}{T}\sum_{l=1}^{L_c}I(\Lambda_{(l-1)T_s}^{lT_s};Z_{l}).
\ee 
where equality (a) holds since $Z_{l}$ is conditional independent of $(\Lambda^{(l-1)T_s};\mathbf{Z}_{[l-1]})$ given $\Lambda_{(l-1)T_s}^{lT_s}$. Thus, we have 
$C_{T_s,\tau}\leq\max\limits_{\Lambda^{T_s}\in[0,A]}\frac{1}{T_s}I(\Lambda^{T_s};\mathbf{Z}_{1}),$
where the equality holds if $\Lambda_{(l-1)T_s}^{lT_s}$ is independent of each other for different $l$. Consequently, the capacity-achieving distribution requires independent input signals for different sampling intervals, and the capacity is given by,
\be\label{eq.capacall}
C_{T_s,\tau}=\max\limits_{\Lambda^{T_s}\in[0,A]}\frac{1}{T_s}I(\Lambda^{T_s};\mathbf{Z}_{1}).
\ee

\section{The Bounds and Approximate on Achievable Rate for Long Symbol Duration}\label{sect.bound}
The mutual information involves the entropy of mixture distribution with intractable analytical form. Thus, pairwise-distances are adopted to provide lower bound and upper bound on the mutual information \cite{kolchinsky2017estimating}. The results are shown in the following proposition for completeness.
\begin{proposition}\label{prop.uplowbound}
	Define $X$ as the transmitted signal with measurable supports $\{x_1,\cdots,x_n\}$ and $\mathbb{P}(X=x_i)\dff c_i$ for $i=1,\cdots,n$. The channel transition probability $\mathbb{P}(Y|X)$ can be represented by a set of distribution $\{p_1,\cdots,p_n\}$, where $p_i(y)\dff \mathbb{P}(Y=y|X=x_i)$ for $i=1,\cdots,n$. We have the following lower bound and upper bound on mutual information $I(X;Y)$,
	\be 
	-\sum_{i=1}^{n}c_i\ln\sum_{j=1}^{n}c_j\exp(-C_{\alpha}(p_i||p_j))\leq I(X;Y)\leq-\sum_{i=1}^{n}c_i\ln\sum_{j=1}^{n}c_j\exp(-KL(p_i||p_j)),
	\ee 
	where Chernoff $\alpha$-divergence $C_{\alpha}(p||q)=-\ln\int p^{\alpha}(y)q^{1-\alpha}(y)\mathrm{d}y$ and Kullback-Leibler divergence $KL(p||q)=\int p(y)\ln\frac{p(y)}{q(y)}\mathrm{d}y$.

\end{proposition}

Consider OOK modulation at the transmitter and photon-counting detection at the receiver. As $T_s\geq\tau$, the samples are mutually independent and photon-counting detection is performed via examining whether each sample is higher than a certain threshold. Assume negligible shot and thermal noise such that each sample can distinguish whether photons arrived or not perfectly. Let $p_i=1-e^{-(iA+\Lambda_0)\tau}$ for $i = 0$ and $1$. Recalling that $\hat{N}\sim \mathbb{B}(L,p_i)\dff P_i^B(\cdot)$, the Chernoff $\alpha$-divergence and KL divergence are given by
\be 
C_{\alpha}(P_1^B||P_0^B)&=&-\ln\sum_{i=1}^{n}\binom{L}{i}(p_1^{\alpha}p_0^{1-\alpha})^i[(1-p_1)^{\alpha}(1-p_0)^{1-\alpha}]^{L-i}\nonumber\\&=&-L\ln\Big(p_1^{\alpha}p_0^{1-\alpha}+(1-p_1)^{\alpha}(1-p_0)^{1-\alpha}\Big)=C_{1-\alpha}(P_0^B||P_1^B),\\
KL(P_1^B||P_0^B)&=&\sum_{i=0}^{L}\binom{L}{i}p_1^i(1-p_1)^{L-i}\big(i\ln\frac{p_1}{p_0}+(L-i)\ln\frac{1-p_1}{1-p_0}\big)\nonumber\\&=&L\Big(p_1\ln\frac{p_1}{p_0}+(1-p_1)\ln\frac{1-p_1}{1-p_0}\Big).
\ee 

Note that mutual information $I(X;\hat{N})$ depends on $\Lambda_0$, $A$, $L$, $\mu$ and $\tau$. Since $I(X;\hat{N})=0$ for $\mu=0$ or $\mu=1$, we focus on the maximum mutual information $I(X;\hat{N})$ over $\mu\in[0,1]$ given fixed dead time $\tau$. Define  $I_{max}(\Lambda_0,A,L)\dff\max\limits_{\mu\in[0,1]}I(X;\hat{N})$. In the remainder of this Section, we investigate the lower and upper bounds on $I_{max}(\Lambda_0, A, L)$ and the asymptotic properties for large $L$ and $A$.
\subsection{Lower Bound on Mutual Information}
According to Proposition \ref{prop.uplowbound}, the lower bound on $I(X;\hat{N})$ is given as
\be\label{eq.lower}
I(X;\hat{N})&\geq&-\Big\{\mu\ln[(1-\mu)\exp(-C_{\alpha}(P_1^B||P_0^B))+\mu]\nonumber\\&&\qquad+(1-\mu)\ln[\mu\exp(-C_{\alpha}(P_0^B||P_1^B))+(1-\mu)]\Big\}.
\ee 
Note that the right-hand side of equation (\ref{eq.lower}) increases with respect to $C_{\alpha}(P_1^B||P_0^B)$ and $C_{\alpha}(P_0^B||P_1^B)$, where the optimal $\alpha$ maximizing the right-hand side is intractable. We resort to a suboptimal solution to $\alpha$, given as follows,
\be\label{eq.optalpha}
\alpha^{*}\dff\arg\max\limits_{0\leq\alpha\leq1}\min\{C_{\alpha}(P_1^B||P_0^B),C_{\alpha}(P_0^B||P_1^B)\}.
\ee 

We have the following Lemma \ref{lemma.suboptalpha} on optimal $\alpha^*$.
\begin{lemma}\label{lemma.suboptalpha}
	The optimal solution to problem (\ref{eq.optalpha}), denoted as $\alpha^{*}$, is $\frac{1}{2}$.
	\begin{proof}
		Please refer to Appendix \ref{appen.suboptalpha}.
	\end{proof}
\end{lemma}

Define $\beta\dff e^{-C_{\alpha^{*}}(P_0^B||P_1^B)}=e^{-C_{\alpha^{*}}(P_1^B||P_0^B)}=(\sqrt{p_0p_1}+\sqrt{(1-p_0)(1-p_1)})^{L}<1$ and function $F_l(\mu,\beta)\dff-\{\mu\ln[(1-\mu)\beta+\mu]+(1-\mu)\ln[\mu\beta+(1-\mu)]\}$. We aim to maximize $F_l(\mu,\beta)$ with respect to $\mu$ to tighten the lower bound on $I(X;\hat{N})$. Since
\be\label{eq.prime}
\frac{\partial F_l}{\partial \mu}=\ln[1-(1-\beta)\mu]-\frac{\beta}{1-(1-\beta)\mu}-\{\ln[\beta+(1-\beta)\mu]-\frac{\beta}{\beta+(1-\beta)\mu}\},
\ee 
we have
\be 
\frac{\partial F_l}{\partial \mu}\Big|_{(0,\beta)}=-\ln\beta+1-\beta>0, \qquad\frac{\partial F_l}{\partial \mu}\Big|_{(1,\beta)}=\ln\beta-1+\beta<0, \qquad\\
\frac{\partial^2 F_l }{\partial \mu^2}=-\frac{1-\beta}{1-(1-\beta)\mu}-\frac{\beta(1-\beta)}{\big(1-(1-\beta)\mu\big)^2}-\frac{1-\beta}{\beta+(1-\beta)\mu}-\frac{\beta(1-\beta)}{\big(\beta+(1-\beta)\mu\big)^2}<0.
\ee 
Thus, the optimal $\mu$ maximizing $F_l(\mu)$ uniquely exists and satisfies $\frac{\partial F_l}{\partial \mu}=0$. Define monotonic increasing function $G(x)\dff\ln x-\frac{\beta}{x}$. Since $\frac{\partial F_l}{\partial \mu}\Big|_{(\mu^{*},\beta)}=G(1-(1-\beta)\mu^{*})-G(\beta+(1-\beta)\mu^{*})=0$, we have $1-(1-\beta)\mu^{*}=\beta+(1-\beta)\mu^{*}$ and $\mu^{*}=\frac{1}{2}$. Thus, we have the following lower bound,
\be\label{eq.lowerbound}
I_{max}(\Lambda_0,A,L)\geq\max\limits_{\mu\in[0,1]}F_l(\mu,\beta)=-\ln\frac{1+\beta}{2}.
\ee 
\subsection{Upper Bound on Mutual Information}
The upper bound can be obtained using similar method as that of obtaining the lower bound. Defining $\beta_1\dff\exp\big(-KL(P_1^B||P_0^B)\big)$ and $\beta_2\dff\exp\big(-KL(P_0^B||P_1^B)\big)$, we have the following,
\be 
F_u(\mu,\beta_1,\beta_2)=-\{\mu\ln[(1-\mu)\beta_1+\mu]+(1-\mu)\ln[\mu\beta_2+(1-\mu)]\},\label{eq.upper}\qquad\\
KL(P_1^B||P_0^B)-KL(P_0^B||P_1^B)=(p_1-p_0)\ln\frac{p_1(1-p_1)}{p_0(1-p_0)}\lesseqqgtr0,\quad\text{if}\quad p_0+p_1\gtreqqless1.
\ee 
Define $\mu^{*}(\beta_1,\beta_2)\dff\arg\max\limits_{0\leq\mu\leq1}F_u(\mu,\beta_1,\beta_2)$. Although closed form of $\mu^{*}(\beta_1,\beta_2)$ is intractable, we have the following properties on $\mu^{*}(\beta_1,\beta_2)$,
\begin{lemma}\label{lemma.beta}
	Cycle $\mu^{*}(\beta_1,\beta_2)$ must satisfy the following properties,
	\begin{itemize}
		\item[(1)] $\mu^{*}(\beta_1,\beta_2)+\mu^{*}(\beta_2,\beta_1)=1$. Particularly, $\mu^{*}(\beta_1,\beta_2)=\frac{1}{2}$ if $\beta_1=\beta_2$.
		\item[(2)] $\mu^{*}(\beta_1,\beta_2)\gtrless\frac{1-\beta_1}{2-\beta_1-\beta_2}$ if $\beta_1\gtrless\beta_2$.
	\end{itemize}
	\begin{proof}
		Please refer to Appendix \ref{appen.beta}.
	\end{proof}	
\end{lemma}
\begin{lemma}\label{lemma.beta2}
	We have that
	\be \max\limits_{0\leq\mu\leq1}F_u(\mu,\beta_1,\beta_2)\leq\frac{|\beta_1-\beta_2|(1-\min\{\beta_1,\beta_2\})}{1-\beta_1\beta_2}-\ln\frac{1-\beta_1\beta_2}{2-\beta_1-\beta_2},
	\ee
	where equality holds if and only if $\beta_1=\beta_2$.
	\begin{proof}
		Please refer to Appendix \ref{appen.beta2}.
	\end{proof}	
\end{lemma}

According to Lemma \ref{lemma.beta2}, an upper bound on the maximal mutual information is given by,
\be\label{eq.upperbound} 
I_{max}(\Lambda_0,A,L)\leq\max\limits_{\mu\in[0,1]}F_u(\mu,\beta_1,\beta_2)\leq\frac{|\beta_1-\beta_2|(1-\min\{\beta_1,\beta_2\})}{1-\beta_1\beta_2}-\ln\frac{1-\beta_1\beta_2}{2-\beta_1-\beta_2}.
\ee  

The above discussions can be summarized into the following result.

\begin{theorem}
	We have that lower and upper bounds on $I_{max}(\Lambda_0, A, L)$ are given by Equations (\ref{eq.lowerbound}) and (\ref{eq.upperbound}), respectively.
\end{theorem}
\subsection{Asymptotic Mutual Information}\label{subsection.interp}
We first provide an interpretation to show the tightness of the upper and lower bounds. By applying Jensen’s inequality to Chernoff $\alpha$-divergence, we have
\be 
C_{\alpha}( P^B_0||P^B_1)&=&-\ln\mathbb{E}_{P^B_0}\big[(\frac{P^B_1}{P^B_0})^{1-\alpha}\big]\leq-\int P^B_0\ln(\frac{P^B_1}{P^B_0})^{1-\alpha}\mathrm{d}x= (1-\alpha)KL(P^B_0||P^B_1),\\
C_{\alpha}(P^B_0||P^B_1)&=&-\ln\mathbb{E}_{P^B_1}\big[(\frac{P^B_0}{P^B_1})^{\alpha}\big]\leq-\int P^B_1\ln(\frac{P^B_0}{P^B_1})^{\alpha}\mathrm{d}x= \alpha KL(P^B_1||P^B_0),
\ee 
i.e., $C_{\frac{1}{2}}(P^B_0||P^B_1)\leq\frac{1}{2}\min\{KL(P^B_0||P^B_1),KL(P^B_1||P^B_0)\}$. Thus we have 
\be
\exp(-C_{\frac{1}{2}}(P^B_0||P^B_1))>\exp(-2C_{\frac{1}{2}}(P^B_0||P^B_1))\geq\exp(-\min\{KL(P^B_0||P^B_1),KL(P^B_1||P^B_0)\}),
\ee
i.e., $\beta>\beta^2\geq\max\{\beta_1,\beta_2\}$. We consider two cases, large $C_{\alpha}(P^B_0||P^B_1)$ and negligible $\max\{KL(P^B_1||P^B_0),KL(P^B_0||P^B_1)\}$. \textbf{Define high SNR for negligible $\beta$ and low SNR if $\beta_1$ and $\beta_2$ approach $1$, which agrees with the true scenarios of high SNR and low SNR in the physical communication channel.} Note that for high SNR regime, $\beta$, $\beta_1$ and $\beta_2$ approach $0$; and for low SNR regime, $\beta$, $\beta_1$ and $\beta_2$ approach $1$, i.e., $\beta$ and $(\beta_1,\beta_2)$ contribute similarly to the lower and upper bounds. Thus, lower bound (\ref{eq.lowerbound}) and upper bound (\ref{eq.upperbound}) are valid in both high and low SNR regimes. 

As the asymptotic maximum mutual information approaches $0$ in low SNR regime, we focus on high SNR regime, including large $L$ and $A$. For large $L$, we have the following Theorem \ref{theorem.asymI1} on the asymptotic results of the maximum mutual information.
\begin{theorem}\label{theorem.asymI1}
	For large $L$, the asymptotic maximum mutual information is given by 
	\be
	I_{max}(\Lambda_0,A,L)\left\{\begin{array}{ll}
		\geq\ln2-\beta+o(\beta),&\forall \beta_1,\beta_2; \\
		\leq\ln2-\beta_1,&\beta_1=\beta_2; \\
		\leq\ln2+\frac{\max\{\beta_1,\beta_2\}}{2}+o(\max\{\beta_1,\beta_2\}),& \beta_1\neq\beta_2;
	\end{array}\right.
	\ee
	where $\beta=\exp\Big(L\ln\big(\sqrt{p_0p_1}+\sqrt{(1-p_0)(1-p_1)}\big)\Big)$,
	$\beta_1=\exp\Big(-L\big(p_1\ln\frac{p_1}{p_0}+(1-p_1)\ln\frac{1-p_1}{1-p_0}\big)\Big)$ and $\beta_2=\exp\Big(-L\big(p_0\ln\frac{p_0}{p_1}+(1-p_0)\ln\frac{1-p_0}{1-p_1}\big)\Big)$.
	\begin{proof}
		Please refer to Appendix \ref{appen.asymI1}.
	\end{proof}	
\end{theorem}

Theorem \ref{theorem.asymI1} implies that the asymptotic maximum mutual information $\lim\limits_{L\to\infty}I_{max}(\Lambda_0,A,L)=\ln2$. For large peak power $A$, we have the following expansions on $\beta$, $\beta_1$, $\beta_2$.
\begin{lemma}\label{lemma.asymexpan}
	For large $A$, the expansions on $\beta$, $\beta_1$ and $\beta_2$ are given by 
	\be
	\beta&=&\big(\sqrt{p_0p_1}+\sqrt{(1-p_0)(1-p_1)}\big)^L\nonumber\\
	&=&p_0^\frac{L}{2}-p_0^\frac{L-1}{2}\big(\frac{\sqrt{p_0}}{2}(1-p_1)-\sqrt{(1-p_0)}(1-p_1)^{\frac{1}{2}}\big)+o(1-p_1);\\
	\beta_1&=&(\frac{p_0}{p_1})^{p_1L}(\frac{1-p_0}{1-p_1})^{(1-p_1)L}\nonumber\\
	&=&p_0^L-p_0^L\Big(-L(1-p_1)+(1-p_1)L\ln\frac{1-p_1}{1-p_0}\Big)+o(1-p_1);\\
	\beta_2&=&(\frac{p_1}{p_0})^{p_0L}(\frac{1-p_1}{1-p_0})^{(1-p_0)L}\nonumber\\
	&=&(\frac{1}{p_0})^{p_0L}(\frac{1}{1-p_0})^{(1-p_0)L}(1-p_1)^{(1-p_0)L}+o(1-p_1).
	\ee
	\begin{proof}
		Please refer to Appendix \ref{appen.asymexpan}.
	\end{proof}	
\end{lemma}

Noting that $1-p_1=\exp(-(A+\Lambda_0)\tau)$, Lemma \ref{lemma.asymexpan} shows the expansions of $\beta$, $\beta_1$ and $\beta_2$ with exponential convergence for large $A$. Furthermore, we have the following Theorem \ref{theorem.asymI2} on the asymptotic maximum mutual information. 
\begin{theorem}\label{theorem.asymI2}
	For large $A$, the asymptotic maximum mutual information is given by 
	\be
	I_{max}(\Lambda_0,A,L)\geq\ln\frac{2}{1+p_0^\frac{L}{2}}+\frac{p_0^\frac{L-1}{2}}{1+p_0^\frac{L}{2}}\big(\frac{\sqrt{p_0}}{2}(1-p_1)-\sqrt{(1-p_0)}(1-p_1)^{\frac{1}{2}}\big)+o(1-p_1),\\
	I_{max}(\Lambda_0,A,L)\leq p_0^L+\ln(2-p_0^L)+O(\max\{(1-p_1)\ln(1-p_1),(1-p_1)^{(1-p_0)L}\}).\qquad\quad
	\ee
	\begin{proof}
		Please refer to Appendix \ref{appen.asymI2}.
	\end{proof}	
\end{theorem}

Theorem \ref{theorem.asymI2} shows the upper and lower bounds on the maximum mutual information as $\ln\frac{2}{1+p_0^\frac{L}{2}}\leq\lim\limits_{A\to\infty}I_{max}(\Lambda_0,A,L)\leq p_0^L+\ln(2-p_0^L)$ for fixed $\Lambda_0$. Specifically, we have the following on the asymptotic maximum mutual information for zero $\Lambda_0$,
\be \lim\limits_{A\to\infty}I_{max}(0,A,L)=\ln2=\lim\limits_{L\to\infty}I_{max}(\Lambda_0,A,L).
\ee
\subsection{Approximate Method}\label{sect.approx}
For most scenarios of UV communication, background radiation arrival intensity $\Lambda_0$ are negligible. Since the proposed lower and upper bounds on $I(X;\hat{N})$ is loose in medium SNR regime, we propose an approximation method to characterize $I(X;\hat{N})$ in medium SNR regime. The approximated mutual information $I(X;\hat{N})$ based on low $\Lambda_0$ is shown in Theorem~\ref{theorem.appromut}.
\begin{theorem}\label{theorem.appromut}
	For low background radiation arrival intensity $\Lambda_0$, we have the following expansion on $I(X;\hat{N})$,
	\be\label{eq.appromut}
	I(X;\hat{N})&=&-[\mu(1-p_1)^{L}+1-\mu]\ln[\mu(1-p_1)^{L}+1-\mu]+\mu L(1-p_1)^{L}\ln(1-p_1)\nonumber\\
	&&-\mu[1-(1-p_1)^{L}]\ln\mu+(1-\mu)Lp_0\{\ln[\mu(1-p_1)^{L}+1-\mu]-\ln(\mu Lp_1)\nonumber\\
	&&-(L-1)\ln(1-p_1)\}-(1-\mu)h_b(Lp_0)+o(Lp_0)+O(\frac{1}{L}).
	\ee	
	\begin{proof}
		Please refer to Appendix \ref{appen.appromut}.
	\end{proof}
\end{theorem}

The approximation mutual information can be obtained from Equation (\ref{eq.appromut}) via omitting the terms with small $o$ and big $O$. For reliable communication system, the sampling numbers $L$ is typically large and background radiation arrival intensity $\Lambda_0$ is low. Thus, the proposed approximate mutual information can be adopted especially in the medium SNR regime.
\section{Asymptotic Tightness of Upper and Lower bounds}\label{sect.asym}
Section \ref{subsection.interp} provides an interpretation on the tightness of bounds and shows the asymptotic maximum mutual information for large $L$ and $A$. However, the convergence rate of upper and lower bounds is still unknown. In this Section, we proceed to investigate the convergence rate on the upper and lower bounds. 

Defining bound gap $\Delta(\beta,\beta_1,\beta_2)\dff\max\limits_{\mu\in[0,1]} F_u(\mu,\beta_1,\beta_2)-F_l(\mu,\beta)$, we have the following Theorem \ref{theorem.boundgap} on the upper and lower bounds on $\Delta(\beta,\beta_1,\beta_2)$.
\begin{theorem}\label{theorem.boundgap}
	For low SNR, we have the following upper bound on $\Delta(\beta,\beta_1,\beta_2)$,
	\be 
	\Delta(\beta,\beta_1,\beta_2)\leq\frac{1}{108}(\frac{\beta}{\beta_1}-1)(16\frac{\beta}{\beta_1}+11)+\frac{1}{108}(\frac{\beta}{\beta_2}-1)(16\frac{\beta}{\beta_2}+11);
	\ee 
	and for high SNR, we have the following upper bound on $\Delta(\beta,\beta_1,\beta_2)$,
	\be 
	\Delta(\beta,\beta_1,\beta_2)\leq(\beta-\beta_1)+(\beta-\beta_2).
	\ee
	For general $\beta,\beta_1,\beta_2$, we have the following lower bound on $\Delta(\beta,\beta_1,\beta_2)$,
	\be 
	\Delta(\beta,\beta_1,\beta_2)\geq\frac{1}{2}\ln\frac{1+\beta}{1+\beta_1}+\frac{1}{2}\ln\frac{1+\beta}{1+\beta_2}.
	\ee
	\begin{proof}
		Please refer to Appendix \ref{appen.boundgap}.
	\end{proof}	
\end{theorem}

To characterize the convergence of $\Delta(\beta,\beta_1,\beta_2)$, we consider the \textit{exponential rate} of convergence \cite{cover2012elements}. In summary, we consider five scenarios, where the convergence rates of the bound gap are shown in Table~\ref{tab.Table1}.
\begin{table}
	\caption{The convergence rate of bound gap for $5$ scenario.}\label{tab.Table1}
	\centering
	\begin{tabular}{|p{3cm}<{\centering} |p{9cm}<{\centering} |p{3cm}<{\centering} |}
		\hline
		Scenario & Convergence& Asymptotic tightness  \\
		\hline
		Large $L$ & $O\Bigg(\exp\Big(L\ln\big(\sqrt{p_0p_1}+\sqrt{(1-p_0)(1-p_1)}\big)\Big)\Bigg)$ & \Large{\checkmark}\\
		\hline
		Large $A$ fixed $\Lambda_0$ & $\geq\ln\big(1+p_0^{\frac{L}{2}}\big)-\frac{1}{2}\ln\big(1+p_0^{L}\big)+O(\exp\Big(\min\{\frac{1}{2},(1-p_0)L\}A\tau\Big))$
		
		$\leq2p_0^{\frac{L}{2}}-p_0^{L}+O(\exp\Big(\min\{\frac{1}{2},(1-p_0)L\}A\tau\Big))$ & \Large{$\times$}\\
		\hline
		Low $\Lambda_0$ fixed $A$ & $\geq\ln\big(1+(1-p_1)^{\frac{L}{2}}\big)-\frac{1}{2}\ln\big(1+(1-p_1)^{L}\big)+O(\min\{\frac{1}{2},p_1L\}\Lambda_0\tau)$
		
		$\leq2(1-p_1)^{\frac{L}{2}}-(1-p_1)^{L}+O(\min\{\frac{1}{2},p_1L\}\Lambda_0\tau)$ & \Large{$\times$} \\
		\hline
		Large $A$ fixed $\Lambda_0=0$ & $O\Big(\exp\big(-\frac{L\tau}{2}A\big)\Big)$ & \Large{\checkmark} \\
		\hline
		Low $A$ fixed $\Lambda_0$ & $O\Big(\frac{3L(1-p_0)}{16p_0}\tau^2A^2\Big)$ & \Large{\checkmark}\\
		\hline
	\end{tabular}
\end{table}

\subsection{Asymptotic Tightness of Bound Gap for Large $L$}
As $L$ approaches infinity, $\beta$, $\beta_1$ and $\beta_2$ approach $0$, which corresponds to high SNR regime. Then, we have the following Theorem \ref{theorem.boundgap1} on the convergence rate of bound gap $\Delta(\beta,\beta_1,\beta_2)$.
\begin{theorem}\label{theorem.boundgap1}
	As $L$ approaches infinity, the convergence rate of gap $\Delta(\beta,\beta_1,\beta_2)$ is given by,
	\be\label{eq.exponential1} 
	-\lim_{L \rightarrow \infty} \frac{\ln \Delta(\beta, \beta_1, \beta_2)}{L}=-\ln\big(\sqrt{p_0p_1}+\sqrt{(1-p_0)(1-p_1)}\big).
	\ee
	\begin{proof}
		Please refer to Appendix \ref{appen.boundgap1}.
	\end{proof}
\end{theorem}

Theorem \ref{theorem.boundgap2} demonstrates that the proposed bounds are asymptotically tight, where bound gap $\Delta(\beta,\beta_1,\beta_2)$ approaches zero with exponential rate $-\ln\big(\sqrt{p_0p_1}+\sqrt{(1-p_0)(1-p_1)}\big)$ as $L$ approaches infinity.

\subsection{Bound Gap for Large Peak Power $A$}
As peak power $A$ approaches infinity, probability $p_1$ approaches $1$ and $\beta,\beta_1,\beta_2$ approach $0$, which also corresponds to high SNR regime. We have the following upper and lower bounds on the bound gap $\Delta(\beta,\beta_1,\beta_2)$.
\begin{theorem}\label{theorem.boundgap2}
	For large peak power and fixed background radiation arrival intensity, we have the following upper and lower bounds on $\Delta(\beta,\beta_1,\beta_2)$,
	\be
	\Delta(\beta,\beta_1,\beta_2)&\leq&2p_0^{\frac{L}{2}}-p_0^{L}+\epsilon_u+o(\epsilon_u),\label{eq.offsetA1}\\
	\Delta(\beta,\beta_1,\beta_2)&\geq&\ln\big(1+p_0^{\frac{L}{2}}\big)-\frac{1}{2}\ln\big(1+p_0^{L}\big)+\epsilon_l+o(\epsilon_l),\label{eq.offsetA2}
	\ee
	where
	\be 
	\epsilon_u&=&\left\{
	\begin{array}{lcl}
		2p_0^{\frac{L-1}{2}}\sqrt{1-p_0}(1-p_1)^{\frac{1}{2}}, & & {(1-p_0)L>\frac{1}{2}};\\
		\Big\{2p_0^{\frac{L-1}{2}}\sqrt{1-p_0}-p_0^{-L+\frac{1}{2}}(1-p_0)^{-\frac{1}{2}}\Big\}(1-p_1)^{\frac{1}{2}}, & & {(1-p_0)L=\frac{1}{2}};\\
		-p_0^{-Lp_0}(1-p_0)^{-L(1-p_0)}(1-p_1)^{L(1-p_0)}, & & {(1-p_0)L<\frac{1}{2}};
	\end{array} \right.\label{eq.epsilonu}\\
	\epsilon_l&=&\left\{
	\begin{array}{lcl}
		(1+p_0^{\frac{L}{2}})^{-1}p_0^{\frac{L-1}{2}}\sqrt{1-p_0}(1-p_1)^{\frac{1}{2}}, &  {(1-p_0)L>\frac{1}{2}};\\
		\Big\{(1+p_0^{\frac{L}{2}})^{-1}p_0^{\frac{L-1}{2}}\sqrt{1-p_0}-\frac{1}{2}p_0^{-L+\frac{1}{2}}(1-p_0)^{-\frac{1}{2}}\Big\}(1-p_1)^{\frac{1}{2}}, &  {(1-p_0)L=\frac{1}{2}};\\
		-\frac{1}{2}p_0^{-Lp_0}(1-p_0)^{-L(1-p_0)}(1-p_1)^{L(1-p_0)}, &  {(1-p_0)L<\frac{1}{2}}.
	\end{array} \right.\label{eq.epsilonl}
	\ee 
	
	\begin{proof}
		Please refer to Appendix \ref{appen.boundgap2}.
	\end{proof}
\end{theorem}

Theorem \ref{theorem.boundgap2} demonstrates that the offset items $\epsilon_u$ and $\epsilon_l$ converge to $0$ as peak power $A$ approaches infinity. Furthermore, the exponential rates of $\epsilon_u$ and $\epsilon_l$ with respect to $A$ are given as follows,
\be 
-\lim\limits_{A\to\infty}\frac{\ln\epsilon_u}{A}&=&\min\{\frac{1}{2},(1-p_0)L\}\tau,\\
-\lim\limits_{A\to\infty}\frac{\ln\epsilon_l}{A}&=&\min\{\frac{1}{2},(1-p_0)L\}\tau.
\ee 
When peak power $A$ approaches infinity, the offset items are negligible for low $p_0$, and the following is approximately satisfied,
\be\label{eq.deltacoonvB}
\ln\big(1+p_0^{\frac{L}{2}}\big)-\frac{1}{2}\ln\big(1+p_0^{L}\big)\leq\Delta(\beta,\beta_1,\beta_2)\leq2p_0^{\frac{L}{2}}-p_0^{L}.
\ee 
\subsection{Bound Gap for Low Background Noise $\Lambda_0$}
For low background radiation arrival intensity, probability $p_0$ approaches $0$ and $\beta,\beta_1,\beta_2$ approach $0$, which corresponds to high SNR regime. We have the following upper and lower bounds on bound gap $\Delta(\beta,\beta_1,\beta_2)$.
\begin{theorem}\label{theorem.boundgap3}
	For low background radiation arrival intensity given fixed peak power, we have the following upper and lower bounds on $\Delta(\beta,\beta_1,\beta_2)$,
	\be
	\Delta(\beta,\beta_1,\beta_2)&\leq&2(1-p_1)^{\frac{L}{2}}-(1-p_1)^{L}+\epsilon^{'}_u+o(\epsilon^{'}_u),\label{eq.offsetB1}\\
	\Delta(\beta,\beta_1,\beta_2)&\geq&\ln\big(1+(1-p_1)^{\frac{L}{2}}\big)-\frac{1}{2}\ln\big(1+(1-p_1)^{L}\big)+\epsilon^{'}_l+o(\epsilon^{'}_l),\label{eq.offsetB2}
	\ee
	where
	\be 
	\epsilon^{'}_u&=&\left\{
	\begin{array}{lcl}
		2(1-p_1)^{\frac{L-1}{2}}\sqrt{p_1}p_0^{\frac{1}{2}}, & & {p_1L>\frac{1}{2}};\\
		\Big\{2(1-p_1)^{\frac{L-1}{2}}\sqrt{p_1}-(1-p_1)^{-L+\frac{1}{2}}p_1^{-\frac{1}{2}}\Big\}p_0^{\frac{1}{2}}, & & {p_1L=\frac{1}{2}};\\
		-(1-p_1)^{-L(1-p_1)}p_1^{-Lp_1}p_0^{Lp_1}, & & {p_1L<\frac{1}{2}};
	\end{array} \right.\\
	\epsilon^{'}_l&=&\left\{
	\begin{array}{lcl}
		(1+(1-p_1)^{\frac{L}{2}})^{-1}(1-p_1)^{\frac{L-1}{2}}\sqrt{p_1}p_0^{\frac{1}{2}}, &  {p_1L>\frac{1}{2}};\\
		\Big\{(1+(1-p_1)^{\frac{L}{2}})^{-1}(1-p_1)^{\frac{L-1}{2}}\sqrt{p_1}-\frac{1}{2}(1-p_1)^{-L+\frac{1}{2}}p_1^{-\frac{1}{2}}\Big\}p_0^{\frac{1}{2}}, &  {p_1L=\frac{1}{2}};\\
		-\frac{1}{2}(1-p_1)^{-L(1-p_1)}p_1^{-Lp_1}p_0^{Lp_1}, & {p_1L<\frac{1}{2}}.
	\end{array} \right.
	\ee 
	
	\begin{proof}
		According to reciprocities $p_0\longleftrightarrow1-p_1$, $p_1\longleftrightarrow1-p_0$ and Theorem \ref{theorem.boundgap2}, we can readily obtain the results in Theorem \ref{theorem.boundgap3}. The detailed procedure is omitted here.
	\end{proof}
\end{theorem}

Theorem \ref{theorem.boundgap3} demonstrates that offset items $\epsilon^{'}_u$ and $\epsilon^{'}_l$ converge $0$ as the background radiation arrival intensity $\Lambda_b$ approaches $0$. Furthermore, the linear convergence rate of $\epsilon^{'}_u$ and $\epsilon^{'}_l$ with respect to $\Lambda_b$ can be obtained as follows,
\be 
\lim\limits_{\Lambda_b\to0}\frac{\epsilon^{'}_u}{\Lambda_b}&=&\min\{\frac{1}{2},p_1L\}\tau,\\
\lim\limits_{\Lambda_b\to0}\frac{\epsilon^{'}_l}{\Lambda_b}&=&\min\{\frac{1}{2},p_1L\}\tau.
\ee 
As background radiation arrival intensity $\Lambda_b$ approaches $0$, the gap is negligible for small $\Lambda_b$ and the following is approximately satisfied,
\be 
\ln\big(1+(1-p_1)^{\frac{L}{2}}\big)-\frac{1}{2}\ln\big(1+(1-p_1)^{L}\big)\leq\Delta(\beta,\beta_1,\beta_2)\leq2(1-p_1)^{\frac{L}{2}}-(1-p_1)^{L}.
\ee 
\subsection{Bound Gap for Large Peak Power $A$ and $\Lambda_0=0$}
For zero background radiation arrival intensity, we have probability $p_0=0$ and $\beta=(1-p_1)^{\frac{L}{2}}$, $\beta_1 = 0$, $\beta_2 = (1-p_1)^{L}$, which corresponds to high SNR regime. We have the following on bound gap $\Delta(\beta,\beta_1,\beta_2)$.
\begin{theorem}\label{theorem.boundgap3b}
	For large peak power $A$ and zero background radiation arrival intensity $\Lambda_0$, we have the following on $\Delta(\beta,\beta_1,\beta_2)$,
	\be \label{eq.asym4}
	-\lim_{A \rightarrow \infty} \frac{\ln \Delta(\beta, \beta_1, \beta_2)}{A}=\frac{L\tau}{2}.
	\ee 
	\begin{proof}
		Please refer to Appendix \ref{appen.boundgap3b}.
	\end{proof}
\end{theorem}

Theorem~\ref{theorem.boundgap3b} demonstrates that the upper and lower bounds are asymptotically tight for sufficiently large peak power $A$ if background radiation arrival intensity $\Lambda_0=0$, with exponential rate $\frac{L\tau}{2}$. 
\subsection{Bound Gap for Low Peak Power $A$}
For low peak power, probability $p_1$ approaches $p_0$ and $\beta,\beta_1,\beta_2$ approach $1$, which corresponds to low SNR regime. We have the following result on bound gap $\Delta(\beta,\beta_1,\beta_2)$.
\begin{theorem}\label{theorem.boundgap4}
	For low peak power $A$ given fixed background radiation arrival intensity $\Lambda_0$, we have the following on $\Delta(\beta,\beta_1,\beta_2)$,
	\be \label{eq.asym5}
	\Delta(\beta,\beta_1,\beta_2)=\frac{3L(1-p_0)}{16p_0}\tau^2A^2+o(A^2).
	\ee 
	\begin{proof}
		Please refer to Appendix \ref{appen.boundgap4}.
	\end{proof}
\end{theorem}

Theorem \ref{theorem.boundgap4} demonstrates that bound gap $\Delta(\beta,\beta_1,\beta_2)$ converges to $0$ with order $A^2$.

\section{Capacity for Arbitrarily Symbol Duration }\label{sect.sisocapa}

Assuming low to medium sampling rate, we investigate the capacity for two cases, $T_s = \tau$ and $T_s > \tau$. According to Equation (\ref{eq.capacall}), the capacity is given by $C_{T_s,\tau}\dff\max\limits_{\Lambda^{T_s}\in[0,A]}\frac{1}{T_s}I(\Lambda^{T_s};Z)$. 
\subsection{Capacity for Sampling Time $T_s = \tau$}\label{subsect.samptau}
Assuming $T_s = \tau$, the main result on the Poisson capacity with non-perfect receiver is summarized in Theorem~\ref{theor.siso}.
\begin{theorem}{\label{theor.siso}}
	For $T_s = \tau$, the optimal input signal is constrained within binary level $\{0,A\}$, and $C_{\tau,\tau}$ can be obtained by solving the following problem:
	\be\label{eq.sisocapa}
	C_{\tau,\tau}=\frac{1}{\tau}\max\limits_{0\leq\mu\leq 1}h_b\big(\hat{p}(\mu)\big)-(1-\mu)h_b\big(p(\Lambda_0)\big)-\mu h_b\big(p(A+\Lambda_0)\big),
	\ee 
	where $\hat{p}(\mu)\dff(1-\mu)p(\Lambda_0)+\mu p(A+\Lambda_0)$, $h_b(x)=-x\ln x-(1-x)\ln(1-x)$, $p(x)\dff1-e^{-x\tau}$, and $\mu$ denotes the duty cycle. Furthermore, the optimal duty cycle $\mu^{*}$ satisfies $$h_b\big(p(A+\Lambda_0)\big)-h_b\big(p(\Lambda_0)\big)=h_b^{'}(\hat{p}(\mu))\Big(p(A+\Lambda_0)-p(\Lambda_0)\Big)$$, and is given by
	\be\label{eq.optsing}
	\mu^{*}=\frac{\frac{a}{1+a}-p(\Lambda_0)}{p(A+\Lambda_0)-p(\Lambda_0)}\in[0,1],
	\ee 
	where $a=\exp(-\frac{h_b\big(p(A+\Lambda_0)\big)-h_b\big(p(\Lambda_0)\big)}{p(A+\Lambda_0)-p(\Lambda_0)})$. The capacity $C_{\tau,\tau}=\frac{1}{\tau}F(\mu^{*})$, where
	\be 
	F(\mu)\dff h_b\big(\hat{p}(\mu)\big)-(1-\mu)h_b\big(p(\Lambda_0)\big)-\mu h_b\big(p(A+\Lambda_0)\big).
	\ee  
\end{theorem}
\begin{remark}
	The same as the scenario of continuous Poisson channel, the optimal input distribution is also binary-level. However, for continuous Poisson channel, the optimal input signal requires infinite transmitter bandwidth; while for the non-perfect receiver under consideration, the optimal input signal distribution requires finite transmitter bandwidth related to the receiver dead time.
\end{remark}

Here we provide two major steps on the proof.

In \textbf{Step 1}, we prove that the optimal input distribution must be constrained within two levels $\{0, A\}$, given by the following Proposition.
\begin{proposition}\label{prop.siso1}
	The optimal input signal is constrained within two binary levels $\{0,A\}$.
	\begin{proof}
		Please refer to Appendix \ref{appen.siso1}.
	\end{proof}
\end{proposition}

In \textbf{Step 2}, We provide the optimal duty cycle, given by the following proposition.
\begin{proposition}\label{prop.siso2}
	The optimal duty cycle is 
	$\mu^{*}=\frac{\frac{a}{1+a}-p(\Lambda_0)}{p(A+\Lambda_0)-p(\Lambda_0)}$,
	where $a=\exp(-\frac{h_b\big(p(A+\Lambda_0)\big)-h_b\big(p(\Lambda_0)\big)}{p(A+\Lambda_0)-p(\Lambda_0)})$. The capacity with non-perfect receiver $C_{\tau,\tau}=\frac{1}{\tau}F(\mu^{*})$, where $F(\mu)\dff h_b(\hat{p}(\mu))-(1-\mu)h_b\big(p(\Lambda_0)\big)-\mu h_b\big(p(A+\Lambda_0)\big)$.
	\begin{proof}
		Please refer to Appendix \ref{appen.siso2}.
	\end{proof}
\end{proposition}

\subsection{Capacity for Sampling Time $T_s > \tau$}\label{subsect.capavssampling}

Define $\alpha=\frac{\tau}{T_s}$ such that $0<\alpha<1$. Following the proof procedure of Section~\ref{subsect.samptau}, the optimal duty cycle does not depend on $T_s$ and the capacity is given by $C^{T_s}=\alpha C_{\tau,\tau}$. This result implies that $\alpha$ is an attenuation factor related to the sampling rate of the non-perfect receiver.

\section{Asymptotic Properties on the Capacity}\label{sect.asympcapa}
Section \ref{sect.sisocapa} provides a rigorous proof on the capacity of a sample-based receiver and shows that the optimal input distribution is binary, the same as the continuous Poisson channel. In this Section, we further investigate the asymptotic properties of the non-perfect receiver compared with the continuous Poisson channel.

\subsection{Asymptotic Property of Capacity for $\tau\to0$}\label{sect.asymmucapa}
We consider sampling time $T_s = \tau$ and both approach zero. The main results are summarized in Theorem~\ref{theo.sisolowtau}.
\begin{theorem}\label{theo.sisolowtau}
	The optimal duty cycle and capacity of the non-perfect receiver approach those of continuous Poisson channel for any $A$ and $\Lambda_0$, respectively, as $\tau\to0$.
	\begin{proof}
		Please refer to Appendix \ref{appen.sisolowtau}.
	\end{proof}
\end{theorem}

Theorem \ref{theo.sisolowtau} studies the asymptotic property of the non-perfect receiver for $T_s=\tau\to0$. It shows that Theorem \ref{theor.siso} extends the result of continuous Poisson channel \cite{wyner1988capacity}, and provides a more general and practical results. 

Furthermore, we have the following results on the asymptotic property on the convergence of the optimal duty cycle with respect to $\tau$.
\begin{theorem}\label{theo.sisoasym}
	For fixed $\Lambda_0$, as $\tau$ approaches $0$, the optimal duty cycle of the non-perfect receiver point-wisely, but not uniformly, converge to that of continuous Poisson channel. 
	\begin{proof}
		Please refer to Appendix \ref{appen.sisoasym}.
	\end{proof}
\end{theorem}

\subsection{Asymptotic Property of the Optimal Duty Cycle for $A\to0$ and $A\to\infty$}\label{subsect.asympcycle}
We investigate the asymptotic property of the optimal duty cycle for the non-perfect receiver. The asymptotic property consists of $4$ cases: $A\to\infty$ given $\Lambda_0=0$, $A\to0$ given $\Lambda_0=0$, $A\to\infty$ given $\Lambda_0>0$ and $A\to0$ given $\Lambda_0>0$, as shown in Theorem~\ref{theo.asymlargelowpeak}. 
\begin{theorem}\label{theo.asymlargelowpeak}
	The optimal duty cycles of the non-perfect receiver for $A\to0$ and $A\to\infty$ are summarized in Table~\ref{table.asymmucapacity}.
	\begin{proof}
		Please refer to Appendix \ref{appen.asymlargelowpeak}.
	\end{proof}
\end{theorem}

Theorem \ref{theo.asymlargelowpeak} investigates the optimal duty cycle of the non-perfect receiver and show the difference with that of continuous Poisson channel for large peak power $A$, since larger peak power $A$ leads to larger photon-counting loss for the non-perfect photon-counting receiver. The optimal duty cycle for low peak power demonstrates negligible difference with that of continuous Poisson channel, since there is almost no photon-counting loss for low peak power $A$.

\subsection{Asymptotic Property of Non-perfect Poisson Capacity for $A\to0$ and $A\to\infty$}
Similar to Section~\ref{subsect.asympcycle}, the asymptotic property analysis of the capacity with non-perfect receiver consists of $4$ cases: $A\to\infty$ given $\Lambda_0=0$, $A\to0$ given $\Lambda_0=0$, $A\to\infty$ given $\Lambda_0>0$ and $A\to0$ given $\Lambda_0>0$. The results on the above four cases are summarized in Table~\ref{table.asymmucapacity}.

\begin{table}
	\caption{The asymptotic property of non-perfect receiver Poisson channel and continuous Poisson channel.}\label{table.asymmucapacity}
	\centering
	\begin{tabular}{|c|c|c|c|c|c|}
		\hline
		\multicolumn{2}{|c|}{} & \multicolumn{2}{|c|}{Practical Receiver}
		& \multicolumn{2}{|c|}{Continuous Poisson}\\
		\hline
		\multicolumn{2}{|c|}{Peak Power}
		& $\to0$& $\to\infty$ & $\to0$& $\to\infty$\\
		\hline
		\multirow{2}*{Duty cycle}& $\Lambda_0=0$ & $\frac{1}{e}$& $\frac{1}{2}$ & $\frac{1}{e}$ & $\frac{1}{e}$\\
		\cline{2-6}
		~&$\Lambda_0>0$ & $\frac{1}{2}$ & $1-\frac{1}{\big(1+\exp\big(e^{\Lambda_0\tau}h_b\big(p(\Lambda_0)\big)\big)\big)(1-p(\Lambda_0))}$ & $\frac{1}{2}$ & $\frac{1}{e}$\\
		\hline
		\multirow{2}*{Capacity}& $\Lambda_0=0$ & $\frac{A}{e}+o(A)$ & $\frac{1}{\tau}$ & $\frac{A}{e}$ & $\frac{A}{e}$\\
		\cline{2-6}
		~&$\Lambda_0>0$ & $d_{\tau}A^2+o(A^2)$ & $c_{\Lambda_0}\frac{1}{\tau}$ & $d_{Poi}A^2+o(A^2)$ & $\frac{A}{e}+o(A)$\\
		\hline		
	\end{tabular}
\end{table}

Recall that the capacity $C_{\tau,\tau}=\frac{1}{\tau}F(\mu^{*})$, where $F(\mu)= h_b\big(\hat{p}(\mu)\big)-(1-\mu)h_b\big(p(\Lambda_0)\big)-\mu h_b\big(p(A+\Lambda_0)\big)$, $\hat{p}(\mu)=(1-\mu)p(\Lambda_0)+\mu p(A+\Lambda_0)$, $a=\exp\Big(-\frac{h_b\big(p(A+\Lambda_0)\big)-h_b\big(p(\Lambda_0)\big)}{p(A+\Lambda_0)-p(\Lambda_0)}\Big)$ and $\mu^{*}=\frac{\frac{a}{1+a}-p(\Lambda_0)}{p(A+\Lambda_0)-p(\Lambda_0)}$. We demonstrate the asymptotic results of the four cases. 

\textbf{Case 1}: $A\to\infty$ given $\Lambda_0=0$.

According to \cite{wyner1988capacity}, for $\Lambda_0=0$ and any $A$, the asymptotic Poisson capacity is given by $C_{Poi}=\frac{1}{e}A$. Such linear capacity properties motivate us to investigate the asymptotic capacity for non-perfect receiver with dead time $\tau$. It is easy to check that $\lim\limits_{A\to\infty}a=\lim\limits_{A\to\infty}\exp(-\frac{h_b(p(A))}{p(A)})=1$, $\lim\limits_{A\to\infty}\mu^{*}=\frac{1}{2}$ and $\lim\limits_{A\to\infty}\hat{p}(\mu^{*})=\frac{1+p(\Lambda_0)}{2}=\frac{1}{2}$. Thus, we have 
\be\label{eq.asymcapa1}
\lim\limits_{A\to\infty}C_{\tau,\tau}=\lim\limits_{A\to\infty}\frac{1}{\tau}F(\mu^{*})=\frac{1}{\tau}h_b(\frac{1}{2})=\frac{1}{\tau},
\ee 
which shows that \textbf{the capacity for non-perfect receiver with dead time $\tau$ approaches $\frac{1}{\tau}$ for large peak power $A$.} The loss compared with the continuous Poisson channel stems from photon-counting loss for large peak power. 

\textbf{Case 2}: $A\to0$ given $\Lambda_0=0$.

It is obvious that the capacity with perfect or non-perfect photon receiver approaches $0$ when $A\to0$. Work \cite{wyner1988capacity} shows that $C_{Poi}=\frac{1}{e}A$ for continuous Poisson channel, i.e., the convergence rate is linear for low $A$, while the convergence rate of non-perfect photon receiver for low $A$ still needs to be investigated. 

For $\Lambda_0=0$, it is easy to check that $\lim\limits_{A\to\infty}a=0$. Noting that $h_b(x)=x(1-\ln x)+o(x)$, we have 
\be 
\mu^{*}&=&\frac{a}{p(A)}=\exp\big(-\frac{p(A)(1-\ln p(A))+o(p(A))}{p(A)}-\ln p(A)\big)=\frac{1}{e}+o(A),\\
\hat{p}(\mu^{*})&=&\frac{1}{e}p(A)+o(A).
\ee
Thus, the capacity with non-perfect photon receiver for low $A$ is given by
\be\label{eq.asymcapa2} 
C_{\tau,\tau}&=&\frac{1}{\tau}\{h_b(\hat{p})-\mu^{*}h_b(p(A))\}=\frac{1}{\tau}\{\hat{p}-\hat{p}\ln\hat{p}-\mu^{*}\big(p(A)-p(A)\ln p(A)\big)+o(A)\}\nonumber\\
&=&\frac{1}{e\tau}p(A)+o(A)=\frac{1}{e}A+o(A),
\ee 
which shows that the capacity for non-perfect receiver with dead time $\tau$ approaches $0$ with the same linear convergence rate as that of continuous Poisson channel, i.e., finite dead time receiver causes negligible capacity loss for low $A$.

\textbf{Case 3}: $A\to\infty$ given $\Lambda_0>0$.

For $\Lambda_0>0$ and large $A$, the asymptotic continuous Poisson capacity is given by $C_{Poi}=\frac{1}{e}A+o(A)$. It is seen that the asymptotic capacity loss given $\Lambda_0>0$ compared with that given $\Lambda_0=0$ is negligible for large $A$. Thus, there is a problem on the asymptotic Poisson capacity loss for non-perfect receiver given dead time $\tau$. Theorem~\ref{theo.asymcapalargepeak} provides the answer as follows.

\begin{theorem}\label{theo.asymcapalargepeak}
	The non-perfect receiver capacity for $A\to\infty$ is given by $\lim\limits_{A\to\infty}C_{\tau,\tau}=c_{\Lambda_0}\frac{1}{\tau}$, where $c_{\Lambda_0}=h_b\Big(\frac{\exp\big(e^{\Lambda_0\tau}h_b\big(p(\Lambda_0)\big)\big)}{1+\exp\big(e^{\Lambda_0\tau}h_b\big(p(\Lambda_0)\big)\big)}\Big)-\frac{h_b\big(p(\Lambda_0)\big)e^{\Lambda_0\tau}}{\Big(1+\exp\big(e^{\Lambda_0\tau}h_b\big(p(\Lambda_0)\big)\big)\Big)}$.
	\begin{proof}
		Please refer to Appendix \ref{appen.asymcapalargepeak}.
	\end{proof}
\end{theorem}

Coefficient $c_{\Lambda_0}$ characterizes the asymptotic capacity with non-perfect receiver for nonzero background radiation $\Lambda_0$. It is seen that $c_{\Lambda_0}=1$ iff $\Lambda_0=0$ and $c_{\Lambda_0}<1$ for $\Lambda_0>0$. However, the monotonicity properties of $c_{\Lambda_0}$ with respect to $\Lambda_0$ needs to be investigated, which is the main argument of Theorem \ref{theo.sisoasymclambda}.

\begin{theorem}\label{theo.sisoasymclambda}
	$c_{\Lambda_0}$ monotonically decreases with $\Lambda_0$ and $c_{\Lambda_0}\in(0,1]$ for $\Lambda_0\in[0,+\infty)$. 
	\begin{proof}
		Please refer to Appendix \ref{appen.sisoasymclambda}.
	\end{proof}
\end{theorem}

\textbf{Case 4}: $A\to0$ given $\Lambda_0>0$.

For the asymptotic capacity for low $A$ given $\Lambda_0>0$, the main results are shown in Theorem~\ref{theo.sisoasymclambda2}.
\begin{theorem}\label{theo.sisoasymclambda2}
	For $\Lambda>0$, the asymptotic capacity for continuous Poisson channel and non-perfect receiver are  $C_{Poi}=d_{\Lambda_0}^{Poi}A^2+o(A^2)$ and $C_{\tau,\tau}=d_{\Lambda_0}^{\tau}A^2+o(A^2)$ for small $A$, respectively, where $d_{\Lambda_0}^{Poi}=\frac{1}{8\Lambda_0}$ and $d_{\Lambda_0}^{\tau}=\frac{\tau\big(1-p(\Lambda_0)\big)}{8p(\Lambda_0)}$. 
	\begin{proof}
		Please refer to Appendix \ref{appen.sisoasymclambda2}.
	\end{proof}
\end{theorem} 

Theorem~\ref{theo.sisoasymclambda2} demonstrates the asymptotic capacity with non-perfect receiver and continuous Poisson capacity both as $O(A^2)$ for low $A$ given $\Lambda_0>0$. Furthermore, we have Theorem~\ref{theo.sisoasymefficient} on $d_{\Lambda_0}^{Poi}$ and $d_{\Lambda_0}^{\tau}$.
\begin{theorem}\label{theo.sisoasymefficient}
	$d_{\Lambda_0}^{Poi}>d_{\Lambda_0}^{\tau}$ holds for any $\Lambda_0>0$ and $\tau>0$. In addition, $d_{\Lambda_0}^{\tau}$ approaches $d_{\Lambda_0}^{Poi}$ for any $\Lambda_0>0$ when $\tau\to 0$, i.e., $\lim\limits_{\tau\to 0}d_{\Lambda_0}^{\tau}=d_{\Lambda_0}^{Poi}$.
	\begin{proof}
		Please refer to Appendix \ref{appen.sisoasymefficient}.
	\end{proof}
\end{theorem} 

Theorem~\ref{theo.sisoasymefficient} implies that the capacity with non-perfect receiver is strictly lower than that of continuous Poisson channel for low $A$ given $\Lambda_0>0$ asymptotically, where the two capacities converge asymptotically for small $A$ given $\Lambda_0=0$.
\subsection{The Monotonicity of Non-perfect Receiver Capacity}
Theorem~\ref{theor.siso} characterizes the non-perfect receiver capacity given dead time $\tau$, sampling interval $T_s$, background radiation $\Lambda_0$ and peak power $A$. According to Section~\ref{subsect.capavssampling}, the non-perfect receiver capacity is proportional to the sampling rate $T_s^{-1}$. The relationship between the non-perfect receiver capacity and other parameters still needs to be investigated.
\subsubsection{The Monotony with peak power $A$}
We still consider $T_s=\tau$ and provide the following result on the monotonicity of the non-perfect receiver capacity $C_{\tau,\tau}$ and the non-perfect receiver capacity per power $\frac{C_{\tau,\tau}}{A}$, as shown in Theorem~\ref{theo.sisomonopower}.
\begin{theorem}\label{theo.sisomonopower}
	The non-perfect receiver capacity $C_{\tau,\tau}(A,\Lambda_0)$ increases with peak power $A$ for any $\Lambda_0$. In addition, there exists $A_{th}$, $A_{th_1}$ and $A_{th_2}$ such that the non-perfect receiver capacity $C_{\tau,\tau}(A,\Lambda_0)$ is concave with $A$ for $A\geq A_{th_{1}}$ and the non-perfect receiver capacity per power $\frac{C_{\tau,\tau}}{A}$ decreases with peak power $A$ for any $A\geq A_{th_{2}}$. 
	\begin{proof}
		Please refer to Appendix \ref{appen.sisomonopower}.
	\end{proof}
\end{theorem} 

Theorem~\ref{theo.sisomonopower} provides a strict proof that larger $A$ corresponds larger capacity with non-perfect receiver. Theorem~\ref{theo.sisomonopower} shows the capacity with non-perfect receiver is concave for large $A$,  and the capacity with non-perfect receiver per power decreases with peak power $A$ due to capacity saturation characteristics for large power. 

\subsubsection{The Monotonicity with dead time $\tau$}
Section~\ref{sect.asymmucapa} shows the asymptotic property of the non-perfect receiver for low $\tau$ and reveals the connection between non-perfect receiver and continuous Poisson channel. We further provide the monotonicity results on two special cases, for large $\tau$ and $\Lambda_0=0$ in Theorem~\ref{theo.sisomonosmall} and Theorem~\ref{theo.sisomonolam0}, respectively.

\begin{theorem}\label{theo.sisomonosmall}
	For $\tau\geq\frac{\ln2}{\Lambda_0}$, the capacity with non-perfect receiver $C_{T_s,\tau}$ for fixed $T_s$ increases with $\tau$.
	\begin{proof}
		Please refer to Appendix \ref{appen.sisomonosmall}.
	\end{proof}
\end{theorem} 
\begin{theorem}\label{theo.sisomonolam0}
	For $\Lambda_0=0$, the capacity with non-perfect receiver $C_{\tau,\tau}$ for $T_s=\tau$ decreases with $\tau$ for any $\tau\geq \frac{A_{th_{2}}}{A}$, where $A_{th_2}$ is given by Theorem~\ref{theo.sisomonopower}. 
	\begin{proof}
		Please refer to Appendix \ref{appen.sisomonolam0}.
	\end{proof}
\end{theorem}

\section{Numerical Results}\label{sect.numer}
\subsection{Numerical Results on the Achievable Rate for Long Symbol Duration}\label{sect.numer1}
\begin{figure}
	\setlength{\abovecaptionskip}{-0.1cm}
	\centering
	{\includegraphics[angle=0, width=0.8\textwidth]{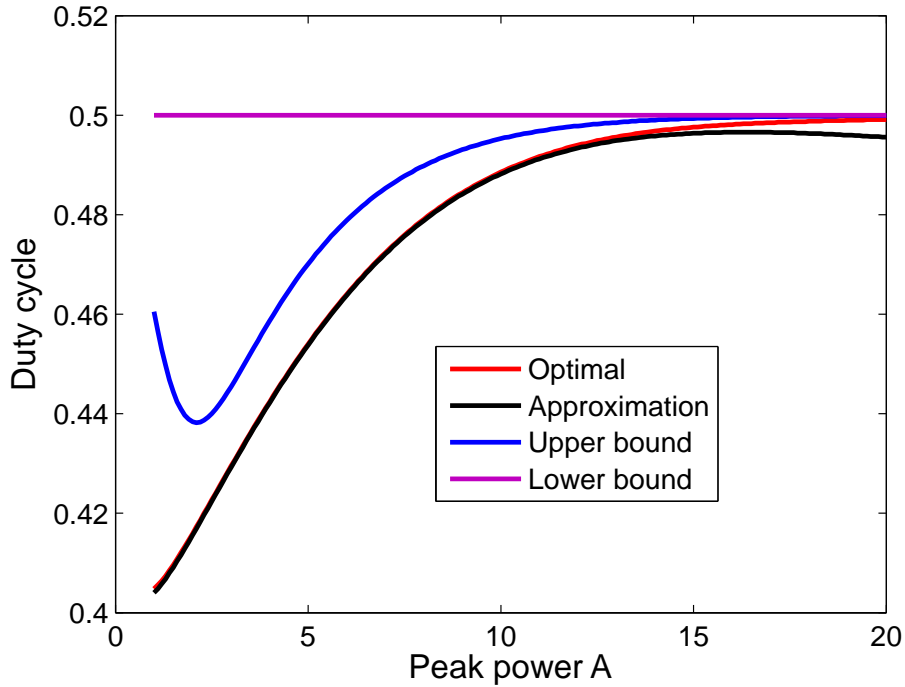}}
	\caption{The optimal/suboptimal duty cycle $\mu$ versus peak power $A$ from the brute-force approach, the derived bounds and approximation for $L=20$.}
	\label{fig.optimal_duty_max}
\end{figure}

\begin{figure}
	\setlength{\abovecaptionskip}{-0.1cm}
	\centering
	{\includegraphics[angle=0, width=0.8\textwidth]{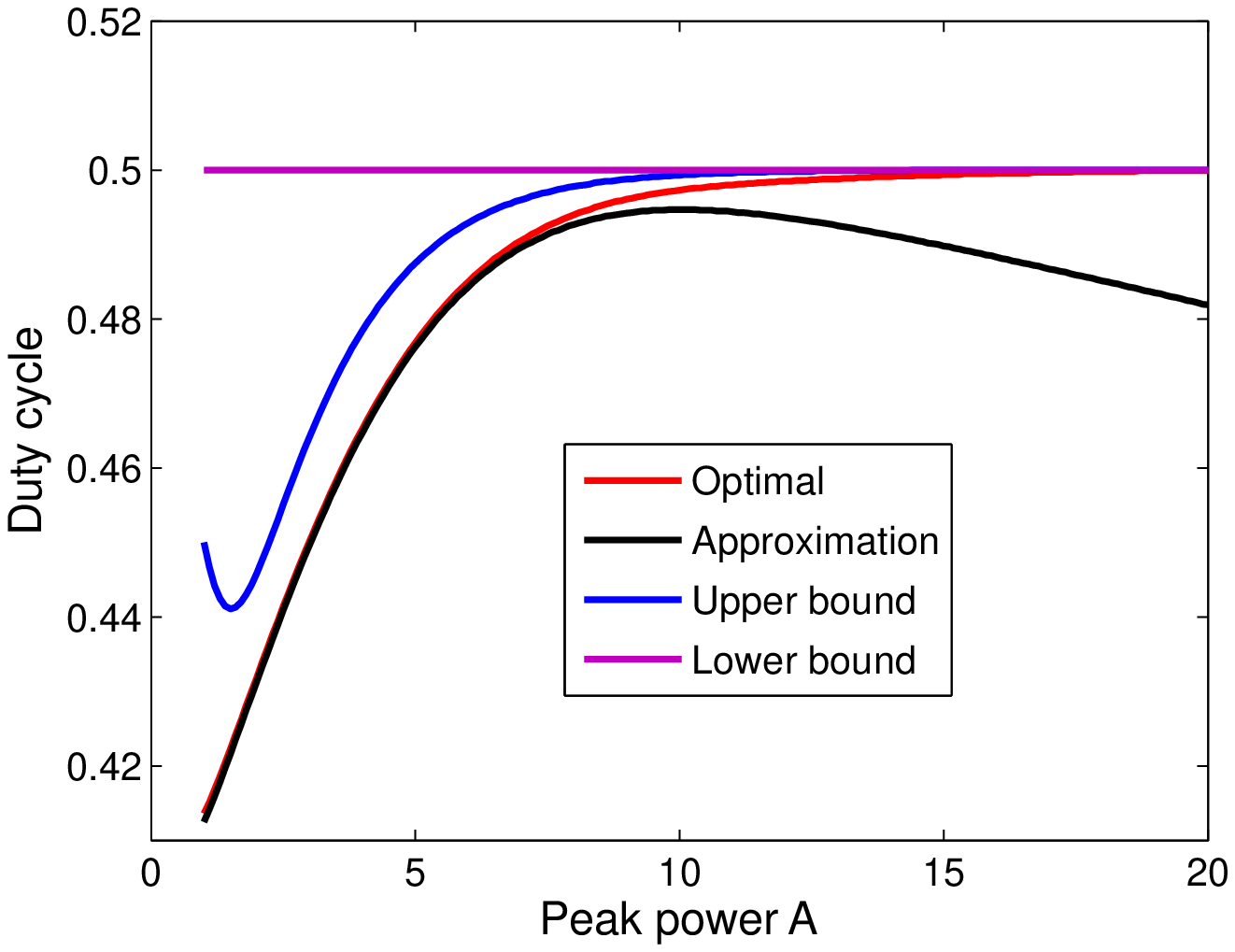}}
	\caption{The optimal/suboptimal duty cycle $\mu$ versus peak power $A$ from the brute-force approach, the derived bounds and the approximation for $L=30$.}
	\label{fig.optimal_duty_max2}
\end{figure}

Assume photon-counting receiver with OOK modulation. We adopt the following system parameters: symbol rate is set to $1$Msps; dead time $20$ns \cite{wang20181mbps}; background radiation arrival intensity $20000$s$^{-1}$, such that the normalized dead time is $0.02$ and the normalized background photon rate is $0.02$. For simplicity, we adopt normalized dead time, peak power, background radiation arrival intensity. For practical system, the symbol duration is typically $200$ns to $1000$ns and far exceeds the dead time that is typically $10$ns to $20$ns. We first investigate the optimal duty cycle for binominal channel by brute-force method (red full line), the suboptimal duty cycle by approximation based on Equation (\ref{eq.appromut}) (black full line), and the lower and upper bounds (blue and purple full line) with respect to peak power $A$, for $L=20$ and $L=30$, as shown in Figure~\ref{fig.optimal_duty_max} and Figure~\ref{fig.optimal_duty_max2}, respectively. It is seen that the optimal duty cycle and proposed suboptimal duty cycle from the derived lower and upper bounds approach $0.5$ as the peak power approaches infinity, i.e., the proposed suboptimal duty cycle from the derived lower and upper bounds is asymptotically optimal for large peak power. In addition, the proposed suboptimal duty cycle converges to optimal duty cycle faster for a larger sampling number $L$. For large peak power and large $L$, the suboptimal duty cycle by approximation method is less accurate due to the omitted larger coefficient one-order term in Equation (\ref{eq.binentropy}) \cite{jacquet1999}.

\begin{figure}
	\setlength{\abovecaptionskip}{-0.1cm}
	\centering
	{\includegraphics[angle=0, width=0.8\textwidth]{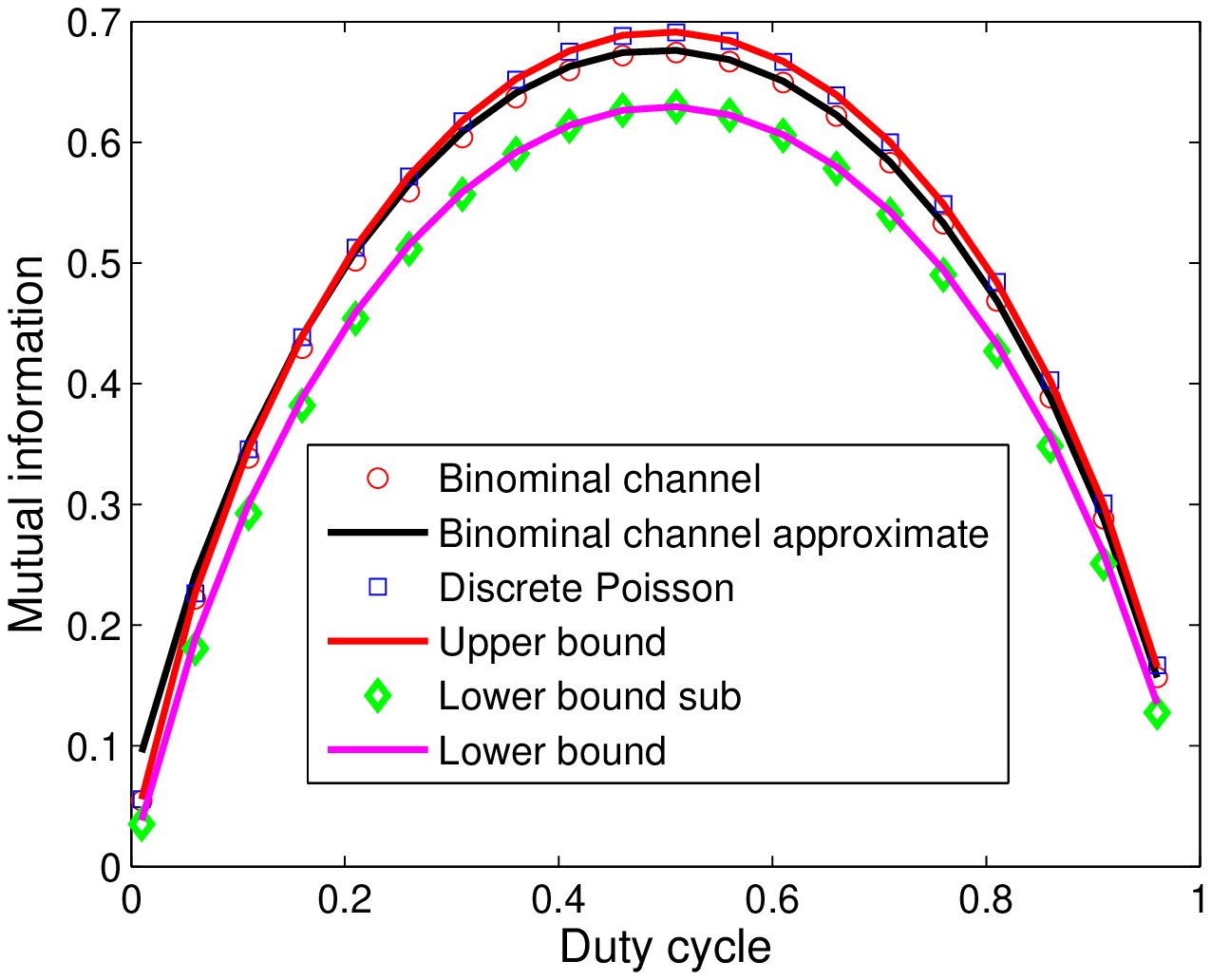}}
	\caption{The mutual information versus duty cycle $\mu$ from simulation, the derived bounds and approximation.}
	\label{fig.optimal_mutual}
\end{figure}

\begin{figure}
	\setlength{\abovecaptionskip}{-0.1cm}
	\centering
	{\includegraphics[angle=0, width=0.8\textwidth]{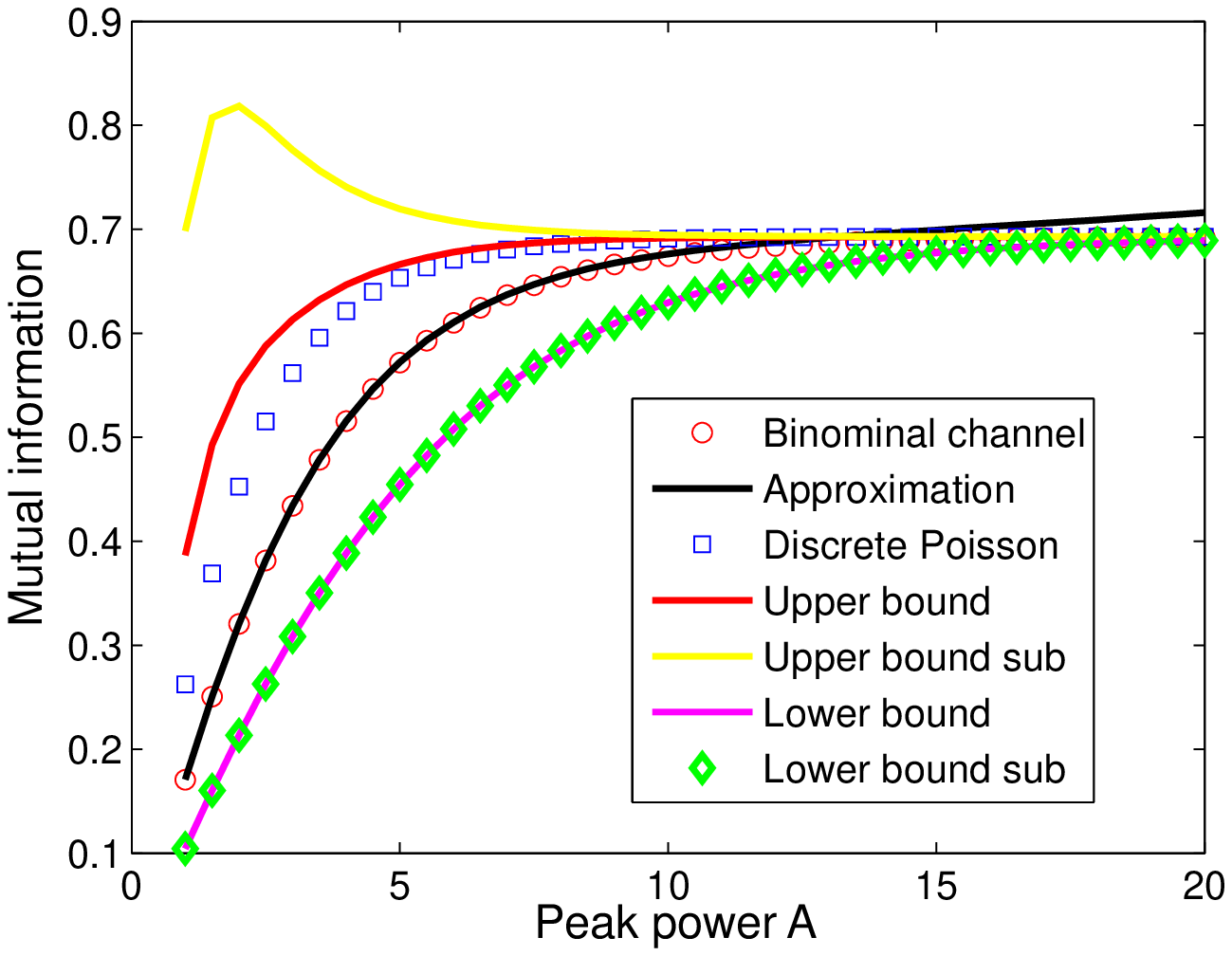}}
	\caption{The maximum mutual information over duty cycle $\mu$ versus peak power $A$ from simulation, the derived bounds and approximation.}
	\label{fig.optimal_mutual_max}
\end{figure}

For mutual information, Figure~\ref{fig.optimal_mutual} shows the mutual information of binominal channel, discrete Poisson channel, along with the derived upper and lower bounds and the approximation based on Equation (\ref{eq.appromut}) with respect to the duty cycle. The normalized dead time, background radiation, peak power and sampling numbers are set to $0.02$, $0.02$, $10$ and $30$, respectively. It is seen that the proposed upper bound and lower bound are more accurate in low or large duty cycle and the approximation is more accurate for medium and large duty cycle. The mutual information of discrete Poisson channel is also plotted as a benchmark to show the small loss due to imperfect receiver. ``Lower bound" and ``Lower bound sub" curves are obtained by brute-force search on $\alpha$ and suboptimal $\alpha$ in Lemma \ref{lemma.suboptalpha}, respectively. Figure~\ref{fig.optimal_mutual_max} shows the maximum mutual information over duty cycle $\mu$ with respect to peak power. The maximum mutual information with respect to duty cycle $\mu$ for binominal channel, approximation method, discrete Poisson channel, the lower bound and the upper bound are obtained by brute-force search, and that for ``lower bound sub" and ``upper bound sub" are obtained from Lemma \ref{lemma.suboptalpha} and Lemma \ref{lemma.beta2}, respectively. It is seen that proposed upper bound and lower bound become more accurate as peak power $A$ increases, and the approximation is more accurate in low and medium peak power regimes.

\begin{figure}
	\setlength{\abovecaptionskip}{-0.1cm}
	\centering
	{\includegraphics[angle=0, width=0.8\textwidth]{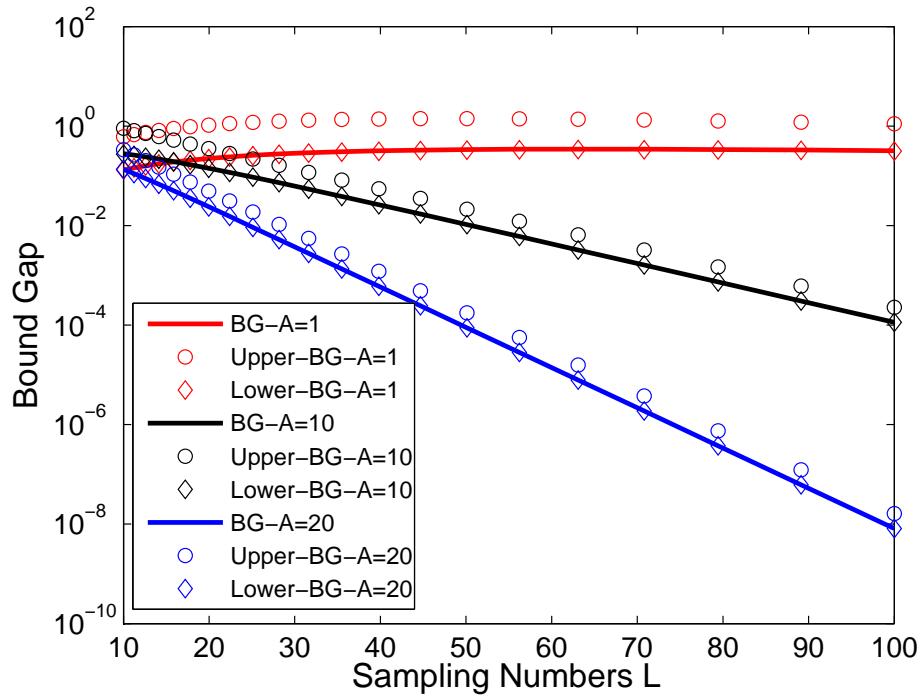}}
	\caption{The derived upper and lower bounds on $\Delta(\beta,\beta_1,\beta_2)$ versus sampling numbers $L$ for different peak power.}
	\label{fig.Boundgap1a}
\end{figure}

\begin{figure}
	\setlength{\abovecaptionskip}{-0.1cm}
	\centering
	{\includegraphics[angle=0, width=0.8\textwidth]{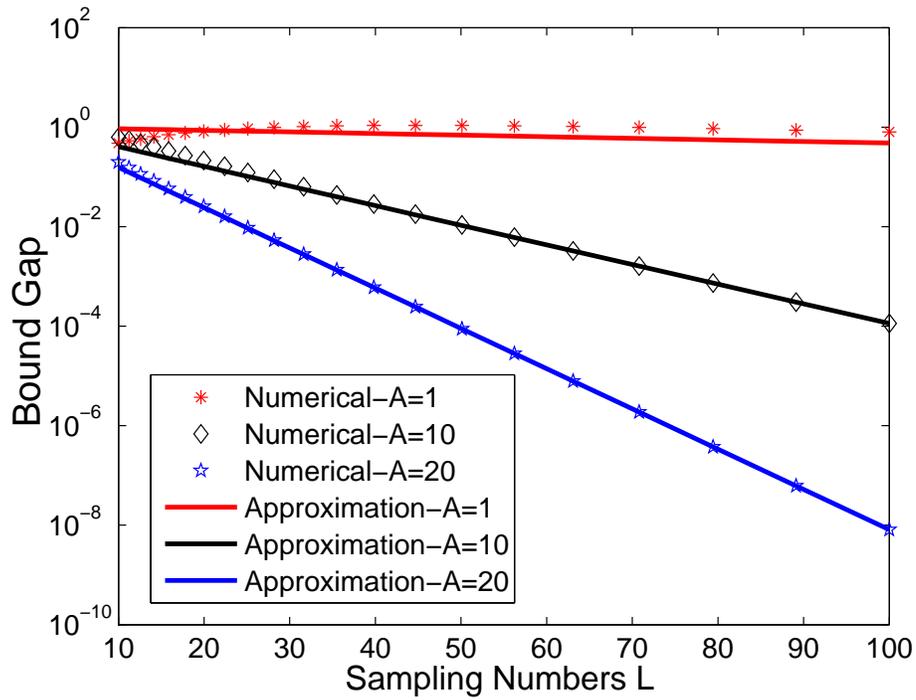}}
	\caption{The gap of derived upper and lower bounds on $\Delta(\beta,\beta_1,\beta_2)$ versus sampling numbers $L$ from numerical computation and theoretical derivations for different peak power values $A$.}
	\label{fig.Boundgap1}
\end{figure}

Consider the asymptotic tightness of the proposed upper and lower bounds. The normalized dead time and background radiation are both
set to $0.02$. We  focus on the five scenarios addressed in Section \ref{sect.asym}. For large sampling numbers $L$, Figure~\ref{fig.Boundgap1a} plots the bound gap by numerical method and the derived upper and lower bounds against sampling numbers $L$ for different peak power values $A$. It is seen that the proposed upper and lower bounds on gap become tighter as the peak power increases. Figure~\ref{fig.Boundgap1} shows the numerical values and the exponential term from Equation (\ref{eq.exponential1}) of $\Delta(\beta,\beta_1,\beta_2)$ against sampling numbers $L$ for different peak power values $A$. It is seen that the proposed upper and lower bounds converge to $0$ with exponential rate as predicted by Equation (\ref{eq.exponential1}). The normalized dead time and background radiation are set to $0.02$. 

\begin{figure}
	\setlength{\abovecaptionskip}{-0.1cm}
	\centering
	{\includegraphics[angle=0, width=0.8\textwidth]{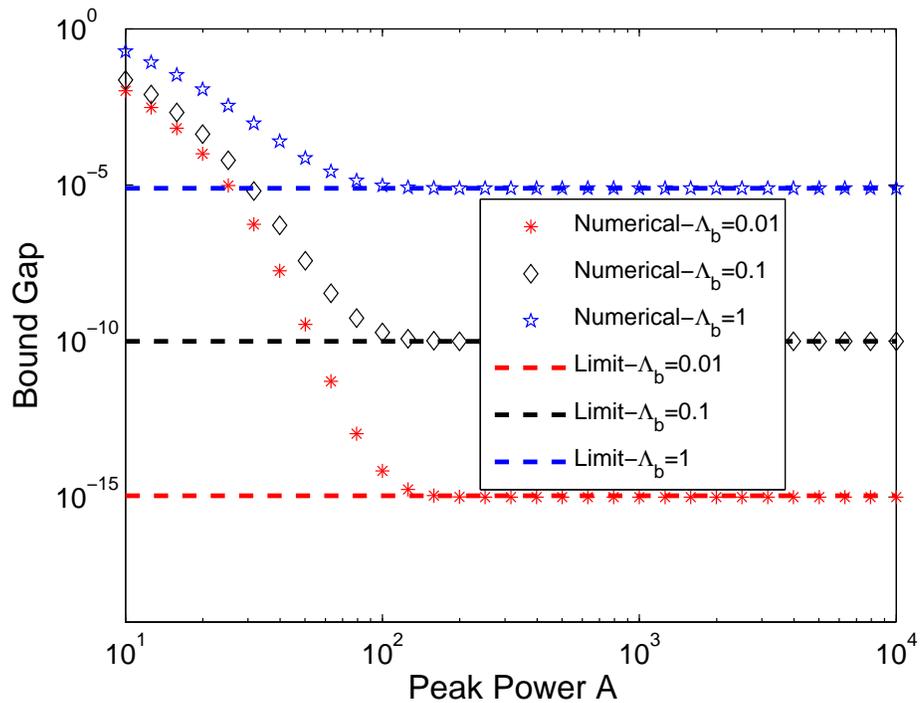}}
	\caption{The bound gap versus large peak power $A$ from simulation and the limit for different background radiation arrival intensities $\Lambda_0$.}
	\label{fig.Boundgap2}
\end{figure}

\begin{figure}
	\setlength{\abovecaptionskip}{-0.1cm}
	\centering
	{\includegraphics[angle=0, width=0.8\textwidth]{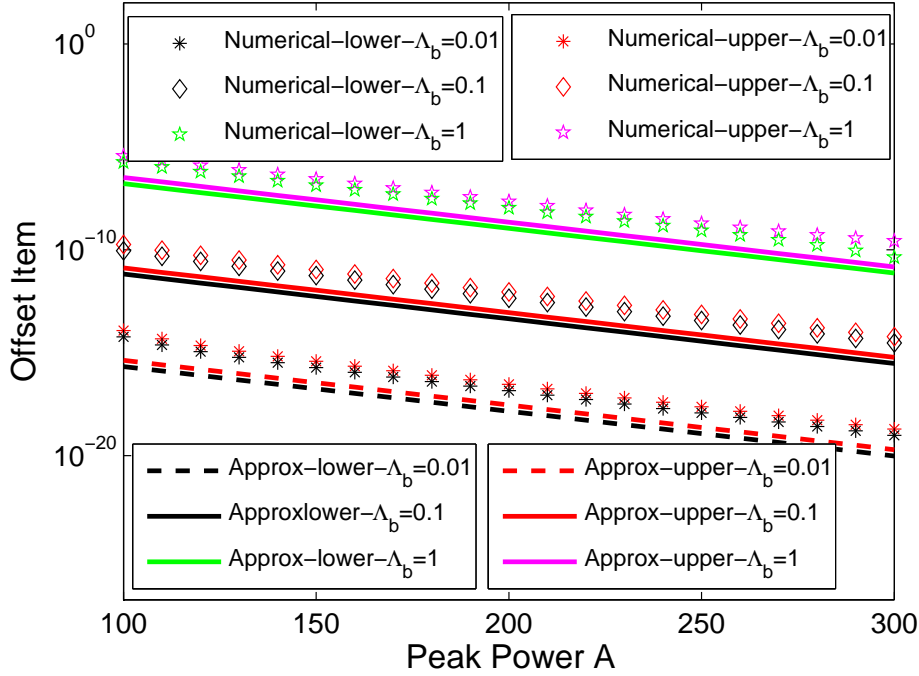}}
	\caption{The offset item $\epsilon_u+o(\epsilon_u)$ and $\epsilon_l+o(\epsilon_l)$ versus large peak power $A$ from numerical computation and exponential approximation for different background radiation arrival intensities $\Lambda_0$.}
	\label{fig.Boundgap2b}
\end{figure}

Set the normalized dead time and sampling numbers to $0.1$ and $10$, respectively. For large peak power $A$ given fixed background radiation arrival intensity $\Lambda_0$, Figure~\ref{fig.Boundgap2} plots the difference of derived upper and lower bounds on $\Delta(\beta,\beta_1,\beta_2)$ against peak power $A$ for different background radiation arrival intensity $\Lambda_0$, from both numerical computations and the limit from Equation (\ref{eq.deltacoonvB}) via omitting the vanishing terms. It is seen that the gap converges as $A$ increases beyond $100$. Figure~\ref{fig.Boundgap2b} plots the offset items in the derived upper and lower bounds from Equations (\ref{eq.offsetA1}) and (\ref{eq.offsetA2}), respectively, against peak power $A$ for different background radiation arrival intensity $\Lambda_0$. The approximation values are obtained from the exponential terms. It is seen that the derived upper and lower bounds on the offset terms can well predict the true value with the same attenuation rate. In addition, the gap converges to $0$ exponentially with the peak power.

\begin{figure}
	\setlength{\abovecaptionskip}{-0.1cm}
	\centering
	{\includegraphics[angle=0, width=0.8\textwidth]{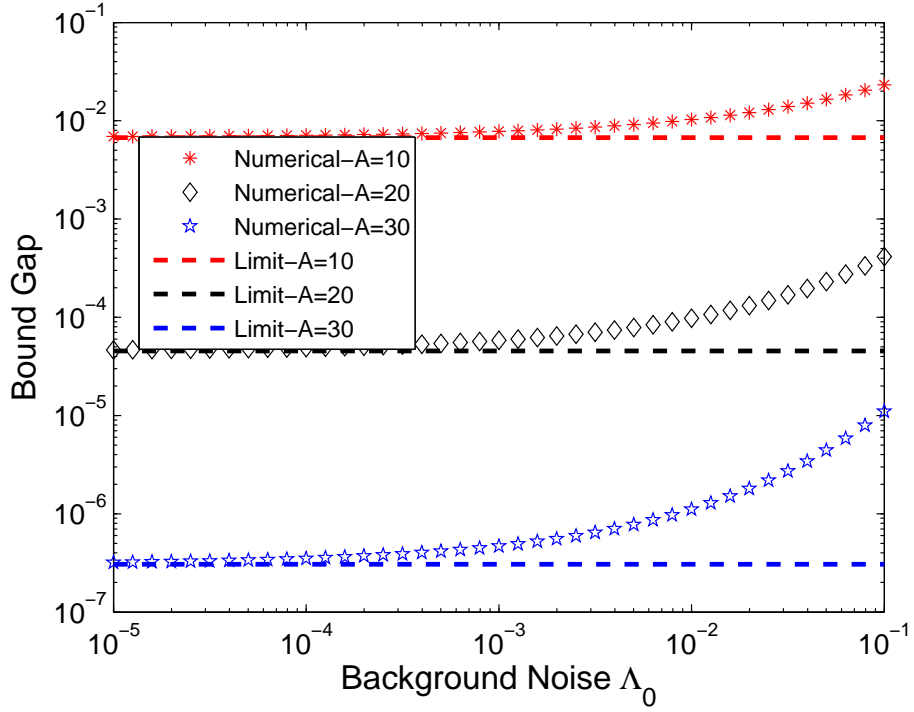}}
	\caption{The bound gap versus low background radiation arrival intensity $\Lambda_0$ from numerical computation and theoretical limit for different peak power $A$.}
	\label{fig.Boundgap3}
\end{figure}

\begin{figure}
	\setlength{\abovecaptionskip}{-0.1cm}
	\centering
	{\includegraphics[angle=0, width=0.8\textwidth]{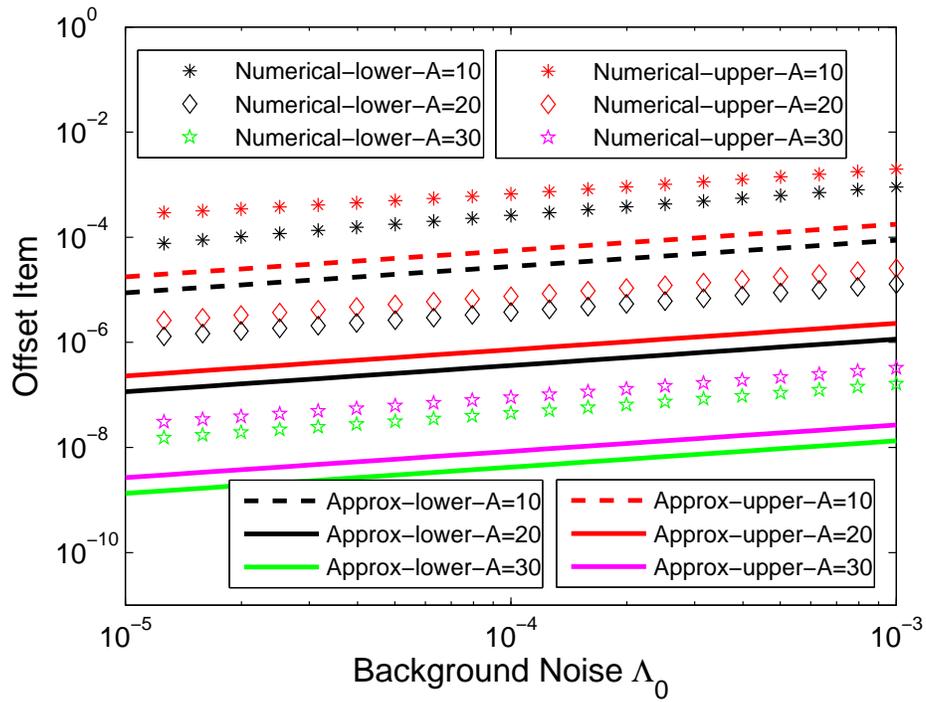}}
	\caption{The offset item $\epsilon^{'}_u+o(\epsilon^{'}_u)$ and $\epsilon^{'}_l+o(\epsilon^{'}_l)$ versus low background radiation arrival intensity $\Lambda_0$ from numerical computation and exponential approximation for different peak power $A$.}
	\label{fig.Boundgap3b}
\end{figure}

Consider low background radiation arrival intensity $\Lambda_0$ given fixed peak power $A$, where the normalized dead time and sampling numbers are
set to 0.1 and 10, respectively. Figure~\ref{fig.Boundgap3} plots the difference of derived upper and lower bounds on $\Delta(\beta,\beta_1,\beta_2)$ against background radiation arrival intensity $\Lambda_0$ for different peak power $A$. It is seen that the limit of the gap can well predict the true value. Figure~\ref{fig.Boundgap3b} plots the offset item in the derived upper and lower bounds from Equations (\ref{eq.offsetB1}) and (\ref{eq.offsetB2}), against background radiation arrival intensity $\Lambda_0$ for different peak power $A$. It is seen that the offset items in the derived upper and lower bounds can well predict those from numerical computation. In addition, the gap between the numerical computation and theoretical approximation converges to $0$ with linear rate for low peak power. 

\begin{figure}
	\setlength{\abovecaptionskip}{-0.1cm}
	\centering
	{\includegraphics[angle=0, width=0.8\textwidth]{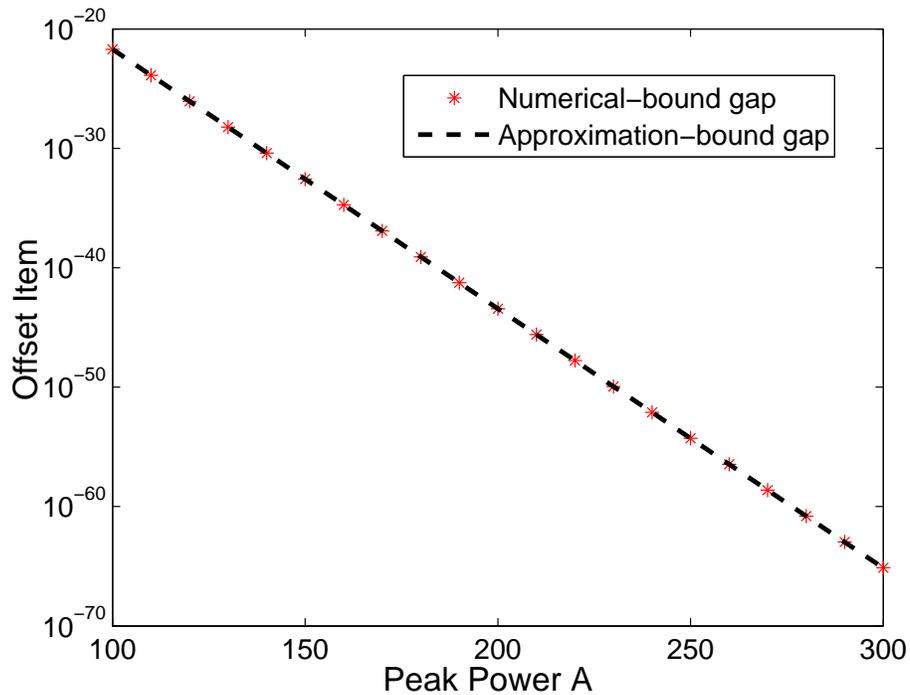}}
	\caption{The bound gap versus large peak power $A$ from numerical computation and theoretical approximation for background radiation arrival intensity $\Lambda_0=0$.}
	\label{fig.Boundgap4b}
\end{figure}

\begin{figure}
	\setlength{\abovecaptionskip}{-0.1cm}
	\centering
	{\includegraphics[angle=0, width=0.8\textwidth]{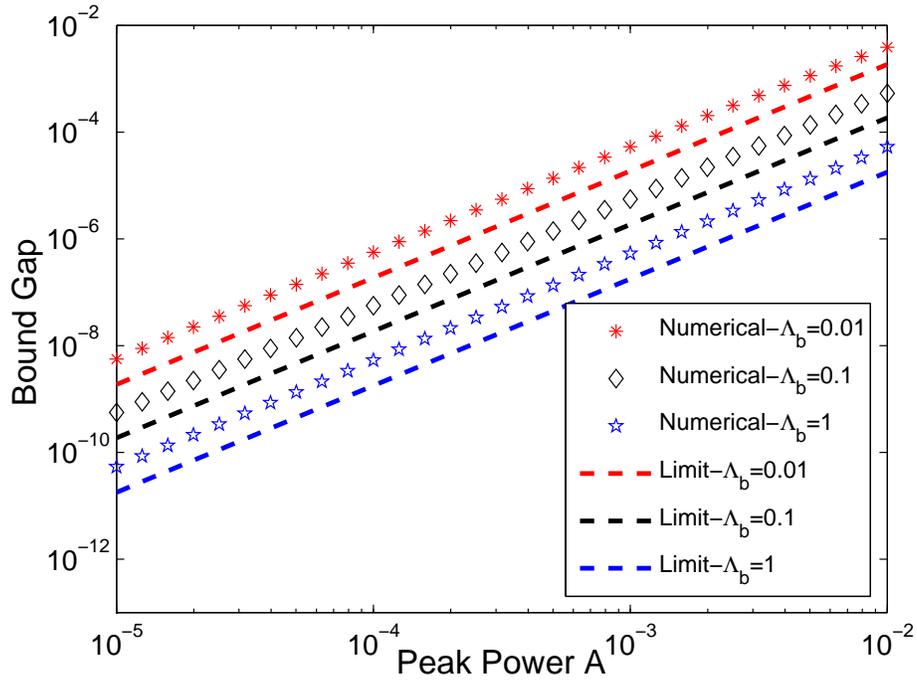}}
	\caption{The difference of derived upper and lower bounds on $\Delta(\beta,\beta_1,\beta_2)$ versus low peak power $A$ from numerical computation and theoretical approximation for different background radiation arrival intensities $\Lambda_0$.}
	\label{fig.Boundgap5}
\end{figure}

Consider large peak power $A$ given background radiation arrival intensity $\Lambda_0=0$ where the normalized dead time and sampling numbers are set to $0.1$, $10$, respectively. Figure~\ref{fig.Boundgap4b} plots the gap between derived upper and lower bounds on $\Delta(\beta,\beta_1,\beta_2)$ from Equations (\ref{eq.deltaconvD1}) and (\ref{eq.deltaconvD2}), respectively, against peak power $A$. It is seen that the gap from theoretical derivations can well predict the numerical results. The normalized dead time and sampling numbers are set to $0.1$, $10$, respectively. For low peak power $A$ with the same normalized dead time and sampling numbers, Figure~\ref{fig.Boundgap5} plots the numerical values and theoretical approximations of the derived bounds gap on $\Delta(\beta,\beta_1,\beta_2)$ against peak power $A$ for different background radiation arrival intensity $\Lambda_0$. It is seen that the approximation via dropping $o(A^2)$ item (denoted as ``Limit'') from Equation (\ref{eq.asym5}) can well predict that from numerical computation, which converges to $0$ in the rate of order two predicted by Equation (\ref{eq.asym5}) for low peak power.
\subsection{Numerical Results on the Capacity for Arbitrary Symbol Duration}
It has been concluded that the case of non-perfect receiver for $T_s\geq\tau$ can be converted to that of non-perfect receiver for $T_s=\tau$. Hence, we investigate the case for $T_s=\tau$.

\begin{figure}
	\setlength{\abovecaptionskip}{-0.1cm}
	\centering
	{\includegraphics[angle=0, width=0.8\textwidth]{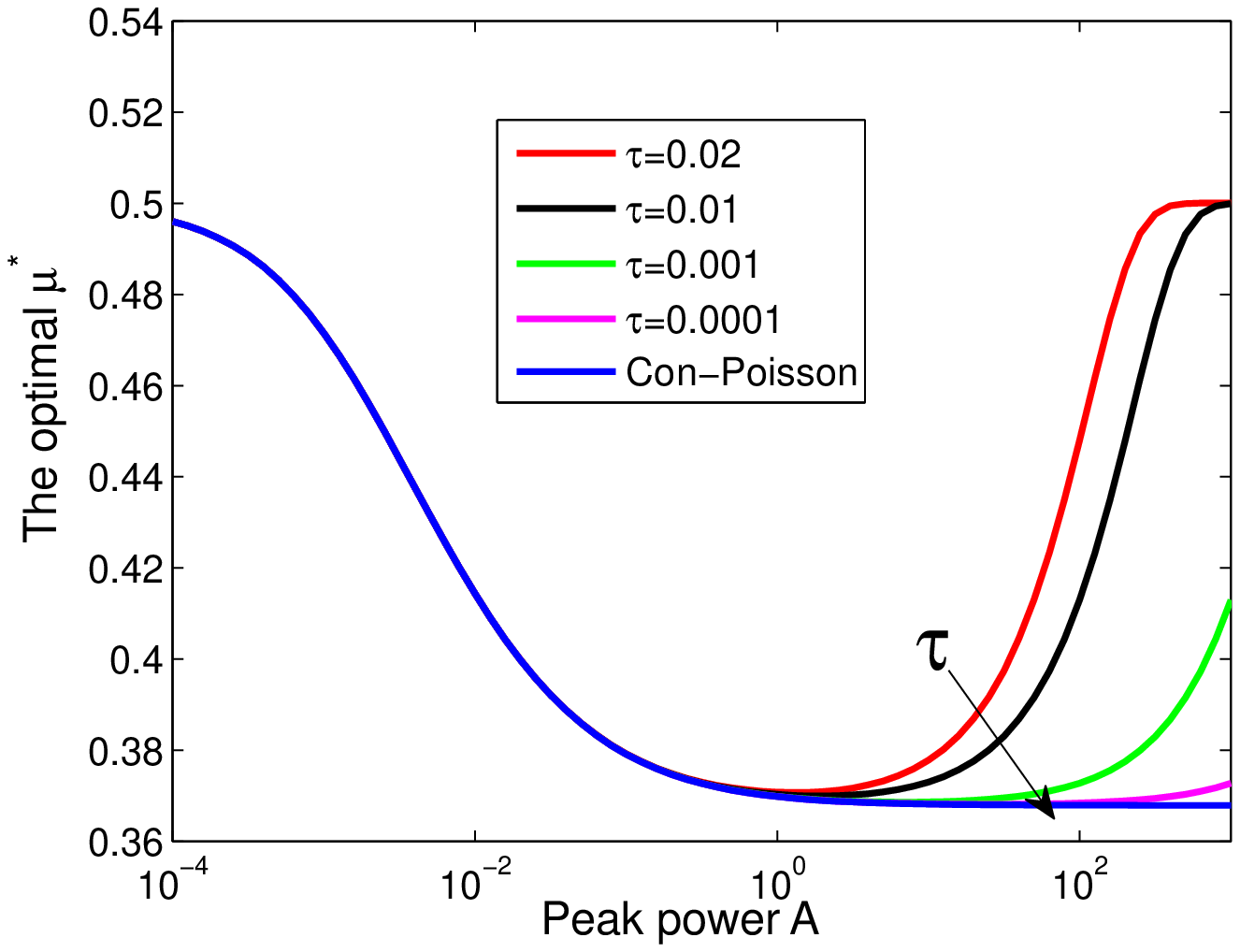}}
	\caption{The optimal duty cycle versus peak power $A$ given $\Lambda_0=0.001$ for different normalized dead time $\tau$.}
	\label{fig.SISOoptmu}
\end{figure}

\begin{figure}
	\setlength{\abovecaptionskip}{-0.1cm}
	\centering
	{\includegraphics[angle=0, width=0.8\textwidth]{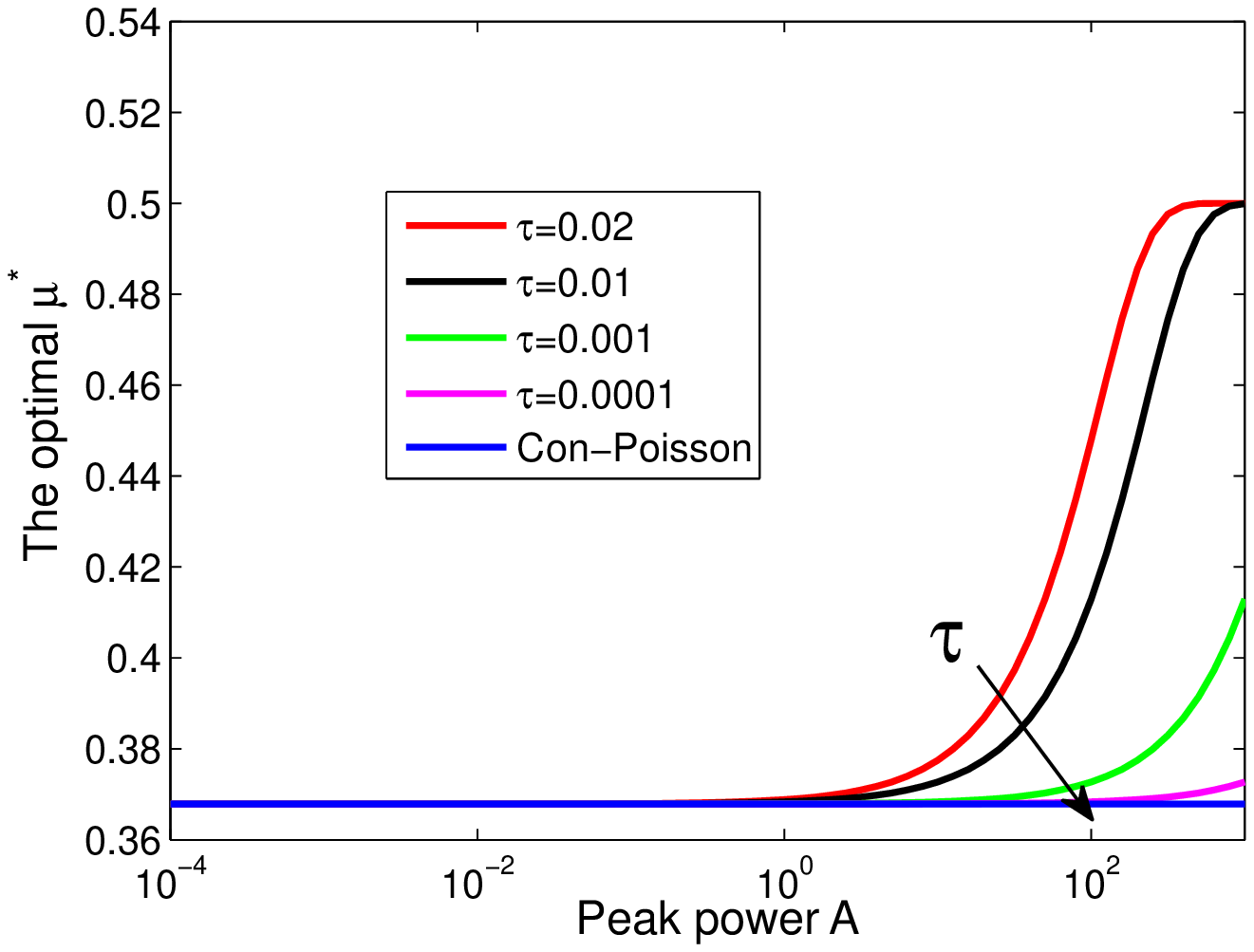}}
	\caption{The optimal duty cycle versus peak power $A$ given $\Lambda_0=0$ for different dead time $\tau$.}
	\label{fig.SISOoptmu2}
\end{figure}

\begin{figure}
	\setlength{\abovecaptionskip}{-0.1cm}
	\centering
	{\includegraphics[angle=0, width=0.8\textwidth]{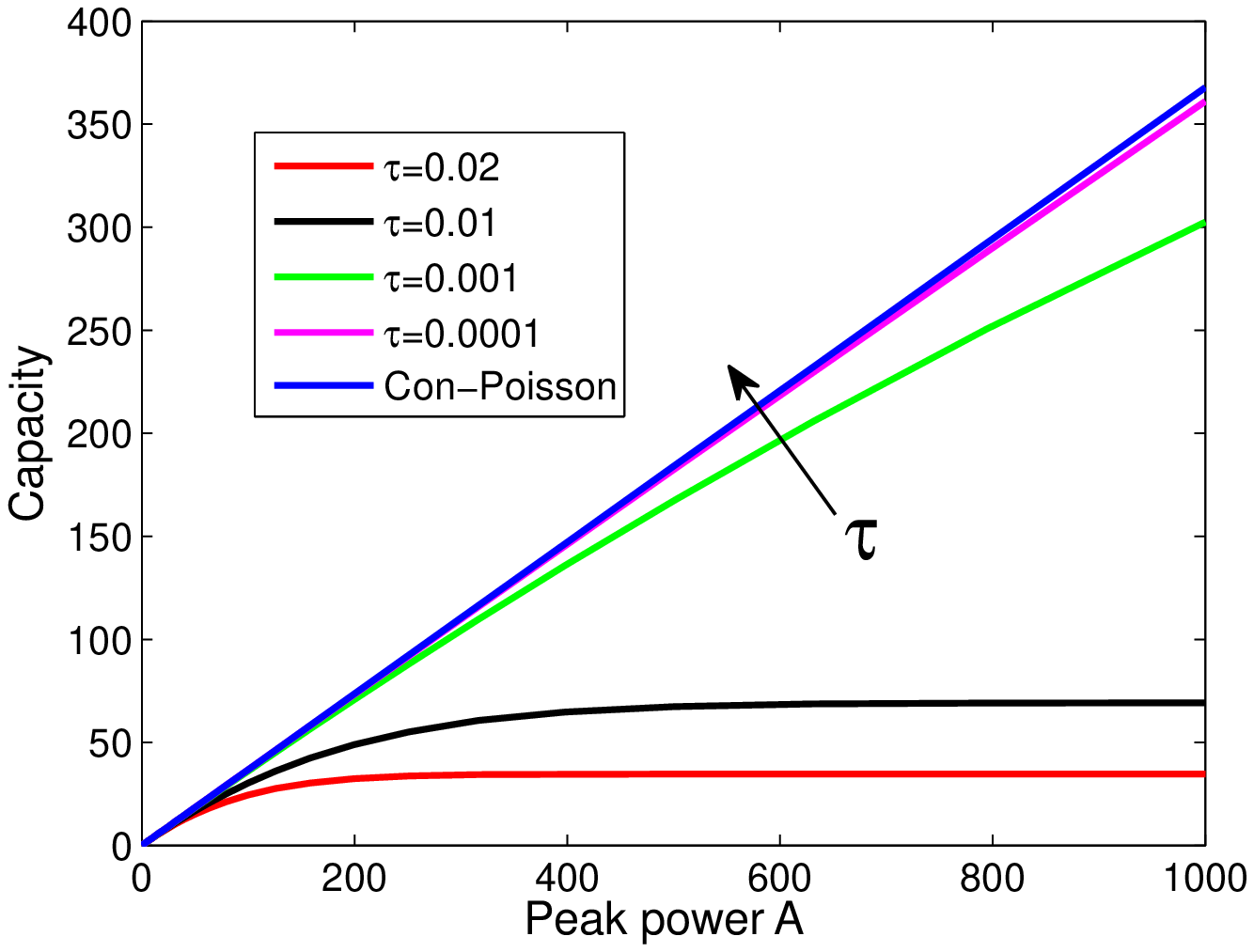}}
	\caption{The non-perfect receiver capacity versus peak power $A$ given $\Lambda_0=0.001$ for different dead time $\tau$.}
	\label{fig.SISOoptmutu}
\end{figure}

\begin{figure}
	\setlength{\abovecaptionskip}{-0.1cm}
	\centering
	{\includegraphics[angle=0, width=0.8\textwidth]{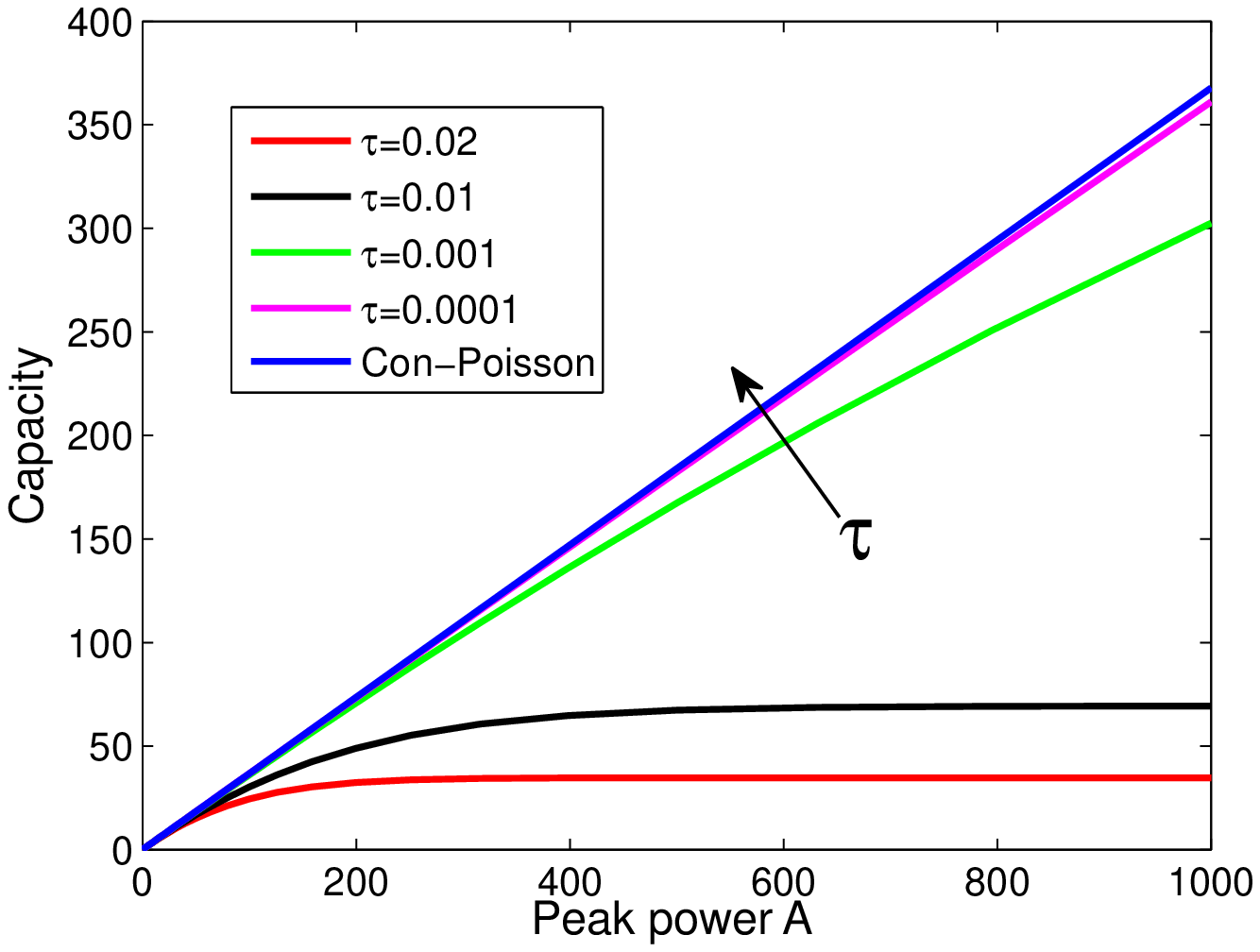}}
	\caption{The non-perfect receiver capacity versus peak power $A$ given $\Lambda_0=0$ for different dead time $\tau$.}
	\label{fig.SISOoptmutu2}
\end{figure}

Consider the same receiver parameters as those in Section~\ref{sect.numer1}. The optimal duty cycle versus $A$ for different dead time and $\Lambda_0=0.001$ and $\Lambda_0=0$ are shown in Figure~\ref{fig.SISOoptmu} and Figure~\ref{fig.SISOoptmu2}, respectively. It is seen that the optimal duty cycle converges to that of continuous Poisson channel, while asymptotic duty cycles for large peak power are more different. Similarly, the non-perfect receiver capacity versus peak power for different dead time given $\Lambda_0=0.001$ and $\Lambda_0=0$ are shown in Figure~\ref{fig.SISOoptmutu} and Figure~\ref{fig.SISOoptmutu2}, respectively. It is seen that the capacity with non-perfect receiver converges to that of continuous Poisson channel. Moreover, the capacity with non-perfect receiver converges for large peak power given dead time $\tau$, while the capacity of continuous Poisson channel linearly increases with peak power. The gap in large peak power regime stems from the photon-counting loss.

\begin{figure}
	\setlength{\abovecaptionskip}{-0.1cm}
	\centering
	{\includegraphics[angle=0, width=0.8\textwidth]{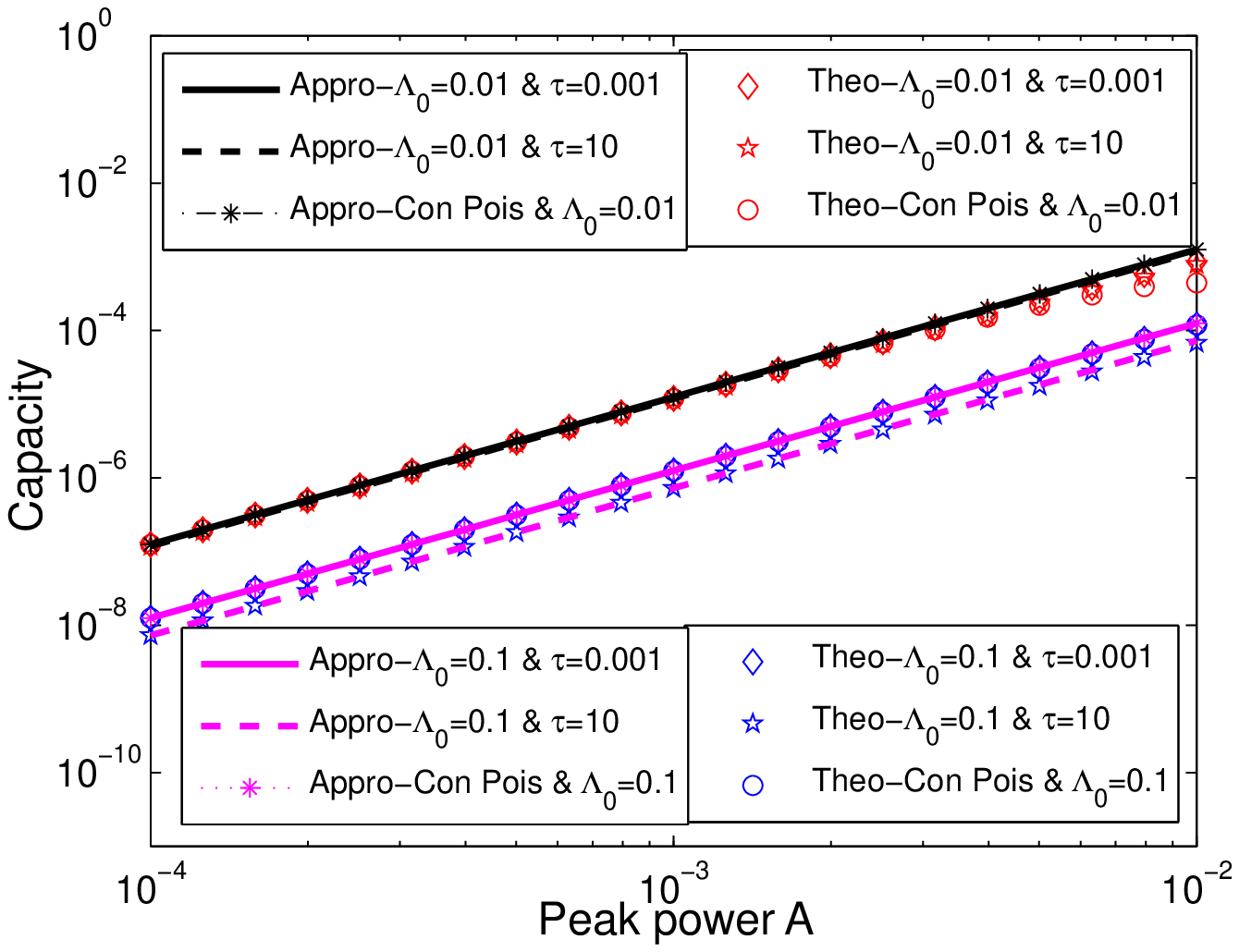}}
	\caption{The non-perfect receiver capacity, continuous Poisson channel capacity and the corresponding approximation versus low peak power $A$ for different $\Lambda_0>0$.}
	\label{fig.capalowp}
\end{figure}

\begin{figure}
	\setlength{\abovecaptionskip}{-0.1cm}
	\centering
	{\includegraphics[angle=0, width=0.8\textwidth]{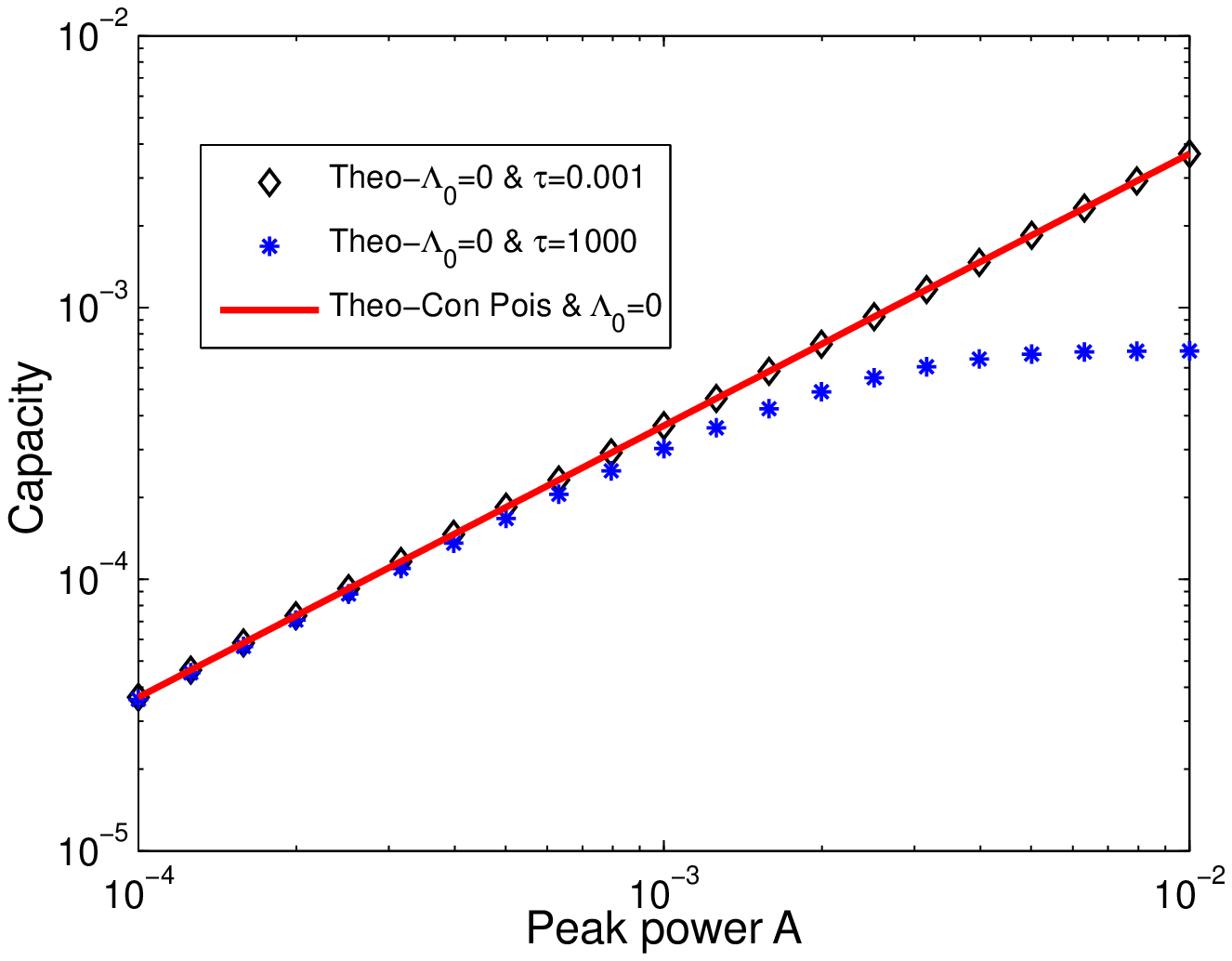}}
	\caption{The non-perfect receiver capacity, continuous Poisson channel capacity and the corresponding approximation versus low peak power for different $\Lambda_0=0$.}
	\label{fig.capalowp2}
\end{figure}

\begin{figure}
	\setlength{\abovecaptionskip}{-0.1cm}
	\centering
	{\includegraphics[angle=0, width=0.8\textwidth]{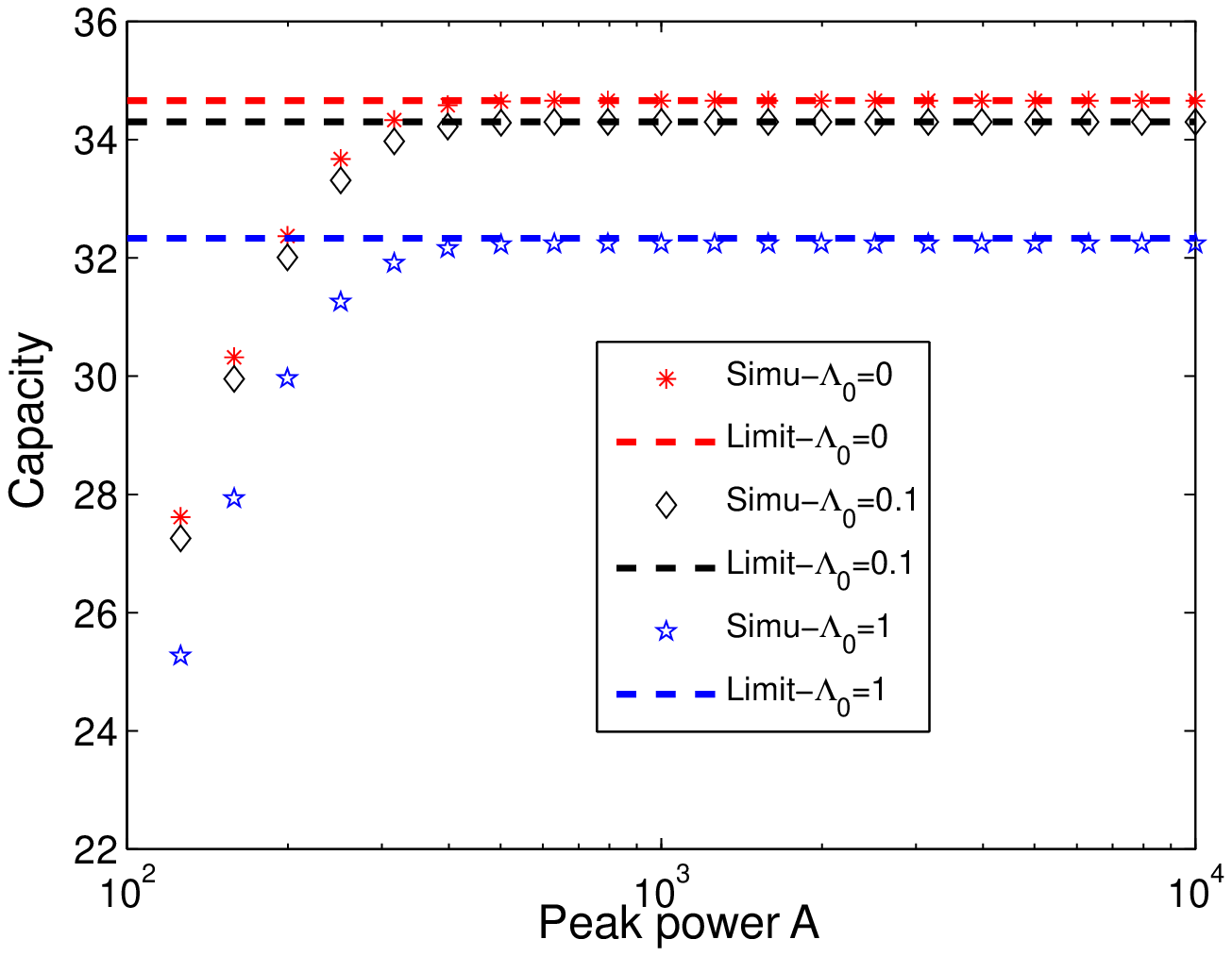}}
	\caption{The non-perfect receiver capacity, continuous Poisson channel capacity and the theoretical limitation versus large peak power $A$ for different $\Lambda_0$.}
	\label{fig.capalargep}
\end{figure}

We then analyze the asymptotic property for the capacity with non-perfect receiver. Figure~\ref{fig.capalowp} and Figure~\ref{fig.capalowp2} show the non-perfect receiver capacity, continuous Poisson capacity and the approximation versus low peak power $A$ for different $\Lambda_0>0$ and $\Lambda_0=0$, respectively. Prefix ``Theo-" denotes the exact capacity with non-perfect receiver shown in Theorem~\ref{theor.siso} and prefix ``Appro-" represents the dominant term approximation of non-perfect receiver given by Equation (\ref{eq.asymcapa2}) and Theorem~\ref{theo.sisoasymclambda2}. It is seen that the dominant term approximation is close to the exact value for low peak power. Figure~\ref{fig.capalargep} shows the capacity of non-perfect receiver and the corresponding capacity limit given in Equation (\ref{eq.asymcapa1}) and Theorem~\ref{theo.asymcapalargepeak} for dead time $\tau=0.02$. Numerical results shows that the capacity with non-perfect receiver is close to the saturation capacity for peak power $A>10^3$.



\section{Conclusion}\label{sect.conc}
We have investigated the achievable rate and capacity of a practical photon counting receiver with positive dead time and finite sampling rate. For the symbol duration that cannot be sufficiently small, we have proposed upper and lower bounds on the achievable rate based on Kullback-Leibler (KL) divergence and Chernoff $\alpha$-divergence, and shown the tightness of the proposed bounds. The convergence rate of proposed bounds is investigated for five scenarios. Moreover, an approximation on the achievable rate is proposed, which is more accurate compared with the proposed upper and lower bounds in the medium signal to noise ratio (SNR) regime. For the symbol duration that can be arbitrarily small, we investigate the capacity and the optimal signal distribution for the non-perfect receiver. We demonstrate that the continuous Poisson capacity equals that of non-perfect receiver with $T_s=\tau\to0$. Furthermore, the asymptotic capacity and the capacity loss from continuous Poisson channel for low and large peak power are characterized. The results on the achievable rate, the capacity, the signal distribution, the gap between the upper and lower bounds, and the loss from the continuous Poisson channel are validated by the numerical results.

\appendices
\renewcommand{\baselinestretch}{1.4}
\section{The proof of main results on achievable rate for Long Symbol Duration}
\subsection{Proof of Lemma \ref{lemma.suboptalpha}}\label{appen.suboptalpha}
Note that $C_{1-\alpha}(P_0^B||P_1^B)=C_{\alpha}(P_1^B||P_0^B)\dff-L\ln f(\alpha)$, where $f(\alpha|p_0,p_1)=p_1^{\alpha}p_0^{1-\alpha}+(1-p_1)^{\alpha}(1-p_0)^{1-\alpha}$, we have $f^{'}(\alpha|p_0,p_1)=p_0(\frac{p_1}{p_0})^{\alpha}\ln\frac{p_1}{p_0}+(1-p_0)(\frac{1-p_1}{1-p_0})^{\alpha}\ln\frac{1-p_1}{1-p_0}$,
$f^{''}(\alpha|p_0,p_1)>0$, $f^{'}(0|p_0,p_1)=-KL(p_0||p_1)$ and $f^{'}(1|p_0,p_1)=KL(p_1||p_0)>0$. Thus, the optimal $\alpha^{\vartriangle}$ to maximize $C_{\alpha}(P_1^B||P_0^B)$ uniquely exists and satisfies $f^{'}(\alpha^{\vartriangle})=0$, i.e.,
\be\label{eq.alpha}
\alpha^{\vartriangle}(p_0,p_1)=\frac{\ln\frac{1-p_0}{p_0}+\ln\ln\frac{1-p_0}{1-p_1}-\ln\ln\frac{p_1}{p_0}}{\ln\frac{p_1(1-p_0)}{p_0(1-p_1)}}.
\ee 
Since the symmetry $f(\alpha|p_0,p_1)=f(1-\alpha|1-p_1,1-p_0)$, we have $1-\alpha^{\vartriangle}(p_0,p_1)=\alpha^{\vartriangle}(1-p_1,1-p_0)$ and
\be \label{eq.alphar}
\alpha^{\vartriangle}-(1-\alpha^{\vartriangle})&=&\frac{[\ln\frac{1-p_0}{p_0}+\ln\ln\frac{1-p_0}{1-p_1}-\ln\ln\frac{p_1}{p_0}]-[\ln\frac{p_1}{1-p_1}+\ln\ln\frac{p_1}{p_0}-\ln\ln\frac{1-p_0}{1-p_1}]}{\ln\frac{p_1(1-p_0)}{p_0(1-p_1)}}\nonumber\\
&=&\frac{\ln\frac{(1-p_0)(1-p_1)}{p_0p_1}+2\big(\ln\ln\frac{1-p_0}{1-p_1}-\ln\ln\frac{p_1}{p_0}\big)}{\ln\frac{p_1(1-p_0)}{p_0(1-p_1)}}\lesseqqgtr0,\quad \text{if}\quad p_0+p_1\gtreqqless1,
\ee
where the last inequality follows from the fact that the right term of the numerator of Equation (\ref{eq.alphar}) decreases with $p_0$ and becomes $0$ for $p_0=1-p_1$. Based on the above statement, we can readily obtain
$
\alpha^{*}=\arg\max\limits_{0\leq\alpha\leq1}\min\{C_{\alpha}(P_1^B||P_0^B),C_{1-\alpha}(P_1^B||P_0^B)\}=\frac{1}{2}.$

\subsection{Proof of Lemma \ref{lemma.beta}}\label{appen.beta}
Based on symmetry $F_u(\mu,\beta_1,\beta_2)=F_u(1-\mu,\beta_2,\beta_1)$, we have
\be 
\frac{\partial F_u(\cdot,\beta_1,\beta_2)}{\partial\mu}|_{\mu^{*}(\beta_1,\beta_2)}=-\frac{\partial F_u(\cdot,\beta_2,\beta_1)}{\partial\mu}|_{1-\mu^{*}(\beta_2,\beta_1)}=0,
\ee 
i.e., $\mu^{*}(\beta_1,\beta_2)+\mu^{*}(\beta_2,\beta_1)=1$. 
Defining $G(x,\beta)=\ln x-\frac{\beta}{x}$, we have
\be \frac{\partial F_u}{\partial \mu}\big|_{\big(\mu^{*}(\beta_1,\beta_2),\beta_1,\beta_2\big)}&=&G\big(1-(1-\beta_2)\mu,\frac{\beta_1+\beta_2}{2}\big)-G\big(\beta_1+(1-\beta_1)\mu,\frac{\beta_1+\beta_2}{2}\big)\nonumber
\\&&+\frac{\beta_1-\beta_2}{2}\{\frac{1}{1-(1-\beta_2)\mu}+\frac{1}{\beta_1+(1-\beta_1)\mu}\}=0.
\ee
Thus we have $G\big(1-(1-\beta_2)\mu,\frac{\beta_1+\beta_2}{2}\big)-G\big(\beta_1+(1-\beta_1)\mu,\frac{\beta_1+\beta_2}{2}\big)\lessgtr0$ if $\beta_1\gtrless\beta_2$. As $G(x,\beta)$ decreases with $x$, we can obtain $\mu^{*}(\beta_1,\beta_2)\gtrless\frac{1-\beta_1}{2-\beta_1-\beta_2}$ if $\beta_1\gtrless\beta_2$.

\subsection{Proof of Lemma \ref{lemma.beta2}}\label{appen.beta2}
Consider the following three cases.\\
	\textbf{Case 1}: $\beta_1=\beta_2$. According to Equation (\ref{eq.prime}), we have $\max\limits_{0\leq\mu\leq1}F_u(\mu,\beta_1,\beta_2)=-\ln\frac{1+\beta_2}{2}$, i.e., the equality holds.\\
	\textbf{Case 2}: $\beta_1<\beta_2$. According to Lemma \ref{lemma.beta}, we have
	\be 
	\ln\frac{\beta_1+(1-\beta_1)\mu^{*}(\beta_1,\beta_2)}{1-(1-\beta_2)\mu^{*}(\beta_1,\beta_2)}=\frac{\beta_1}{\beta_1+(1-\beta_1)\mu^{*}(\beta_1,\beta_2)}-\frac{\beta_2}{1-(1-\beta_2)\mu^{*}(\beta_1,\beta_2)}<0.
	\ee 
	As $\mu^{*}(\beta_1,\beta_2)<\frac{1-\beta_1}{2-\beta_1-\beta_2}$, we have the following upper bound on $\max\limits_{0\leq\mu\leq1}F_u(\mu,\beta_1,\beta_2)$,
	\be 
	F_u(\mu^{*}(\beta_1,\beta_2),\beta_1,\beta_2)&=&-\ln[1-(1-\beta_2)\mu^{*}(\beta_1,\beta_2)]-\mu^{*}(\beta_1,\beta_2)\nonumber\\
	&&\cdot\{\frac{\beta_1}{\beta_1+(1-\beta_1)\mu^{*}(\beta_1,\beta_2)}-\frac{\beta_2}{1-(1-\beta_2)\mu^{*}(\beta_1,\beta_2)}\}\nonumber\\
	&<&-\ln\frac{1-\beta_1\beta_2}{2-\beta_1-\beta_2}-\frac{\beta_1-\beta_2}{(1-\beta_1\beta_2)/(2-\beta_1-\beta_2)}\mu^{*}(\beta_1,\beta_2)\nonumber\\
	&<&-\ln\frac{1-\beta_1\beta_2}{2-\beta_1-\beta_2}+\frac{(\beta_2-\beta_1)(1-\beta_1)}{(1-\beta_1\beta_2)}.
	\ee 
	\textbf{Case 3}: $\beta_1>\beta_2$. Similarly to Case 2, we have
	\be 
	F_u(\mu^{*}(\beta_1,\beta_2),\beta_1,\beta_2)<-\ln\frac{1-\beta_1\beta_2}{2-\beta_1-\beta_2}+\frac{(\beta_1-\beta_2)(1-\beta_2)}{(1-\beta_1\beta_2)}.
	\ee 

\subsection{Proof of Theorem \ref{theorem.asymI1}}\label{appen.asymI1}
Note that $\beta$, $\beta_1$ and $\beta_2$ approach $0$ as $L$ approaches infinity.
According to Equation (\ref{eq.lowerbound}) and Taylor expansion $\ln(a+x)=\ln a+\frac{1}{a}x+o(x)$, we have 
\be 
I_{max}(\Lambda_0,A,L)\geq-\ln\frac{1+\beta}{2}=\ln2-\beta+o(\beta).
\ee 

For $\beta_1>\beta_2$, since $\frac{1-\beta_1\beta_2}{2-\beta_1-\beta_2}-\frac{1}{2}=\frac{\beta_1+\beta_2-2\beta_1\beta_2}{2(2-\beta_1-\beta_2)}=\frac{\beta_1}{4}+o(\beta_1)$, we have
\be 
I_{max}(\Lambda_0,A,L)&\leq&\frac{|\beta_1-\beta_2|(1-\min\{\beta_1,\beta_2\}\})}{1-\beta_1\beta_2}-\ln\frac{1-\beta_1\beta_2}{2-\beta_1-\beta_2}\nonumber\\
&=&\beta_1+o(\beta_1)+\ln2-\frac{\beta_1}{2}+o(\beta_1)=\ln2+\frac{\beta_1}{2}+o(\beta_1).
\ee 
Similarly, for $\beta_1<\beta_2$, we have $I_{max}(\Lambda_0,A,L)\leq\ln2+\frac{\beta_2}{2}+o(\beta_2)$. Thus, $I_{max}(\Lambda_0,A,L)\leq\ln2+\frac{\max\{\beta_1,\beta_2\}}{2}+o(\max\{\beta_1,\beta_2\})$ for $\beta_1\neq\beta_2$.

For $\beta_1=\beta_2$, we have
\be 
I_{max}(\Lambda_0,A,L)&=&\frac{|\beta_1-\beta_2|(1-\min\{\beta_1,\beta_2\}\})}{1-\beta_1\beta_2}-\ln\frac{1-\beta_1\beta_2}{2-\beta_1-\beta_2}\nonumber\\
&=&-\ln\frac{1+\beta_1}{2}=\ln2-\beta_1+o(\beta_1).
\ee

\subsection{Proof of Lemma \ref{lemma.asymexpan}}\label{appen.asymexpan}
As $\beta=\big(\sqrt{p_0p_1}+\sqrt{(1-p_0)(1-p_1)}\big)^L$ and $1-\sqrt{x}=\frac{1}{2}(1-x)+o(1-x)$ for $x\to1$, we have
\be 
&&p_0^\frac{L}{2}-\big(\sqrt{p_0p_1}+\sqrt{(1-p_0)(1-p_1)}\big)^L\nonumber\\&=&\big(\sqrt{p_0}-\sqrt{p_0p_1}-\sqrt{(1-p_0)(1-p_1)}\big)\sum_{i=0}^{L-1}(p_0^\frac{L}{2})^i\Big(\sqrt{p_0p_1}+\sqrt{(1-p_0)(1-p_1)}\Big)^{L-1-i}\nonumber\\&=&p_0^\frac{L-1}{2}\big(\frac{\sqrt{p_0}}{2}(1-p_1)-\sqrt{(1-p_0)}(1-p_1)^{\frac{1}{2}}\big)+o(1-p_1).
\ee
Since $\beta_1=(\frac{p_0}{p_1})^{p_1L}(\frac{1-p_0}{1-p_1})^{(1-p_1)L}$ and $1-x^{-ax}=ax\ln x+o(x\ln x)=-a(1-x)+o(1-x)$ for $x\to1$, we have
\be 
p_0^L-\beta_1&=&p_0^L\Big(1-(\frac{1}{p_1})^{p_1L}(\frac{1-p_0}{1-p_1})^{(1-p_1)L}\Big)\nonumber\\&=&p_0^L\Big(\big(1-(\frac{1}{p_1})^{p_1L}\big)+(\frac{1}{p_1})^{p_1L}\big(1-(\frac{1-p_0}{1-p_1})^{(1-p_1)L}\big)\Big)\nonumber\\
&=&p_0^L\Big(-L(1-p_1)+(1-p_1)L\ln\frac{1-p_1}{1-p_0}\Big)+o(1-p_1).
\ee
Noting that $\beta_2=(\frac{p_1}{p_0})^{p_0L}(\frac{1-p_1}{1-p_0})^{(1-p_0)L}$, we have
\be 
&&(\frac{1}{p_0})^{p_0L}(\frac{1}{1-p_0})^{(1-p_0)L}(1-p_1)^{(1-p_0)L}-\beta_2\nonumber\\
&=&(\frac{1}{p_0})^{p_0L}(\frac{1}{1-p_0})^{(1-p_0)L}(1-p_1)^{(1-p_0)L}\big(1-p_1^{Lp_0}\big)\nonumber\\
&=&(\frac{1}{p_0})^{p_0L}(\frac{1}{1-p_0})^{(1-p_0)L}(1-p_1)^{(1-p_0)L}Lp_0\big(1-p_1\big)+o(1-p_1)=o(1-p_1).
\ee 
\subsection{Proof of Theorem \ref{theorem.asymI2}}\label{appen.asymI2}
For large $A$, $p_1$ and $\beta_2$ approach $1$ and $0$, respectively.
According to Lemma \ref{lemma.asymexpan}, Equation (\ref{eq.lowerbound}) and $\ln(a+x)=\ln a+\frac{x}{a}+o(x)$, we have
\be 
I_{max}(\Lambda_0,A,L)&\geq&-\ln\frac{1+\beta}{2}\nonumber\\&=&\ln\frac{2}{1+p_0^{\frac{L}{2}}}+\frac{p_0^\frac{L-1}{2}}{1+p_0^\frac{L}{2}}\big(\frac{\sqrt{p_0}}{2}(1-p_1)-\sqrt{(1-p_0)}(1-p_1)^{\frac{1}{2}}\big)+o(1-p_1).
\ee 
For the upper bound, since
\be 
\frac{1-\beta_1\beta_2}{2-\beta_1-\beta_2}-\frac{1}{2-\beta_1}=\frac{-(1+2\beta_1-\beta_1^2)\beta_2}{(2-\beta_1-\beta_2)(2-\beta_1)}=\frac{-(1+2\beta_1-\beta_1^2)\beta_2}{(2-\beta_1)^2}+o(\beta_2),
\ee 
the maximal mutual information is given by
\be 
I_{max}(\Lambda_0,A,L)&\leq&\frac{|\beta_1-\beta_2|(1-\min\{\beta_1,\beta_2\})}{1-\beta_1\beta_2}-\ln\frac{1-\beta_1\beta_2}{2-\beta_1-\beta_2}\nonumber\\
&=&\beta_1-(1+\beta_1)\beta_2+o(\beta_2)+\ln(2-\beta_1)+\frac{-(1+2\beta_1-\beta_1^2)\beta_2}{2-\beta_1)}+o(\beta_2)\nonumber\\
&=&p_0^L+\ln(2-p_0^L)+O(\max\{(1-p_1)\ln(1-p_1),(1-p_1)^{(1-p_0)L}\}).
\ee 
\subsection{Proof of Theorem \ref{theorem.appromut}}\label{appen.appromut}
Note that for binomial distribution $P_i^B$, we have the following approximation on entropy \cite[Theorem 2]{jacquet1999}, 
\be\label{eq.binentropy}
H(P_i^B)=\frac{1}{2}\ln2\pi eLp_i(1-p_i)+O(\frac{1}{L}), i=0,1.
\ee 
Since $\mathbb{P}(\hat{N}=0|X=0)=(1-p_0)^{L}=1-Lp_0+o(Lp_0)$, defining random variable $\hat{Y}\sim \mathbb{B}(1,Lp_0)$, we have $H(P_0^B)-H(\hat{Y})=o(Lp_0)$ and 
\be 
H(\hat{N}|X)=\frac{\mu}{2}\ln[2\pi eLp_1(1-p_1)]+(1-\mu)h_b(Lp_0)+O(\frac{1}{L})+o(Lp_0).
\ee 
Considering the mixture distribution of $\hat{N}$, we have
\be 
\mathbb{P}(\hat{N}=0)&=&\mu(1-p_1)^{L}+(1-\mu)(1-Lp_0)+o(Lp_0)\dff q_0+o(Lp_0);\\ \mathbb{P}(\hat{N}=1)&=&\mu Lp_1(1-p_1)^{L-1}+(1-\mu)Lp_0+o(Lp_0)\dff q_1+o(Lp_0);\\ \mathbb{P}(\hat{N}=i)&=&\mu\binom{L}{i}p_1^i(1-p_1)^{L-i}+o(Lp_0)\dff q_i+o(Lp_0), \text{ for }i\geq2.
\ee
According to the continuity of entropy function, we have $H(\hat{N})=-\sum_{i=0}^{L}q_i\ln q_i+o(Lp_0).$

Based on Taylor expansion, we have
\be 
-q_0\ln q_0&=&-[\mu(1-p_1)^{L}+1-\mu]\ln[\mu(1-p_1)^{L}+1-\mu]\nonumber\\ &&+(1-\mu)Lp_0\{1+\ln[\mu(1-p_1)^{L}+1-\mu]\}+o(Lp_0),\\
-q_1\ln q_1&=&-\mu Lp_1(1-p_1)^{L-1}[\ln(\mu Lp_1)+(L-1)\ln(1-p_1)]\nonumber\\&&-(1-\mu)Lp_0[1+\ln(\mu Lp_1)+(L-1)\ln(1-p_1)]+o(Lp_0),\\
-\sum_{i=2}^{L}q_i\ln q_i&=&-\mu\ln\mu[1-q_0-q_1]+\mu H(P_1^B)+\mu\Big\{L(1-p_1)^{L}\ln(1-p_1)\nonumber\\&&+Lp_1(1-p_1)^{L-1}[\ln (L p_1)+(L-1)\ln(1-p_1)]\Big\}.
\ee 
Since $I(X;\hat{N})=H(\hat{N})-H(\hat{N}|X)$, we can obtain Equation (\ref{eq.appromut}).

\subsection{Proof of Theorem \ref{theorem.boundgap}}\label{appen.boundgap}
Note that $\frac{\partial F_u}{\partial \beta_1}=-\frac{\mu(1-\mu)}{(1-\mu)\beta_1+\mu}<0$, $\frac{\partial F_u}{\partial \beta_2}=-\frac{\mu(1-\mu)}{\mu\beta_2+1-\mu}<0$, $\frac{\partial^2 F_u}{\partial \beta_1^2}=\frac{\mu(1-\mu)^2}{[(1-\mu)\beta_1+\mu]^2}>0$ and $\frac{\partial^2 F_u}{\partial \beta_2^2}=\frac{\mu^2(1-\mu)}{[\mu\beta_2+1-\mu]^2}>0$. \textbf{For low SNR}, according to Taylor Theorem and $\beta>\max\{\beta_1,\beta_2\}$, we have
\be\label{eq.beta0}
F_u(\mu,\beta_1,\beta_2)-F_u(\mu,\beta,\beta_2)&\overset{(a)}{=}&\frac{\mu(1-\mu)}{(1-\mu)\beta_1+\mu}(\beta-\beta_1)+\frac{\partial^2 F_u}{\partial \beta_1^2}\Big|_{(\beta,\xi_1,\beta_2)}(\beta-\beta_1)^2\nonumber\\
&\overset{(b)}{\leq}&\frac{\mu(1-\mu)}{(1-\mu)\beta_1+\mu}(\beta-\beta_1)+\frac{\partial^2 F_u}{\partial \beta_1^2}\Big|_{(\beta,\beta_1,\beta_2)}(\beta-\beta_1)^2,
\ee
where $(a)$ holds due to the Taylor expansion in terms of $\beta_1$, $\xi_1\in(\beta_1,\beta)$ and $(b)$ holds since $\frac{\partial^2 F_u}{\partial \beta_1^2}$ is monotonically decreasing with respect to $\beta_1$. Furthermore, we have
\be\label{eq.delta1}
\max\limits_{\mu\in[0,1]}F_u(\mu,\beta_1,\beta_2)-F_u(\mu,\beta,\beta_2)&\leq&\max\limits_{\mu\in[0,1]}\frac{\mu(1-\mu)}{(1-\mu)\beta_1+\mu}(\beta-\beta_1)+\frac{\mu(1-\mu)^2}{[(1-\mu)\beta_1+\mu]^2}(\beta-\beta_1)^2\nonumber\\
&\overset{(c)}{\leq}& \frac{1}{4\beta_1}(\beta-\beta_1)+\frac{4}{27\beta_1^2}(\beta-\beta_1)^2,
\ee 
where $(c)$ holds since $(1-\mu)\beta_1+\mu\geq\beta_1$, $\mu(1-\mu)\leq(\frac{\mu+(1-\mu)}{2})^2=\frac{1}{4}$ and $\mu(1-\mu)^2\leq\frac{1}{2}(\frac{2\mu+(1-\mu)+(1-\mu)}{3})^2=\frac{4}{27}$. Similar to equation (\ref{eq.delta1}), we have
\be\label{eq.delta2}
\max\limits_{\mu\in[0,1]}F_u(\mu,\beta,\beta_2)-F_u(\mu,\beta,\beta)\leq \frac{1}{4\beta_2}(\beta-\beta_2)+\frac{4}{27\beta_2^2}(\beta-\beta_2)^2.
\ee

As $F_l(\mu,\beta)=F_u(\mu,\beta,\beta)$, we have the upper bound on $\Delta(\beta,\beta_1,\beta_2)$ in low SNR regime,
\be 
\Delta(\beta,\beta_1,\beta_2)&=&\max\limits_{\mu\in[0,1]}F_u(\mu,\beta_1,\beta_2)-F_u(\mu,\beta,\beta)\nonumber\\&\overset{(d)}{\leq}&\max\limits_{\mu\in[0,1]}F_u(\mu,\beta_1,\beta_2)-F_u(\mu,\beta,\beta_2)+\max\limits_{\mu\in[0,1]}F_u(\mu,\beta,\beta_2)-F_u(\mu,\beta,\beta)\nonumber\\
&\overset{(e)}{\leq}&\frac{1}{4\beta_1}(\beta-\beta_1)+\frac{4}{27\beta_1^2}(\beta-\beta_1)^2+\frac{1}{4\beta_2}(\beta-\beta_2)+\frac{4}{27\beta_2^2}(\beta-\beta_2)^2\nonumber\\
&=&\frac{1}{108}(\frac{\beta}{\beta_1}-1)(16\frac{\beta}{\beta_1}+11)+\frac{1}{108}(\frac{\beta}{\beta_2}-1)(16\frac{\beta}{\beta_2}+11),
\ee
where $(d)$ holds due to $\max\limits_{x}f(x)+g(x)\leq\max\limits_{x}f(x)+\max\limits_{x}g(x)$ and $(e)$ holds according to Equations (\ref{eq.delta1}) and (\ref{eq.delta2}).

\textbf{For high SNR}, note that
\be F_u(\mu,\beta_1,\beta_2)-F_u(\mu,\beta,\beta_2)&=&\mu\ln[1+\frac{(1-\mu)(\beta-\beta_1)}{(1-\mu)\beta_1+\mu}]\overset{(f)}{\leq}\frac{\mu(1-\mu)(\beta-\beta_1)}{(1-\mu)\beta_1+\mu}\nonumber\\
&\overset{(g)}{\leq}&(1-\mu)(\beta-\beta_1),
\ee
where $(f)$ and $(g)$ hold due to $\ln(1+x)\leq x$ and $\mu\leq(1-\mu)\beta_1+\mu$, respectively. Thus, we have
\be\label{eq.beta4} 
\max\limits_{\mu\in[0,1]}F_u(\mu,\beta_1,\beta_2)-F_u(\mu,\beta,\beta_2)\leq\beta-\beta_1.
\ee 
Similarly to Equation (\ref{eq.beta4}), we have
\be\label{eq.beta5} 
\max\limits_{\mu\in[0,1]}F_u(\mu,\beta,\beta_2)-F_u(\mu,\beta,\beta)\leq \beta-\beta_2.
\ee
Thus, we have the following upper bound on $\Delta(\beta,\beta_1,\beta_2)$ in high SNR regime,
\be 
\Delta(\beta,\beta_1,\beta_2)&\leq&\max\limits_{\mu\in[0,1]}F_u(\mu,\beta_1,\beta_2)-F_u(\mu,\beta,\beta_2)+\max\limits_{\mu\in[0,1]}F_u(\mu,\beta,\beta_2)-F_u(\mu,\beta,\beta)\nonumber\\&\leq&(\beta-\beta_1)+(\beta-\beta_2).
\ee

\textbf{For general $\beta,\beta_1,\beta_2$}, we have the following lower bound on $\Delta(\beta,\beta_1,\beta_2)$,
\be 
\Delta(\beta,\beta_1,\beta_2)&=&\max\limits_{\mu\in[0,1]}F_u(\beta,\beta_1,\beta_2)-F_u(\beta,\beta,\beta)\nonumber\\
&\overset{(h)}{\geq}& \max\limits_{\mu\in[0,1]}F_u(\beta,\beta_1,\beta_2)-\max\limits_{\mu\in[0,1]}F_u(\beta,\beta,\beta)\nonumber\\
&=&\max\limits_{\mu\in[0,1]}F_u(\beta,\beta_1,\beta_2)+\ln\frac{1+\beta}{2}\nonumber\\
&\overset{\mu=\frac{1}{2}}{\geq}&-\frac{1}{2}(\ln\frac{1+\beta_1}{2}+\ln\frac{1+\beta_2}{2})+\ln\frac{1+\beta}{2}=\frac{1}{2}\ln\frac{1+\beta}{1+\beta_1}+\frac{1}{2}\ln\frac{1+\beta}{1+\beta_2},\nonumber
\ee 
where $(h)$ holds since that for positive function $f(x)$ and $g(x)$,
\be
\max\limits_{x}f(x)-g(x)\geq f(x^{*})-g(x^{*})\geq f(x^{*})-\max\limits_{x}g(x)=\max\limits_{x}f(x)-\max\limits_{x}g(x),\nonumber
\ee
where $x^{*}=\arg\max\limits_{x}f(x)$.
\subsection{Proof of Theorem \ref{theorem.boundgap1}}\label{appen.boundgap1}
As $\beta=\exp\big(-C_{\frac{1}{2}}(P_1^B||P_0^B)\big)\to0$, $\beta_1=\exp\big(-KL(P_1^B||P_0^B)\big)\to0$ and $\beta_2=\exp\big(-KL(P_0^B||P_1^B)\big)\to0$ as $L$ approaches infinity, such scenario corresponds to high SNR regime. According to Theorem \ref{theorem.boundgap}, $\beta_1=o(\beta)$ and $\beta_2=o(\beta)$, we have
\be 
\Delta(\beta,\beta_1,\beta_2)&\leq&(\beta-\beta_1)+(\beta-\beta_2)\nonumber\\
&=&2\exp(-C_{\frac{1}{2}}(P_1^B||P_0^B))+o(\exp(-C_{\frac{1}{2}}(P_1^B||P_0^B))).
\ee
Thus, we have the following lower bound on the exponential rate of $\Delta(\beta,\beta_1,\beta_2)$ with respect to $L$,
\be\label{eq.lowerce} 
-\lim\limits_{L\to\infty}\frac{\ln \Delta(\beta,\beta_1,\beta_2)}{L}\geq\lim\limits_{L\to\infty}\frac{C_{\frac{1}{2}}(P_1^B||P_0^B)}{L}
=-\ln\big(\sqrt{p_0p_1}+\sqrt{(1-p_0)(1-p_1)}\big).
\ee 
Similarly, we have
\be 
\Delta(\beta,\beta_1,\beta_2)\geq\frac{1}{2}\ln\frac{1+\beta}{1+\beta_1}+\frac{1}{2}\ln\frac{1+\beta}{1+\beta_2}=\frac{1}{2}\frac{\beta-\beta_1}{1+\beta_1}+\frac{1}{2}\frac{\beta-\beta_2}{1+\beta_2}+o(\beta)=\beta+o(\beta);
\ee 
and thus an upper bound on exponential rate of $\Delta(\beta,\beta_1,\beta_2)$ with respect to $L$ is given as follows,
\be\label{eq.upperce} 
-\lim\limits_{L\to\infty}\frac{\ln \Delta(\beta,\beta_1,\beta_2)}{L}\leq\lim\limits_{L\to\infty}\frac{C_{\frac{1}{2}}(P_1^B||P_0^B)}{L}
=-\ln\big(\sqrt{p_0p_1}+\sqrt{(1-p_0)(1-p_1)}\big).
\ee
From Equations (\ref{eq.lowerce}) and (\ref{eq.upperce}), we have
\be
-\lim\limits_{L\to\infty}\frac{\ln \Delta(\beta,\beta_1,\beta_2)}{L}=-\ln\big(\sqrt{p_0p_1}+\sqrt{(1-p_0)(1-p_1)}\big).
\ee
It demonstrates the asymptotic tightness of the upper and lower bounds for large $L$, with exponential rate $-\ln\big(\sqrt{p_0p_1}+\sqrt{(1-p_0)(1-p_1)}\big)$.
\subsection{Proof of Theorem \ref{theorem.boundgap2}}\label{appen.boundgap2}
According to Lemma \ref{lemma.asymexpan} and Theorem \ref{theorem.boundgap}, we have the upper bound on $\Delta(\beta,\beta_1,\beta_2)$,
\be 
\Delta(\beta,\beta_1,\beta_2)\leq(\beta-\beta_1)+(\beta-\beta_2)=2p_0^{\frac{L}{2}}-p_0^{L}+\epsilon_u+o(\epsilon_u),
\ee 
where $\epsilon_u$ is shown in Equation (\ref{eq.epsilonu}).
Similarly, according to Theorem~\ref{theorem.boundgap}, we have the following upper bound on $\Delta(\beta,\beta_1,\beta_2)$,
\be 
\Delta(\beta,\beta_1,\beta_2)\geq\ln\big(1+p_0^{\frac{L}{2}}\big)-\frac{1}{2}\ln\big(1+p_0^{L}\big)+\epsilon_l+o(\epsilon_l),
\ee
where $\epsilon_l$ is showed in equation (\ref{eq.epsilonl}).
\subsection{Proof of Theorem \ref{theorem.boundgap3b}}\label{appen.boundgap3b}
According to Theorem \ref{theorem.boundgap}, we have the following upper and lower bounds on bound gap $\Delta(\beta,\beta_1,\beta_2)$,
\be 
\Delta(\beta,\beta_1,\beta_2)&\leq&(\beta-\beta_1)+(\beta-\beta_2),\nonumber\\
&=&2(1-p_1)^{\frac{L}{2}}+o\big((1-p_1)^{\frac{L}{2}}\big),\label{eq.deltaconvD1}\\
\Delta(\beta,\beta_1,\beta_2)&\geq&\frac{1}{2}\ln\frac{1+\beta}{1+\beta_1}+\frac{1}{2}\ln\frac{1+\beta}{1+\beta_2}\nonumber\\
&=&\frac{1}{2}(1-p_1)^{\frac{L}{2}}+\frac{1}{2}\frac{(1-p_1)^{\frac{L}{2}}-(1-p_1)^{L}}{1+(1-p_1)^{L}}+o\big((1-p_1)^{\frac{L}{2}}\big)\nonumber\\&=&(1-p_1)^{\frac{L}{2}}+o\big((1-p_1)^{\frac{L}{2}}\big).\label{eq.deltaconvD2}
\ee 
Thus, the asymptotic tightness is demonstrated as follows,
\be 
0=\lim\limits_{A\to\infty}(1-p_1)^{\frac{L}{2}}+o\big((1-p_1)^{\frac{L}{2}}\big)&\leq&\lim\limits_{A\to\infty}\Delta(\beta,\beta_1,\beta_2)\nonumber\\&\leq&\lim\limits_{A\to\infty}2(1-p_1)^{\frac{L}{2}}+o\big((1-p_1)^{\frac{L}{2}}\big)=0.
\ee 
Furthermore, we have the following on the exponential rate of the bound gap with respect to peak power $A$,
\be 
-\lim_{A \rightarrow \infty} \frac{\ln \Delta(\beta, \beta_1, \beta_2)}{A}&\geq&\lim\limits_{A\to\infty}-\frac{\ln[2(1-p_1)^{\frac{L}{2}}+o\big((1-p_1)^{\frac{L}{2}}\big)]}{A}=\frac{L\tau}{2},\\
-\lim_{A \rightarrow \infty} \frac{\ln \Delta(\beta, \beta_1, \beta_2)}{A}&\leq&\lim\limits_{A\to\infty}-\frac{\ln[(1-p_1)^{\frac{L}{2}}+o\big((1-p_1)^{\frac{L}{2}}\big)]}{A}=\frac{L\tau}{2},
\ee 
i.e., $-\lim_{A \rightarrow \infty} \frac{\ln \Delta(\beta, \beta_1, \beta_2)}{A}=\frac{L\tau}{2}$.
\subsection{Proof of Theorem \ref{theorem.boundgap4}}\label{appen.boundgap4}
For low peak power $A$, we have $p_1\rightarrow p_0$ and $p_1-p_0=e^{-\Lambda_0\tau}(1-e^{-A\tau})=(1-p_0)\tau A+o(A)$. Noting that $\sqrt{x+p_0}=\sqrt{p_0}+\frac{1}{2\sqrt{p_0}}x-\frac{1}{8p_0^{\frac{3}{2}}}x^2+o(x^2)$, we have
\be 
&&1-\sqrt{p_0p_1}-\sqrt{(1-p_0)(1-p_1)}\nonumber\\&=&1-\big(p_0+\frac{p_1-p_0}{2}-\frac{(p_1-p_0)^2}{8p_0}\big)-\big(1-p_0+\frac{p_0-p_1}{2}-\frac{(p_0-p_1)^2}{8(2-p_0)}\big)+o(A^2)\nonumber\\
&=&\frac{(p_1-p_0)^2}{8p_0(1-p_0)}+o(A^2)=\frac{(1-p_0)}{8p_0}\tau^2A^2+o(A^2).
\ee 
Thus, we have the following Taylor expansion on $\beta$,
\be 
\beta&=&\exp\big(-C_{\frac{1}{2}}(P_1^B||P_0^B)\big)=1-C_{\frac{1}{2}}(P_1^B||P_0^B)+o\big(C_{\frac{1}{2}}(P_1^B||P_0^B)\big)\nonumber\\&=&1-L\big(1-\sqrt{p_0p_1}-\sqrt{(1-p_0)(1-p_1)}\big)+o(1-\sqrt{p_0p_1}-\sqrt{(1-p_0)(1-p_1)})\nonumber\\&=&1-\frac{L(1-p_0)}{8p_0}\tau^2A^2+o(A^2).
\ee 
Note that $KL(P_1^B||P_0^B)=L\big(p_1\ln\frac{p_1}{p_0}+(1-p_1)\ln\frac{1-p_1}{1-p_0}\big)$, according to Taylor theorem, we have
\be 
KL(P_1^B||P_0^B)&=&0+L\big(\ln\frac{p_1}{p_0}-\ln\frac{1-p_1}{1-p_0}\big)\Big|_{p_1=p_0}(p_1-p_0)\nonumber\\&&+\frac{L}{p_1(1-p_1)}\Big|_{p_1=p_0}\frac{(p_1-p_0)^2}{2}+o\big((p_1-p_0)^2\big)\nonumber\\
&=&\frac{L(1-p_0)}{2p_0}\tau^2A^2+o(A^2).
\ee 
Thus, we have the following Taylor expansion on the $\beta_1$,
\be 
\beta_1&=&\exp\big(-KL(P_1^B||P_0^B)\big)=1-KL(P_1^B||P_0^B)+o\big(KL(P_1^B||P_0^B)\big)\nonumber\\&=&1-\frac{L(1-p_0)}{2p_0}\tau^2A^2+o(A^2).
\ee 

Similarly, we have Taylor expansion $\beta_2=1-\frac{L(1-p_0)}{2p_0}\tau^2A^2+o(A^2).$ 

According to Theorem \ref{theorem.boundgap}, we have
\be 
\Delta(\beta,\beta_1,\beta_2)&\leq&\frac{1}{108}(\frac{\beta}{\beta_1}-1)(16\frac{\beta}{\beta_1}+11)+\frac{1}{108}(\frac{\beta}{\beta_2}-1)(16\frac{\beta}{\beta_2}+11)\nonumber\\
&=&\frac{1}{108}(\frac{L(1-p_0)}{2p_0}-\frac{L(1-p_0)}{8p_0})\tau^2A^2\times27\times2+o(A^2)\nonumber\\
&=&\frac{3L(1-p_0)}{16p_0}\tau^2A^2+o(A^2),\label{eq.delta1a}\\
\Delta(\beta,\beta_1,\beta_2)&\geq&\frac{1}{2}\ln\frac{1+\beta}{1+\beta_1}+\frac{1}{2}\ln\frac{1+\beta}{1+\beta_2}=\frac{3L(1-p_0)}{16p_0}\tau^2A^2+o(A^2).\label{eq.delta1b}
\ee 
Based on Equations (\ref{eq.delta1a}) and (\ref{eq.delta1b}), we have $\Delta(\beta,\beta_1,\beta_2)=\frac{3L(1-p_0)}{16p_0}\tau^2A^2+o(A^2)$.

\section{The proof of main results on the Capacity for Arbitrarily Symbol Duration}
\subsection{Proof of Proposition \ref{prop.siso1}}\label{appen.siso1}
\textbf{Converse part}: Note that $\Lambda^{T_s}\rightarrow X_{T_s}\rightarrow Z$ forms a Markov chain, where $X_{T_s}=\int_{T_s-\tau}^{T_s}\Lambda(t^{'})\mathrm{d}t^{'}$, according to data processing inequality, we have $I(\Lambda^{T_s};Z)\leq I(X_{T_s};Z)$.
Note that the conditional entropy $H(Z|X_{T_s})=h_b(S)$, where $S=p(X_{T_s}+\Lambda_0)$ and $p(x)=1-e^{-x\tau}$.
Define $\mu(X_{T_s})$ as the probability measure of $X_{T_s}$. Entropy $H(Z)$ is given by $H(Z)=h_b(\hat{p})$,
where $\hat{p}=\int p(X_{T_s}+\Lambda_0)\mathrm{d}\mu(X_{T_s})=\mathbb{E}[S]$. The mutual information $I(X,Z)$ is as follows,
\be 
I(X,Z)=h_b(\hat{p})-\int h_b(p(X_{T_s}+\Lambda_0))\mathrm{d}\mu(X_{T_s}).
\ee 
As mapping $X_{T_s}\rightarrow S$ is a one-to-one mapping, we have 
\be 
I(X_{T_s};Z)=I(S;Z)=h_b(\mathbb{E}[S])-\mathbb{E}[h_b(S)],
\ee
and the following equation holds,
\be
\max\limits_{\mu(X_{T_s})}I(X_{T_s};Z)=\max\limits_{\mu(S)}I(S;Z)&=&\max\limits_{p(\Lambda_0)\leq\hat{p}\leq p(A+\Lambda_0)}\max\limits_{\mu(S):\mathbb{E}[S]=\hat{p}}I(S;Z)\nonumber\\&=&\max\limits_{p(\Lambda_0)\leq\hat{p}\leq p(A+\Lambda_0)}h_b(\hat{p})+\max\limits_{\mu(S):\mathbb{E}[S]=\hat{p}}\mathbb{E}[-h_b(S)].
\ee 
Note that function $-h_b(\cdot)$ is strictly convex and the solution to maximize a strictly convex function over all finite support probability given first
moment is achieved by a distribution of two mass extreme points. Accordingly, defining $\mu\dff\mathbb{P}(X_{T_s}=A),$ we have
\be\label{eq.optsingle}
C_{\tau,\tau}&\leq&\frac{1}{\tau}\max\limits_{\mu(X_{T_s})}I(X_{T_s},Z)\nonumber\\&=&\frac{1}{\tau}\max\limits_{0\leq\mu\leq1}h_b\big(\hat{p}(\mu)\big)-(1-\mu)h_b\big(p(\Lambda_0)\big)-\mu h_b\big(p(A+\Lambda_0)\big),
\ee 
where $\mu$ satisfies $(1-\mu)p(\Lambda_0)+\mu p(A+\Lambda_0)=\hat{p}$, i.e., $\mu=\frac{\hat{p}-p(\Lambda_0)}{p(A+\Lambda_0)-p(\Lambda_0)}$. 

\textbf{Achievability part:} Let waveform $\Lambda^{T_s}$ in $[0,T_s]$ randomly selected from waveform set $\{0,A*(u(t)-u(t-T_s))\}$ with probability $\mu^{*}=\mathbb{P}\{\Lambda^{T_s}=A*(u(t)-u(t-T_s))\}$, where $u(t)$ denotes as a step function, then we have
\be 
C_{\tau,\tau}\geq\frac{1}{\tau}\max\limits_{0\leq\mu\leq1}h_b\big(\hat{p}(\mu)\big)-(1-\mu)h_b\big(p(\Lambda_0)\big)-\mu h_b\big(p(A+\Lambda_0)\big).
\ee 
\subsection{Proof of Proposition \ref{prop.siso2}}\label{appen.siso2}
Recalling $F(\mu)\dff h_b\big(\hat{p}(\mu)\big)-(1-\mu)h_b\big(p(\Lambda_0)\big)-\mu h_b(p(A\tau))$, where $\hat{p}=(1-\mu)p(\Lambda_0)+\mu p(A+\Lambda_0)$, we have
\be 
F^{'}(\mu)&=&-(p(A+\Lambda_0)-p(\Lambda_0))\ln\frac{\hat{p}}{1-\hat{p}}+h_b\big(p(\Lambda_0)\big)-h_b\big(p(A+\Lambda_0)\big),\\
F^{''}(\mu)&=&-\frac{p(A+\Lambda_0)-p(\Lambda_0)}{\hat{p}(1-\hat{p})}<0.
\ee 
Note that $h_b(\cdot)$ is concave, according to Lemma \ref{appenAu.convex}, we have $h_b^{'}(y)>\frac{h_b(x)-h_b(y)}{x-y}>h_b^{'}(x)$ for $0\leq y<x\leq1$, and
\be
F^{'}(0)&=&-(p(A+\Lambda_0)-p(\Lambda_0))\big(\frac{h_b\big(p(A+\Lambda_0)\big)-h_b\big(p(\Lambda_0)\big)}{p(A+\Lambda_0)-p(\Lambda_0)}-h_b^{'}(p(\Lambda_0))\big)>0;\\ F^{'}(1)&=&-(p(A+\Lambda_0)-p(\Lambda_0))\big(\frac{h_b\big(p(A+\Lambda_0)\big)-h_b\big(p(\Lambda_0)\big)}{p(A+\Lambda_0)-p(\Lambda_0)}-h_b^{'}(p(A+\Lambda_0))\big)<0.
\ee
Thus, $\mu^{*}\dff\arg\max F(\mu)$ uniquely exists and satisfies $F^{'}(\mu^{*})=0$, i.e.,
\be\label{eq.optsing2}
\mu^{*}=\frac{\frac{a}{1+a}-p(\Lambda_0)}{p(A+\Lambda_0)-p(\Lambda_0)},
\ee 
where $a=\exp(-\frac{h_b\big(p(A+\Lambda_0)\big)-h_b\big(p(\Lambda_0)\big)}{p(A+\Lambda_0)-p(\Lambda_0)})$. Hence we have $C=\frac{1}{\tau}F(\mu^{*})$.

\subsection{Proof of Theorem \ref{theo.sisolowtau}}\label{appen.sisolowtau}
First we show the Taylor expansion of $a$ given in Theorem \ref{theor.siso}.

Since $h_b(x)=-x\ln(x)-(1-x)\ln(1-x)=x-x\ln(x)+o(x)$ and $\ln\frac{p(x)}{\tau}=\ln(x)+o(\tau)$, we have
\be\label{eq.prea}
\frac{h_b\big(p(A+\Lambda_0)\big)-h_b\big(p(\Lambda_0)\big)}{p(A+\Lambda_0)-p(\Lambda_0)}+\ln\tau&=&\frac{A\tau-A\tau\frac{p(A+\Lambda_0)}{\tau}+p(\Lambda_0)\ln\frac{p(\Lambda_0)}{\tau}+o(\tau)}{p(A+\Lambda_0)-p(\Lambda_0)} \nonumber\\&=&1+\frac{\Lambda_0\ln(\Lambda_0)-(A+\Lambda_0)\ln(A+\Lambda_0)}{A}.
\ee 
Defining $s=\frac{\Lambda_0}{A}$, based on Equation (\ref{eq.prea}), we have
\be \label{eq.pointconv}
\lim\limits_{\tau\to0}\frac{a}{\tau}&=&\exp\big(-\lim\limits_{\tau\to0}(\frac{h_b\big(p(A+\Lambda_0)\big)-h_b\big(p(\Lambda_0)\big)}{p(A+\Lambda_0)-p(\Lambda_0)}+\ln\tau)\big)=\frac{A}{e}\frac{(1+s)^{1+s}}{s^s},\\
\lim\limits_{\tau\to0}\mu^{*}&=&\lim\limits_{\tau\to0}\frac{\frac{A}{e}\frac{(1+s)^{1+s}}{s^s}\tau-\Lambda_0\tau+o(\tau)}{A\tau}=\frac{(1+s)^{1+s}}{es^s}-s.
\ee
Based on the above results, we have the following result, which shows that the capacity is consistent with the scenario of continuous Poisson channel,
\be 
\lim\limits_{\tau\to0}C_{\tau,\tau}&=&\lim\limits_{\tau\to0}\frac{1}{\tau}F(\mu^{*})=\lim\limits_{\tau\to0}\frac{\partial F(\mu^{*})}{\partial \tau}\nonumber\\&=&\lim\limits_{\tau\to0}(1-\mu^{*})\Lambda_0e^{-\Lambda_0\tau}\ln\frac{p(\Lambda_0)(1-\hat{p})}{(1-p(\Lambda_0))\hat{p}}+\mu^{*}(A+\Lambda_0)e^{-(A+\Lambda_0)\tau}\ln\frac{p(A+\Lambda_0)(1-\hat{p})}{(1-p(A+\Lambda_0))\hat{p}}\nonumber\\
&=&(1-\mu^{*})\Lambda_0\ln\frac{\Lambda_0}{\mu^{*}A+\Lambda_0}+\mu^{*}(A+\Lambda_0)\ln\frac{A+\Lambda_0}{\mu^{*}A+\Lambda_0}\nonumber\\
&=&A[-(\mu^{*}+s)\ln(\mu^{*}+s)+\mu^{*}(1+s)\ln(1+s)+(1-\mu^{*})s\ln s].
\ee 

\subsection{Proof of Theorem \ref{theo.sisoasym}}\label{appen.sisoasym}
According to Equation (\ref{eq.pointconv}), point-wise convergence is obvious. Set $A=\frac{1}{\tau}$, then we have
\be 
\lim\limits_{\tau\to0}\mu^{*}=\frac{\exp(-h_b(p(1))/p(1))}{p(1)[1+\exp(-h_b(p(1))/p(1))]}\neq\frac{1}{e},
\ee 
which shows that the convergence is not uniform.

\subsection{Proof of Theorem \ref{theo.asymlargelowpeak}}\label{appen.asymlargelowpeak}
Considering the scenario without background radiation, i.e.,  $\Lambda_0=0$. For $A\to\infty$, we have 
\be 
\lim\limits_{A\to\infty}a=\exp\big(-\lim\limits_{A\to\infty}(\frac{h_b\big(p(A+\Lambda_0)\big)}{p(A+\Lambda_0)})\big)=1,
\ee
and the optimal duty cycle $\lim\limits_{A\to\infty}\mu^{*}=\frac{1}{2}$. When $A\to0$, we have $\lim\limits_{A\to0}a=0$ and
\be \lim\limits_{A\to0}\mu^{*}\xlongequal[]{x=p(A)}\lim\limits_{x\to0}\frac{\exp(-\frac{h_b(x)}{x})}{x}=\lim\limits_{x\to0}\exp(\frac{(1-x)\ln(1-x)}{x})=\frac{1}{e}.
\ee
As the optimal duty cycle for continuous Poisson channel is $\frac{1}{e}$ for $\Lambda_0=0$ and any $A$, the optimal duty cycle for larger $A$ deviates more due to larger counting loss. 

For $\Lambda_0>0$,  as $A\to\infty$, we have 
\be 
\lim\limits_{A\to\infty}a=\exp\big(-\lim\limits_{A\to\infty}(\frac{h_b\big(p(A+\Lambda_0)\big)-h_b\big(p(\Lambda_0)\big)}{p(A+\Lambda_0)-p(\Lambda_0)})\big)=\exp\big(e^{\Lambda_0\tau}h_b\big(p(\Lambda_0)\big)\big),
\ee
and $\lim\limits_{A\to\infty}\mu^{*}=1-\frac{1}{\big(1+\exp\big(e^{\Lambda_0\tau}h_b\big(p(\Lambda_0)\big)\big)\big)(1-p(\Lambda_0))}$.

For $A\to0$ and $\Lambda_0>0$, since $\lim\limits_{A\to0}a=0$, we have
\be 
\lim\limits_{A\to0}\mu^{*}=\lim\limits_{A\to0}\frac{a}{p(A)}=\lim\limits_{A\to0}\frac{\exp\big(\ln p(A)-1+o(1)\big)}{p(A)}=\frac{1}{e}.
\ee 

For` $A\to0$ and $\Lambda_0>0$, we have 
\be
\lim\limits_{A\to0}a=\exp(-h_b^{'}(p(\Lambda_0)))=\frac{p(\Lambda_0)}{1-p(\Lambda_0)}.
\ee
According to Taylor's theorem and $p(A+\Lambda_0)-p(\Lambda_0)=\big(1-p(\Lambda_0)\big)A\tau+o(A)$, we have the following results for sufficiently small $A$,
\be 
\frac{h_b\big(p(A+\Lambda_0)\big)-h_b\big(p(\Lambda_0)\big)}{p(A+\Lambda_0)-p(\Lambda_0)}&=&h_b^{'}(p(\Lambda_0))+\frac{h_b^{''}(p(\Lambda_0))}{2}\big(p(A+\Lambda_0)-p(\Lambda_0)\big)+o\big(p(A+\Lambda_0)-p(\Lambda_0)\big)\nonumber\\
&=&h_b^{'}(p(\Lambda_0))+\frac{h_b^{''}(p(\Lambda_0))}{2}\big(1-p(\Lambda_0)\big)A\tau+o(A),
\ee 
\be
a&=&\frac{p(\Lambda_0)}{1-p(\Lambda_0)}-\exp(-h_b^{'}(p(\Lambda_0)))\frac{h_b^{''}(p(\Lambda_0))}{2}\big(1-p(\Lambda_0)\big)A\tau+o(A),\nonumber\\
&=&\frac{p(\Lambda_0)}{1-p(\Lambda_0)}-\frac{1}{2\big(1-p(\Lambda_0)\big)^2}\big(1-p(\Lambda_0)\big)A\tau+o(A),\\
\lim\limits_{A\to0}\mu^{*}&=&\lim\limits_{A\to0}\frac{\frac{a}{1+a}-p(\Lambda_0)}{\big(1-p(\Lambda_0)\big)A\tau}=(1-p(\Lambda_0))^2[-\frac{p(\Lambda_0)}{1-p(\Lambda_0)}[-2p(\Lambda_0)(1-p(\Lambda_0))]^{-1}\nonumber
\\&=&\frac{1}{2}.
\ee

\subsection{Proof of Theorem \ref{theo.asymcapalargepeak}}\label{appen.asymcapalargepeak}
Note that $\lim\limits_{A\to\infty}\frac{h_b\big(p(A+\Lambda_0)\big)-h_b\big(p(\Lambda_0)\big)}{p(A+\Lambda_0)-p(\Lambda_0)}=\frac{-h_b\big(p(\Lambda_0)\big)}{1-p(\Lambda_0)}$, $\ln\big(1-p(A+\Lambda_0)\big)=-A\tau$ and $h_b(x)=h_b(1-x)=(1-x)-(1-x)\ln(1-x)+o(1-x)$ for $x\to1$. We have
\be\label{eq.tayexpinner} 
&&\frac{h_b\big(p(A+\Lambda_0)\big)-h_b\big(p(\Lambda_0)\big)}{p(A+\Lambda_0)-p(\Lambda_0)}-\frac{-h_b\big(p(\Lambda_0)\big)}{1-p(\Lambda_0)}\nonumber\\&=&\frac{\big(1-p(\Lambda_0)\big)h_b\big(p(A+\Lambda_0)\big)-h_b\big(p(\Lambda_0)\big)\big(1-p(A+\Lambda_0)\big)}{\big(1-p(\Lambda_0)\big)\big(p(A+\Lambda_0)-p(\Lambda_0)\big)}\nonumber\\
&=&\frac{\big(1-p(\Lambda_0)\big)\big(p(A+\Lambda_0)-p(A+\Lambda_0)\ln p(A+\Lambda_0)\big)}{\big(1-p(\Lambda_0)\big)^2}+o(Ae^{-A\tau})\nonumber\\
&=&e^{\Lambda_0\tau}A\tau e^{-A\tau}+o(Ae^{-A\tau}).
\ee 
Since $\exp\big(-(x+\Delta x)\big)=\exp(-x)-\exp(-x)\Delta x+o(\Delta x)$, the Taylor expansion of $a$ can be expressed as follows based on Equation (\ref{eq.tayexpinner}),
\be\label{eq.talora}
a=\exp\Big(e^{\Lambda_0\tau}h_b\big(p(\Lambda_0)\big)\Big)-\exp\Big(e^{\Lambda_0\tau}h_b\big(p(\Lambda_0)\big)\Big)e^{\Lambda_0\tau}A\tau e^{-A\tau}+o(Ae^{-A\tau}).
\ee 
For the optimal duty cycle $\mu^{*}$, based on Equation (\ref{eq.talora}) and the Taylor expansion of $\frac{1}{1+a}$, we have 
\be\label{eq.taylormu}
\mu^{*}&=&\frac{\frac{a}{1+a}-p(\Lambda_0)}{p(A+\Lambda_0)-p(\Lambda_0)}=1-\frac{1-p(A+\Lambda_0)-\frac{1}{1+a}}{p(A+\Lambda_0)-p(\Lambda_0)}=1-\frac{1-p(A+\Lambda_0)-\frac{1}{1+a}}{1-p(\Lambda_0)}+o(Ae^{-A\tau})\nonumber\\
&=&1-\frac{e^{-(A+\Lambda_0)\tau}-[1+\exp\Big(e^{\Lambda_0\tau}h_b\big(p(\Lambda_0)\big)\Big)]^{-1}+[1+\exp\Big(e^{\Lambda_0\tau}h_b\big(p(\Lambda_0)\big)\Big)]^{-2}e^{\Lambda_0\tau}A\tau e^{-A\tau}}{1-p(\Lambda_0)}\nonumber\\&&+o(Ae^{-A\tau}),\nonumber\\
&=&1-[1+\exp\Big(e^{\Lambda_0\tau}h_b\big(p(\Lambda_0)\big)\Big)]^{-1}e^{\Lambda_0\tau}+[1+\exp\Big(e^{\Lambda_0\tau}h_b\big(p(\Lambda_0)\big)\Big)]^{-2}e^{2\Lambda_0\tau}A\tau e^{-A\tau}+o(Ae^{-A\tau}).\nonumber
\ee 
Similarly, the Taylor expansion of $\hat{p}$ is given by
\be\label{eq.taylorhatp}
\hat{p}&=&p(\Lambda_0)+\mu^{*}\big(p(A+\Lambda_0)-p(\Lambda_0)\big)\nonumber\\&=&p(\Lambda_0)+\Big\{1-[1+\exp\Big(e^{\Lambda_0\tau}h_b\big(p(\Lambda_0)\big)\Big)]^{-1}e^{\Lambda_0\tau}\Big\}e^{-\Lambda_0\tau}\nonumber\\&&+[1+\exp\Big(e^{\Lambda_0\tau}h_b\big(p(\Lambda_0)\big)\Big)]^{-2}e^{\Lambda_0\tau}A\tau e^{-A\tau}+o(Ae^{-A\tau}).
\ee 

Based on Equations (\ref{eq.taylormu}) and (\ref{eq.taylorhatp}) and $h_b\big(p(A+\Lambda_0)\big)=O(Ae^{-A\tau})$, the asymptotic capacity is given as follows
\be 
\lim\limits_{A\to\infty}C_{\tau,\tau}&=&\frac{1}{\tau}\lim\limits_{A\to\infty}F(\mu^{*})=\frac{1}{\tau}\lim\limits_{A\to\infty}h_b(\hat{p})-(1-\mu^{*})h_b\big(p(\Lambda_0)\big)-\mu^{*} h_b\big(p(A+\Lambda_0)\big)\nonumber\\&=&\frac{1}{\tau}\Bigg\{\lim\limits_{A\to\infty}\Big\{h_b\Bigg(p(\Lambda_0)+\Big\{1-[1+\exp\Big(e^{\Lambda_0\tau}h_b\big(p(\Lambda_0)\big)\Big)]^{-1}e^{\Lambda_0\tau}\Big\}e^{-\Lambda_0\tau}\Bigg)+O(Ae^{-A\tau})\Big\}\nonumber\\&&-(1-\mu^{*})h_b\big(p(\Lambda_0)\big)\Bigg\} 
\nonumber\\&=&c_{\Lambda_0}\frac{1}{\tau},
\ee 
where $c_{\Lambda_0}=h_b\Big(\frac{\exp\big(e^{\Lambda_0\tau}h_b\big(p(\Lambda_0)\big)\big)}{1+\exp\big(e^{\Lambda_0\tau}h_b\big(p(\Lambda_0)\big)\big)}\Big)-\frac{h_b\big(p(\Lambda_0)\big)e^{\Lambda_0\tau}}{\Big(1+\exp\big(e^{\Lambda_0\tau}h_b\big(p(\Lambda_0)\big)\big)\Big)}$.
\subsection{Proof of Theorem \ref{theo.sisoasymclambda}}\label{appen.sisoasymclambda}
Note that $c_{\Lambda_0}=h_b(\frac{u}{1+u})-\frac{\ln u}{1+u}$, where $u=\exp\big(e^{\Lambda_0\tau}h_b\big(p(\Lambda_0)\big)\big)$, we have 
\be\label{eq.appenclamd}
\frac{\partial c_{\Lambda_0}}{\partial u}=-\frac{1}{u(1+u)}<0.
\ee
Subsequently, we focus on the monotonicity of $u$ with respect to $\Lambda_0$. Define $v(x)=\frac{h_b(x)}{x},x\in(0,1)$, and we have $v^{'}(x)=\frac{\ln(1-x)}{x^2}<0$. Since $e^{\Lambda_0\tau}h_b\big(p(\Lambda_0)\big)=v(e^{-\Lambda_0\tau})$ and $e^{-\Lambda_0\tau}$ monotonically decreases with $\Lambda_0$, we have $\frac{\partial u}{\partial \Lambda_0}>0$ and $c_{\Lambda_0}$ monotonically decreases with $\Lambda_0$. Such monotonically decreasing property aligns with the intuition since larger background radiation $\Lambda_0$ leads to more capacity loss.

For $\Lambda_0=0$, it is easy to check that $c_{\Lambda_0}=1$. According to monotone convergence theorem, the limitation of $c_{\Lambda_0}$ for large background radiation $\Lambda_0$ exists. Since $v(x)=1-\ln x+o(1)$ for small $x$, and due to the continuity of $\exp(\cdot)$ and $v(\cdot)$, we have $\lim\limits_{\Lambda_0\to\infty}u=\exp\big(\lim\limits_{\Lambda_0\to\infty}v(e^{-\Lambda_0\tau})\big)=+\infty$. Similarly, according to monotone convergence theorem and equation (\ref{eq.appenclamd}), we have $\lim\limits_{\Lambda_0\to\infty}c_{\Lambda_0}=\lim\limits_{u\to\infty}h_b(\frac{u}{1+u})-\frac{\ln u}{1+u}=0$.

\subsection{Proof of Theorem \ref{theo.sisoasymclambda2}}\label{appen.sisoasymclambda2}
For continuous Poisson channel and peak power constraint, according to \cite{wyner1988capacity}, the capacity is given by $C_{Poi}=A[q^{*}(1+s)\ln(1+s)+(1-q^{*})s\ln s-(q^{*}+s)\ln(q^{*}+s)]$, As $s=\frac{\Lambda_0}{A}$ and $q^{*}=\frac{(1+s)^{(1+s)}}{s^se}-s$. when $s\to+\infty$ (i.e., low SNR), we have $q^{*}=\frac{1}{2}+O(\frac{1}{s})$. Considering the asymptotic capacity for small $A$, we have
\be 
C_{Poi}&=&As[q^{*}(1+s^{-1})\ln(1+s)+(1-q^{*})\ln s-(1+q^{*}s^{-1})\ln(q^{*}+s)]\nonumber\\
&=&\Lambda_0[-\ln(q^{*}+s)+q^{*}\ln(1+s)+(1-q^{*})\ln s-q^{*}s^{-1}\ln(q^{*}+s)+q^{*}s^{-1}\ln(1+s)]\nonumber\\
&=&\Lambda_0[-\ln(1+\frac{q^{*}}{s})+q^{*}\ln(1+\frac{1}{s})+\frac{q^{*}}{s}\ln (1+\frac{1-q^{*}}{q^{*}+s})]\nonumber\\
&=&\frac{q^{*}(1-q^{*})}{2s^2}+o(s^{-2})=\frac{1}{8\Lambda_0}A^2+o(A^2).
\ee 

Similarly, Taylor expansion is adopt to calculate the asymptotic capacity of non-perfect receiver for small $A$. The main clue is to obtain the Taylor expansion of $\frac{h_b\big(p(A+\Lambda_0)\big)-h_b\big(p(\Lambda_0)\big)}{p(A+\Lambda_0)-p(\Lambda_0)}$, $a$, $\mu^{*}$, $\hat{p}$, and $C_{\tau,\tau}$, one by one.

Since $\frac{f(x)-f(y)}{x-y}=f^{'}(y)+\frac{f^{''}(y)}{2}(x-y)+\frac{f^{'''}(y)}{6}(x-y)^2+o\big((x-y)^2\big)$ for differentiable function $f(\cdot)$, we have
\be\label{eq.appentaylor0}
\frac{h_b\big(p(A+\Lambda_0)\big)-h_b\big(p(\Lambda_0)\big)}{p(A+\Lambda_0)-p(\Lambda_0)}&=&h_b^{'}\big(p(\Lambda_0)\big)+\frac{h_b^{''}\big(p(\Lambda_0)\big)}{2}\big(1-p(\Lambda_0)\big)A\tau\nonumber\\&&+\frac{h_b^{'''}\big(p(\Lambda_0)\big)}{6}\big(1-p(\Lambda_0)\big)^2A^2\tau^2+o(A^2),
\ee
As $h_b^{'}(x)=\ln\frac{1-x}{x}$, $h_b^{''}(x)=-\frac{1}{x(1-x)}$, $h_b^{'''}(x)=\frac{1}{x^2}-\frac{1}{(1-x)^2}$ and 
\be 
&&\exp\big(-(a_0+a_1\Delta x+a_2\Delta^2 x+o(\Delta^2 x))\big)\nonumber\\&=&\exp(-a_0)-\exp(-a_0)a_1\Delta x+\exp(-a_0)\Delta^2 x\big(-a_2+\frac{a_1^2}{2}\big)+o(\Delta^2 x),
\ee
based on equation (\ref{eq.appentaylor0}), the Taylor  expression of $a$ and $\frac{a}{1+a}$ are given by
\be 
a&=&\exp\Big(-\frac{h_b\big(p(A+\Lambda_0)\big)-h_b\big(p(\Lambda_0)\big)}{p(A+\Lambda_0)-p(\Lambda_0)}\Big)\nonumber\\
&=&\exp\Big(-h_b^{'}\big(p(\Lambda_0)\big)\Big)-\exp\Big(-h_b^{'}\big(p(\Lambda_0)\big)\Big)\frac{h_b^{''}\big(p(\Lambda_0)\big)}{2}\big(1-p(\Lambda_0)\big)A\tau\nonumber\\
&&+\exp\Big(-h_b^{'}\big(p(\Lambda_0)\big)\Big)A^2\tau^2\Big\{-\frac{h_b^{'''}\big(p(\Lambda_0)\big)}{6}\big(1-p(\Lambda_0)\big)^2+\frac{1}{2}\big[\frac{h_b^{''}\big(p(\Lambda_0)\big)}{2}\big(1-p(\Lambda_0)\big)\big]^2\Big\}+o(A^2)\nonumber\\
&=&\frac{p(\Lambda_0)}{1-p(\Lambda_0)}+\frac{A\tau}{2\big(1-p(\Lambda_0)\big)}+\frac{p(\Lambda_0)}{1-p(\Lambda_0)}A^2\tau^2\Big\{-\frac{1}{6}[p^{-2}(\Lambda_0)-\big(1-p(\Lambda_0)\big)^{-2}]\big(1-p(\Lambda_0)\big)^{2}\nonumber\\
&&+\frac{1}{8}\frac{\big(1-p(\Lambda_0)\big)^{2}}{p^2(\Lambda_0)\big(1-p(\Lambda_0)\big)^{2}}\Big\}+o(A^2)\nonumber\\
&=&\frac{p(\Lambda_0)}{1-p(\Lambda_0)}+\frac{A\tau}{2\big(1-p(\Lambda_0)\big)}+\frac{8p(\Lambda_0)-1}{24p(\Lambda_0)\big(1-p(\Lambda_0)\big)}A^2\tau^2+o(A^2).
\ee 
Since $\frac{t+\Delta t}{1+t+\Delta t}=\frac{t}{1+t}+(1+t)^{-2}\Delta t-2(1+t)^{-3}\Delta^2 t+o(\Delta^2 t)$, the Taylor expansion of $\frac{a}{1+a}$ and $\mu^{*}$ are given by
\be 
\frac{a}{1+a}&=&p(\Lambda_0)+\big(1-p(\Lambda_0)\big)^2\frac{A\tau}{2\big(1-p(\Lambda_0)\big)}+\big(1-p(\Lambda_0)\big)^2\frac{8p(\Lambda_0)-1}{24p(\Lambda_0)\big(1-p(\Lambda_0)\big)}A^2\tau^2\nonumber\\
&&-\big(1-p(\Lambda_0)\big)^3+\frac{A^2\tau^2}{4\big(1-p(\Lambda_0)\big)^2}\nonumber\\
&=&p(\Lambda_0)+\big(1-p(\Lambda_0)\big)\frac{A\tau}{2}+\frac{2p(\Lambda_0)-1}{24p(\Lambda_0)}\big(1-p(\Lambda_0)\big)A^2\tau^2+o(A^2),\\
\mu^{*}&=&\frac{\frac{a}{1+a}-p(\Lambda_0)}{p(A+\Lambda_0)-p(\Lambda_0)}=\frac{\big(1-p(\Lambda_0)\big)\frac{A\tau}{2}+\frac{2p(\Lambda_0)-1}{24p(\Lambda_0)}\big(1-p(\Lambda_0)\big)A^2\tau^2}{\big(1-p(\Lambda_0)\big)A\tau}+o(A)\nonumber\\
&=&\frac{1}{2}+\frac{2p(\Lambda_0)-1}{24p(\Lambda_0)}A\tau+o(A).\label{eq.taylormua}
\ee 
Based on equation (\ref{eq.taylormua}), we have the Taylor expansion of $\hat{p}$ as follows,
\be 
\hat{p}&=&p(\Lambda_0)+\mu^{*}\big(p(A+\Lambda_0)-p(\Lambda_0)\big)\nonumber\\
&=&p(\Lambda_0)+\frac{1}{2}\big(1-p(\Lambda_0)\big)A\tau+\Big[\frac{2p(\Lambda_0)-1}{24p(\Lambda_0)}\big(1-p(\Lambda_0)\big)-\frac{\big(1-p(\Lambda_0)\big)}{4}\Big]A^2\tau^2+o(A^2)\nonumber\\
&=&p(\Lambda_0)+\frac{1}{2}\big(1-p(\Lambda_0)\big)A\tau+\frac{-4p(\Lambda_0)-1}{24p(\Lambda_0)}\big(1-p(\Lambda_0)\big)A^2\tau^2+o(A^2).
\ee 
To obtain the asymptotic capacity with non-perfect receiver $C_{Poi}$, the Taylor expansion of $h_b(\hat{p})$ is given as follows,
\be\label{eq.taylorhbp} 
h_b(\hat{p})&=&h_b\big(p(\Lambda_0)\big)+h_b^{'}\big(p(\Lambda_0)\big)\frac{1}{2}\big(1-p(\Lambda_0)\big)A\tau+\Big[h_b^{'}\big(p(\Lambda_0)\big)\frac{-4p(\Lambda_0)-1}{24p(\Lambda_0)}\big(1-p(\Lambda_0)\big)\nonumber\\
&&+\frac{1}{8}h_b^{''}\big(p(\Lambda_0)\big)\big(1-p(\Lambda_0)\big)^2\Big]A^2\tau^2+o(A^2).
\ee 
Similarly,  $\mu^{*}h_b\big(p(A+\Lambda_0)\big)+(1-\mu^{*})h_b\big(p(\Lambda_0)\big)$ is given by
\be\label{eq.taylorentro}
&&\mu^{*}h_b\big(p(A+\Lambda_0)\big)+(1-\mu^{*})h_b\big(p(\Lambda_0)\big)\nonumber\\&=&h_b\big(p(\Lambda_0)\big)+\Big(\frac{1}{2}+\frac{2p(\Lambda_0)-1}{24p(\Lambda_0)}A\tau+o(A)\Big)\Big\{h_b^{'}\big(p(\Lambda_0)\big)\big(1-p(\Lambda_0)\big)(A\tau-\frac{1}{2}A^2\tau^2)\nonumber\\&&+h_b^{''}\big(p(\Lambda_0)\big)\frac{1}{2}\big(1-p(\Lambda_0)\big)^2A^2\tau^2+o(A^2)\Big\}\nonumber\\
&=&h_b\big(p(\Lambda_0)\big)+h_b^{'}\big(p(\Lambda_0)\big)\frac{1}{2}\big(1-p(\Lambda_0)\big)A\tau+\Big\{h_b^{'}\big(p(\Lambda_0)\big)\big(1-p(\Lambda_0)\big)\frac{-4p(\Lambda_0)-1}{24p(\Lambda_0)}\nonumber\\&&+h_b^{''}\big(p(\Lambda_0)\big)\frac{1}{4}\big(1-p(\Lambda_0)\big)^2\Big\}A^2\tau^2+o(A^2).
\ee 
Based on Equations (\ref{eq.taylorhbp}) and (\ref{eq.taylorentro}), the asymptotic capacity $C_{\tau,\tau}$ is given as follows,
\be 
C_{\tau,\tau}&=&\frac{1}{\tau}\Big\{h_b(\hat{p})-(1-\mu)h_b\big(p(\Lambda_0)\big)-\mu h_b\big(p(A+\Lambda_0)\big)\Big\}\nonumber\\
&=&\frac{1}{\tau}\Big\{-\frac{h_b^{''}\big(p(\Lambda_0)\big)}{8}\big(1-p(\Lambda_0)\big)^2\Big\}A^2\tau^2+o(A^2)=\frac{\tau\big(1-p(\Lambda_0)\big)}{8p(\Lambda_0)}A^2+o(A^2).
\ee 
\subsection{Proof of Theorem \ref{theo.sisoasymefficient}}\label{appen.sisoasymefficient}
Defining $f(t)\dff t-t\ln t-1$, $t\in(0,1)$, we have $f^{'}(t)=-\ln t>0$. Since $\lim\limits_{t\to 0}f(t)=-1$ and $\lim\limits_{t\to 1}f(t)=0$, we have
$\frac{-t\ln t}{1-t}<1$ holds for $t\in(0,1)$. Let $t=e^{-\Lambda_0\tau}=1-p(\Lambda_0)$, we have $\ln t=-\Lambda_0\tau$ and $d_{\tau}=\frac{\tau\big(1-p(\Lambda_0)\big)}{8p(\Lambda_0)}<\frac{1}{8\Lambda_0}=d_{Poi}$.

For any $\Lambda_0>0$, we have
\be 
\lim\limits_{\tau\to0}\frac{d_{\tau}}{d_{Poi}}=\lim\limits_{\tau\to0}\frac{\Lambda_0\tau\big(1-p(\Lambda_0)\big)}{p(\Lambda_0)}\xlongequal{t=e^{-\Lambda_0\tau}}\lim\limits_{t\to1}\frac{-t\ln t}{1-t}=1.
\ee 
\subsection{Proof of Theorem \ref{theo.sisomonopower}}\label{appen.sisomonopower}
Since $C_{\tau,\tau}(A,\Lambda_0)=\frac{1}{\tau}\Big\{h_b(\hat{p})-(1-\mu^{*})h_b\big(p(\Lambda_0)\big)-\mu^{*} h_b\big(p(A+\Lambda_0)\big)\Big\}$ and $h_b^{'}(\hat{p})=\frac{h_b\big(p(A+\Lambda_0)\big)-h_b\big(p(\Lambda_0)\big)}{p(A+\Lambda_0)-p(\Lambda_0)}$, we have
\be\label{eq.firstdericapa} 
\frac{\partial C_{\tau,\tau}(A,\Lambda_0)}{\partial A}&=&\frac{1}{\tau}\Big\{h_b^{'}(\hat{p})\big[\mu^{*}\big(1-p(A+\Lambda_0)\big)\tau+\frac{\partial \mu^{*}}{\partial A}\big(p(A+\Lambda_0)-p(\Lambda_0)\big)\big]\nonumber\\&&-\mu^{*}h_b^{'}\big(1-p(A+\Lambda_0)\big)\tau-\frac{\partial \mu^{*}}{\partial A}\Big(h_b\big(p(A+\Lambda_0)\big)-h_b\big(p(\Lambda_0)\big)\Big)\Big\}\nonumber\\
&=&\mu^{*}\big(1-p(A+\Lambda_0)\big)\Big(h_b^{'}(\hat{p})-h_b^{'}\big(p(A+\Lambda_0)\big)\Big)>0,
\ee 
where the last inequality is satisfied since $h_b^{''}<0$ and $\hat{p}<p(A+\Lambda_0)$. Thus, $C_{\tau,\tau}(A,\Lambda_0)$ strictly increases with peak power $A$.

Further from Equation (\ref{eq.firstdericapa}), we have
\be 
\frac{\partial^2 C_{\tau,\tau}(A,\Lambda_0)}{\partial A^2}&=&\mu^{*}\big(1-p(A+\Lambda_0)\big)\Big[\frac{h_b\big(p(A+\Lambda_0)\big)-h_b\big(p(\Lambda_0)\big)}{p(A+\Lambda_0)-p(\Lambda_0)}-\ln\frac{1-p(A+\Lambda_0)}{p(A+\Lambda_0)}\Big]\nonumber\\&&+\mu^{*}\big(1-p(A+\Lambda_0)\big)^2\Big[-\frac{1}{p(A+\Lambda_0)-p(\Lambda_0)}\Big(\frac{h_b\big(p(A+\Lambda_0)\big)-h_b\big(p(\Lambda_0)\big)}{p(A+\Lambda_0)-p(\Lambda_0)}\nonumber\\&&\underbrace{-\ln\frac{1-p(A+\Lambda_0)}{p(A+\Lambda_0)}\Big)+\frac{1}{p(A+\Lambda_0)\Big(1-p(A+\Lambda_0)\Big)}\Big]}_{I_1}\nonumber\\&&+\underbrace{\frac{\partial \mu^{*}}{\partial A}\big(1-p(A+\Lambda_0)\big)\Big[\frac{h_b\big(p(A+\Lambda_0)\big)-h_b\big(p(\Lambda_0)\big)}{p(A+\Lambda_0)-p(\Lambda_0)}-\ln\frac{1-p(A+\Lambda_0)}{p(A+\Lambda_0)}\Big]}_{I_2}.\nonumber
\ee 
Since $\lim\limits_{x\to0}xh_b^{'}(x)=\lim\limits_{x\to0}x\ln\frac{1-x}{x}=0$ and $\lim\limits_{A\to\infty}a=\exp\Big(e^{\Lambda_0\tau}h_b\big(p(\Lambda_0)\big)\Big)$, we have
\be 
&&\lim\limits_{A\to\infty}\frac{\partial a}{\partial A}\nonumber\\&=&\lim\limits_{A\to\infty}-a\frac{h_b^{'}\big(p(A+\Lambda_0)\big)\big(p(A+\Lambda_0)-p(\Lambda_0)\big)-\Big(h_b\big(p(A+\Lambda_0)\big)-h_b\big(p(\Lambda_0)\big)\Big)}{[p(A+\Lambda_0)-p(\Lambda_0)]^2}[1-p(A+\Lambda_0)]\nonumber\\&=&0.
\ee 
Note that $\frac{\partial \mu^{*}}{\partial A}=-[\frac{a}{1+a}-p(\Lambda_0)]\frac{1-p(A+\Lambda_0)}{[p(A+\Lambda_0)-p(\Lambda_0)]^2}+\frac{1}{p(A+\Lambda_0)-p(\Lambda_0)}\frac{\frac{\partial a}{\partial A}}{(1+a)^2}$ and $1-p(A+\Lambda_0)=e^{-(A+\Lambda_0)\tau}$,
we have $\lim\limits_{A\to\infty}\frac{\partial \mu^{*}}{\partial A}=0$ and $I_2=o\Big(\big(1-p(A+\Lambda_0)\big)\ln\big(1-p(A+\Lambda_0)\big)\Big)$. For term $I_1$, we have
\be 
I_1&=&-\mu^{*}\big(1-p(A+\Lambda_0)\big)\Big\{\Big[\frac{h_b\big(p(A+\Lambda_0)\big)-h_b\big(p(\Lambda_0)\big)}{p(A+\Lambda_0)-p(\Lambda_0)}-\ln\frac{1-p(A+\Lambda_0)}{p(A+\Lambda_0)}\Big]\nonumber\\&&\cdot\big(1-\frac{1-p(A+\Lambda_0)}{p(A+\Lambda_0)-p(\Lambda_0)}\big)+\frac{1}{p(A+\Lambda_0)}\Big\}\nonumber\\&=&\mu^{*}\big\{\big(1-p(A+\Lambda_0)\big)\ln\big(1-p(A+\Lambda_0)\big)+(1-\frac{h_b\big(p(\Lambda_0)\big)}{p(A+\Lambda_0)})\big(1-p(A+\Lambda_0)\big)\big\}\nonumber\\&&+o\big(1-p(A+\Lambda_0)\big).
\ee 
Since $\big(1-p(A+\Lambda_0)\big)=o\Big(\big(1-p(A+\Lambda_0)\big)\ln\big(1-p(A+\Lambda_0)\big)\Big)$ for $p(A+\Lambda_0)\to1$, we have
\be 
&&\frac{\partial^2 C_{\tau,\tau}(A,\Lambda_0)}{\partial A^2}=I_1+I_2\nonumber\\&=&\mu^{*}\big(1-p(A+\Lambda_0)\big)\ln\big(1-p(A+\Lambda_0)\big)+o(\big(1-p(A+\Lambda_0)\big)\ln\big(1-p(A+\Lambda_0)\big)),
\ee 
and there exists $A_{th_{1}}$ such that $\frac{\partial^2 C_{\tau,\tau}(A,\Lambda_0)}{\partial A^2}<0$ holds for any $A\geq A_{th_{1}}$.

According to Equation (\ref{eq.firstdericapa}) and $\lim\limits_{x\to0}xh_b^{'}(x)=0$, we have that for $A\to\infty$,
\be \label{eq.firstdercapa}
\frac{\partial C_{\tau,\tau}(A,\Lambda_0)}{\partial A}=\mu^{*}h_b^{'}(\hat{p})e^{-(A+\Lambda_0)\tau}+O(Ae^{-A\tau})=O(e^{-A\tau}).
\ee 
Based on Equation (\ref{eq.firstdercapa}) and Theorem~\ref{theo.asymcapalargepeak}, there exists $A_{th_{2}}$ so that $\frac{\partial C_{\tau,\tau}/A}{\partial A}=\frac{1}{A}\big(\frac{\partial C_{\tau,\tau}(A,\Lambda_0)}{\partial A}-C_{\tau,\tau}/A\big)<0$ holds for any $A\geq A_{th_{2}}$. 

\subsection{Proof of Theorem \ref{theo.sisomonosmall}}\label{appen.sisomonosmall}
Recall that the capacity with non-perfect receiver $C_{T_s,\tau}=\frac{1}{T_s}F(\mu^{*})$, where $F(\mu)= h_b(\hat{p}(\mu))-(1-\mu)h_b\big(p(\Lambda_0)\big)-\mu h_b\big(p(A+\Lambda_0)\big)$. Since the capacity with non-perfect receiver depends on $T_s$, $A\tau$ and $\Lambda_0\tau$ and the multiplicative symmetry between $(A,\Lambda_0)$ and $\tau$, we have $C_{T_s,\beta\tau}(A,\Lambda_0)=C_{T_s,\tau}(\beta A,\beta\Lambda_0)$, where $\beta$ is the dead time factor satisfying $ T_s\geq\beta\tau\geq\frac{\ln2}{\Lambda_0}$. According to the capacity of the non-perfect receiver, we have
\be 
&&\frac{\partial C_{T_s,\beta\tau}}{\partial \beta}=\frac{\partial C_{T_s,\tau}(\beta A,\beta\Lambda_0)}{\partial \beta}\nonumber\\
&=&\frac{1}{T_s}\Big\{h_{b}^{'}(\hat{p})\Big((1-\mu^{*})\big(1-p(\beta\Lambda_0)\big)\Lambda_0+\mu^{*}\big(1-p(\beta(A+\Lambda_0))\big)(A+\Lambda_0)\nonumber\\
&&+\frac{\partial \mu^{*}}{\partial \beta}\big(p(\beta(A+\Lambda_0))-p(\beta\Lambda_0)\big)\Big)
-(1-\mu^{*})h_{b}^{'}\big(p(\beta\Lambda_0)\big)\big(1-p(\beta\Lambda_0)\big)\Lambda_0\nonumber\\
&&-\mu^{*}h_{b}^{'}\big(p(\beta(A+\Lambda_0))\big)\big(1-p(\beta\Lambda_0)\big)(A+\Lambda_0)-\frac{\partial \mu^{*}}{\partial \beta}\Big(h_b\big(p(\beta(A+\Lambda_0))\big)-h_b\big(p(\beta\Lambda_0)\big)\Big)\Big\}\nonumber\\
&\overset{(a)}{=}&\frac{1}{T_s}\Big\{h_{b}^{'}(\hat{p})\Big((1-\mu^{*})\big(1-p(\beta\Lambda_0)\big)\Lambda_0+\mu^{*}\big(1-p(\beta(A+\Lambda_0))\big)(A+\Lambda_0)\nonumber\\
&&
-(1-\mu^{*})h_{b}^{'}\big(p(\beta\Lambda_0)\big)\big(1-p(\beta\Lambda_0)\big)\Lambda_0-\mu^{*}h_{b}^{'}\big(p(\beta(A+\Lambda_0))\big)\big(1-p(\beta\Lambda_0)\big)(A+\Lambda_0)\Big\}\nonumber\\
&\overset{(b)}{\geq}&-\mu^{*}(1-\mu^{*})\big[h_b^{'}\big(p(\beta\Lambda_0)\big)-h_b^{'}\big(p(\beta(A+\Lambda_0))\big)\big]\big[\big(1-p(\beta\Lambda_0)\big)\Lambda_0-\big(1-p(\beta(A+\Lambda_0))\big)(A+\Lambda_0)\big]\nonumber\\&>&0,
\ee 
where (a) holds since $h_b\big(p(\beta(A+\Lambda_0))\big)-h_b\big(p(\beta\Lambda_0)\big)=h_b^{'}(\hat{p})\Big(p(\beta(A+\Lambda_0))-p(\beta\Lambda_0)\Big)$ based on Theorem~\ref{theor.siso}, and (b) holds since $h_b{'''}(x)=\frac{1}{x^2}-\frac{1}{(1-x)^2}<0$ for $x\geq\frac{1}{2}$ and $h_b^{'}(\hat{p})\geq(1-\mu^{*})h_b^{'}\big(p(\beta\Lambda_0)\big)+\mu^{*}h_b^{'}\big(p(\beta(A+\Lambda_0))\big)$ for $p(\beta\Lambda_0)\geq1-e^{-(A+\Lambda_0)\frac{\ln 2}{A+\Lambda_0}}=\frac{1}{2}$.

\subsection{Proof of Theorem \ref{theo.sisomonolam0}}\label{appen.sisomonolam0}
For $\Lambda_0=0$ and $T_s=\tau$, according to the multiplicative symmetry between $(A,\Lambda_0)$ and $\tau$, we have $C^{\beta\tau,\beta\tau}(A,0)=\beta^{-1}C_{\tau,\tau}(\beta A,0)$.
According to Theorem \ref{theo.sisomonosmall}, $C_{\tau,\tau}(A,0)/A$ decreases with $A$ for any $A\geq A_{th_{2}}$ and thus, $C_{\tau,\tau}(A,0)$ decreases with $\tau$ for any $\tau\geq \frac{A_{th_{2}}}{A}$.

\renewcommand{\baselinestretch}{1.4}
\section{Auxilary Lemma}
\begin{lemma}\label{appenAu.convex}
	Assume function $f(x)$ is strictly convex and its first-order derivative exists. For $x>y$, then we have function $g(x,y)\dff\frac{f(x)-f(y)}{x-y}$ strictly monotonically increases with $x$, strictly monotonically decreases with $y$. To be specific, we have $f^{'}(y)<\frac{f(x)-f(y)}{x-y}<f^{'}(x)$ 
	\begin{proof}
		According to Lagrange mean value theorem, for $x>y$, we have $f(x)-f(y)=f^{'}(\xi)(x-y)<f^{'}(x)(x-y)$, where $y<\xi<x$. Since $g^{'}_x=\frac{f^{'}(x)(x-y)-[f(x)-f(y)]}{(x-y)^2}>0$, function $g(x,y)$ strictly monotonically increases with $x$. Similarly, we have function $g(x,y)$ strictly monotonically decreases with $y$.
		
		Note that function $g(x,y)$ strictly monotonically increases with $x$, we have $f^{'}(x)=\sup\limits_{y:x>y}\frac{f(x)-f(y)}{x-y}>\frac{f(x)-f(y)}{x-y}$ for any $y<x$. Similarly, we have $f^{'}(y)<\frac{f(x)-f(y)}{x-y}$.
	\end{proof}
\end{lemma}

\renewcommand{\baselinestretch}{1.4}
\small{\baselineskip = 10pt
	\bibliographystyle{IEEEtran}
	\bibliography{./achirate}
	
\end{document}

%% file: pream_bm.tex
\newtheorem{prop}{Proposition}

\newtheorem{cor}{Corollary}

\newtheorem{lm}{Lemma}

\newtheorem{thm}{Theorem}

\newcommand{\be}{\begin{eqnarray}}
\newcommand{\ee}{\end{eqnarray}}
\newcommand{\benn}{\begin{eqnarray*}}
\newcommand{\eenn}{\end{eqnarray*}}
\def\IR{\rm I \kern-0.20em R}

\newcommand{\bthm}{\begin{thm}}
\newcommand{\ethm}{\end{thm}}

\newcommand{\bcor}{\begin{cor}}
\newcommand{\ecor}{\end{cor}}
\newcommand{\bprop}{\begin{prop}}
\newcommand{\eprop}{\end{prop}}
\newcommand{\blm}{\begin{lm}}
\newcommand{\elm}{\end{lm}}
\newcommand{\beq}{\begin{equation}}
\newcommand{\eeq}{\end{equation}}
\newcommand{\ber}{\begin{eqnarray}}
\newcommand{\eer}{\end{eqnarray}}

\newcommand{\bproof}{\begin{proof}}
\newcommand{\eproof}{\end{proof}}

\newcommand{\bit}{\begin{itemize}}
\newcommand{\eit}{\end{itemize}}
\newcommand{\ben}{\begin{enumerate}}
\newcommand{\een}{\end{enumerate}}
\newcommand{\bdesc}{\begin{description}}
\newcommand{\edesc}{\end{description}}
\newcommand{\beqarrn}{\begin{eqnarray*}}
\newcommand{\eeqarrn}{\end{eqnarray*}}
\newcommand{\bproofof}{\begin{proofof}}
\newcommand{\eproofof}{\end{proofof}}
\newenvironment{rem}{\begin{trivlist}\item[]{\bf
Remark:}\hspace{4mm}}{\end{trivlist}}
\newcommand{\brem}{\begin{rem}}
\newcommand{\erem}{\end{rem}}
\newenvironment{rems}{\begin{trivlist}\item[]{\bf
Remarks}\begin{itemize}}{\end{itemize}\end{trivlist}}
\newcommand{\brems}{\begin{rems}}
\newcommand{\erems}{\end{rems}}
\newtheorem{fact}{Fact}
\newcommand{\bfact}{\begin{fact}}
\newcommand{\efact}{\end{fact}}
\newtheorem{examp}{Example}
\newcommand{\bexamp}{\begin{examp}\rm}
\newcommand{\eexamp}{\end{examp}}
\newtheorem{defn}{Definition}
\newcommand{\bdefn}{\begin{defn}\rm}
\newcommand{\edefn}{\end{defn}}

\newtheorem{alg}{Algorithm}
\newcommand{\balg}{\begin{alg}}
\newcommand{\ealg}{\end{alg}}

\newtheorem{prob}{Problem}
\newcommand{\bprob}{\begin{prob}}
\newcommand{\eprob}{\end{prob}}

\newcommand{\bvtm}{\begin{verbatim}}
\newcommand{\bfig}{\begin{figure}}
\newcommand{\efig}{\end{figure}}
\newcommand{\bcen}{\begin{center}}
\newcommand{\ecen}{\end{center}}

\long\def\comment#1{}

\def \n2{{N_0 \over 2}}

\def \h5{\hspace{0.5in}}

\newcommand{\dff}{\stackrel{\triangle}{=}}